\documentclass[acmtog]{acmart}

\usepackage{listings,xcolor}
\usepackage{dsfont}

\AtBeginDocument{%
  \providecommand\BibTeX{{%
    \normalfont B\kern-0.5em{\scshape i\kern-0.25em b}\kern-0.8em\TeX}}}

\setcopyright{acmcopyright}
\copyrightyear{2023}
\acmYear{2023}
\acmDOI{XXXXXXX.XXXXXXX}

\acmConference[Conference acronym 'XX]{Make sure to enter the correct
  conference title from your rights confirmation emai}{June 03--05,
  2023}{Woodstock, NY}
\acmPrice{15.00}
\acmISBN{978-1-4503-XXXX-X/18/06}

\usepackage{color}
\usepackage{soul}
\usepackage{multirow}
\usepackage{xcolor}

\newlength\savewidth\newcommand\shline{\noalign{\global\savewidth\arrayrulewidth\global\arrayrulewidth 1pt}\hline\noalign{\global\arrayrulewidth\savewidth}}

\definecolor{turquoise}{cmyk}{0.65,0,0.1,0.1}
\definecolor{purple}{rgb}{0.65,0,0.65}
\definecolor{darkgreen}{rgb}{0.0, 0.5, 0.0}
\definecolor{darkred}{rgb}{0.5, 0.0, 0.0}
\definecolor{darkblue}{rgb}{0.0, 0.0, 0.5}
\definecolor{blue}{rgb}{0.0, 0.0, 1.0}
\definecolor{orange}{rgb}{1.0,0.5,0.0}

\newcommand{\hide}[1]{{}}

\makeatletter
\renewcommand{\paragraph}{%
  \@startsection{paragraph}{4}%
  {\z@}{0.3ex \@plus 1ex \@minus .1ex}{-1em}%
  {\normalfont\normalsize\bfseries}%
}
\makeatother

\newif\ifproofread

\newcommand{\Name}{{Dr.Bokeh}}

\begin{document}

\title{Dr.Bokeh: DiffeRentiable Occlusion-aware Bokeh Rendering}

\author{Yichen Sheng}
\email{sheng30@purdue.edu}
\affiliation{%
  \institution{Purdue University}
  \streetaddress{305 N University St.,}
  \city{West Lafayette}
  \state{Indiana}
  \country{USA}
  \postcode{47907-2021}
}
\author{Zixun Yu}
\email{yu645@purdue.edu}
\affiliation{%
  \institution{Purdue University}
  \streetaddress{305 N University St.,}
  \city{West Lafayette}
  \state{Indiana}
  \country{USA}
  \postcode{47907-2021}
}
\author{Lu Ling}
\email{luling0506@gmail.com}
\affiliation{%
  \institution{Purdue University}
  \streetaddress{305 N University St.,}
  \city{West Lafayette}
  \state{Indiana}
  \country{USA}
  \postcode{47907-2021}
}
\author{Zhiwen Cao}
\email{zhiwenc@adobe.com}
\affiliation{%
  \institution{Adobe}
  \country{USA}
}
\author{Cecilia Zhang}
\email{cecilia77@berkeley.edu}
\affiliation{%
  \institution{Adobe}
  \country{USA}
}
\author{Xin Lu}
\email{xinlu.psu@gmail.com}
\affiliation{%
  \institution{Adobe}
  \country{USA}
}
\author{Ke Xian}
\email{ke.xian@ntu.edu.sg}
\affiliation{%
  \institution{Nanyang Technological University}
  \country{Singapore}
}
\author{Haiting Lin}
\email{linht122@gmail.com}
\affiliation{%
  \institution{Adobe}
  \country{USA}
}
\author{Bedrich Benes}
\email{bbenes@purdue.edu}
\affiliation{%
  \institution{Purdue University}
  \streetaddress{305 N University St.,}
  \city{West Lafayette}
  \state{Indiana}
  \country{USA}
  \postcode{47907-2021}
}

\begin{abstract}
Bokeh is widely used in photography to draw attention to the subject while effectively isolating distractions in the background. 
Computational methods simulate bokeh effects without relying on a physical camera lens. However, in the realm of digital bokeh synthesis, the two main challenges for bokeh synthesis are color bleeding and partial occlusion at object boundaries. Our primary goal is to overcome these two major challenges using physics principles that define bokeh formation. To achieve this, we propose a novel and accurate filtering-based bokeh rendering equation and a physically-based occlusion-aware bokeh renderer, dubbed \Name{}, which addresses the aforementioned challenges during the rendering stage without the need of post-processing or data-driven approaches. Our rendering algorithm first preprocesses the input RGBD to obtain a layered scene representation. 
\Name{} then takes the layered representation and user-defined lens parameters to render photo-realistic lens blur. By softening non-differentiable operations, we make \Name{} differentiable such that it can be plugged into a machine-learning framework. We perform quantitative and qualitative evaluations on synthetic and real-world images to validate the effectiveness of the rendering quality and the differentiability of our method. We show \Name{} not only outperforms state-of-the-art bokeh rendering algorithms in terms of photo-realism but also improves the depth quality from depth-from-defocus.

\end{abstract}



\begin{teaserfigure}
    \centering
    \setlength{\tabcolsep}{1pt}
    \begin{tabular}{cccc}
    \includegraphics[width=0.24\linewidth]{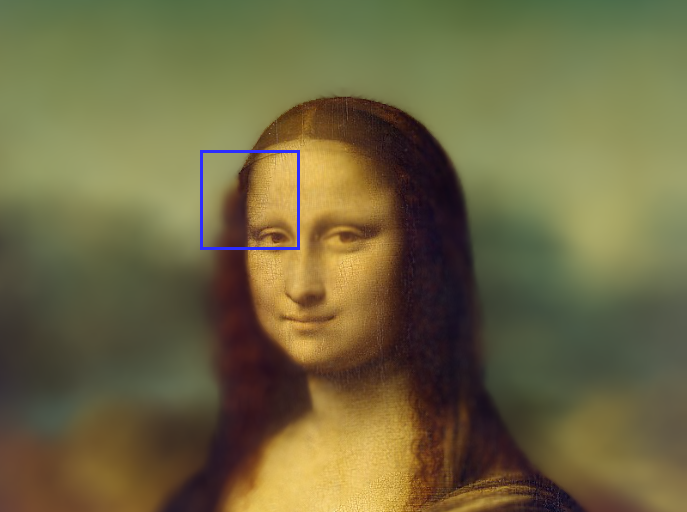} & 
    \includegraphics[width=0.24\linewidth]{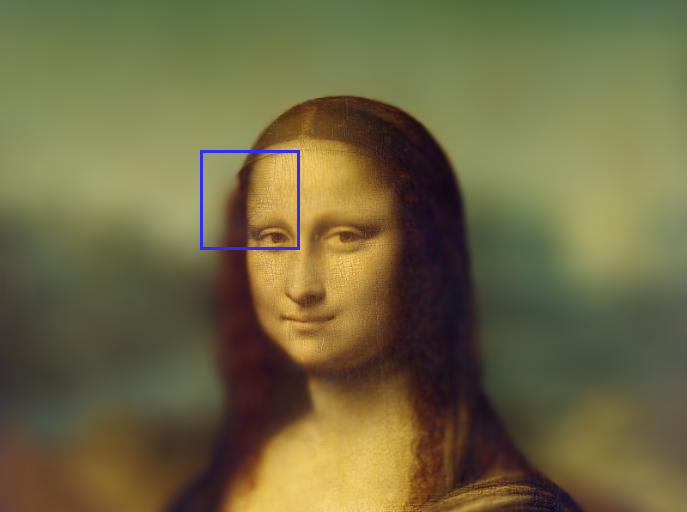} & 
    \includegraphics[width=0.24\linewidth]{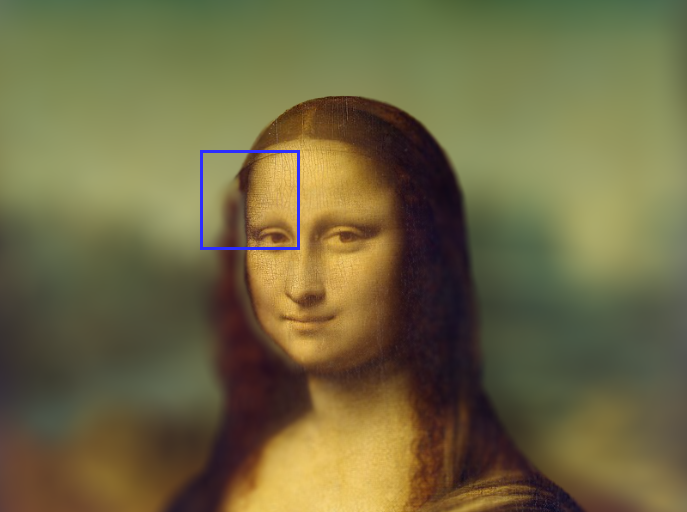} & 
    \includegraphics[width=0.24\linewidth]{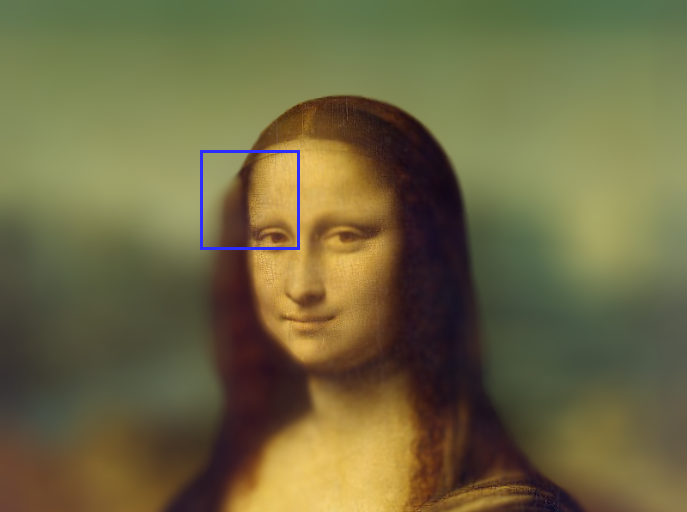} \\
    \includegraphics[width=0.24\linewidth]{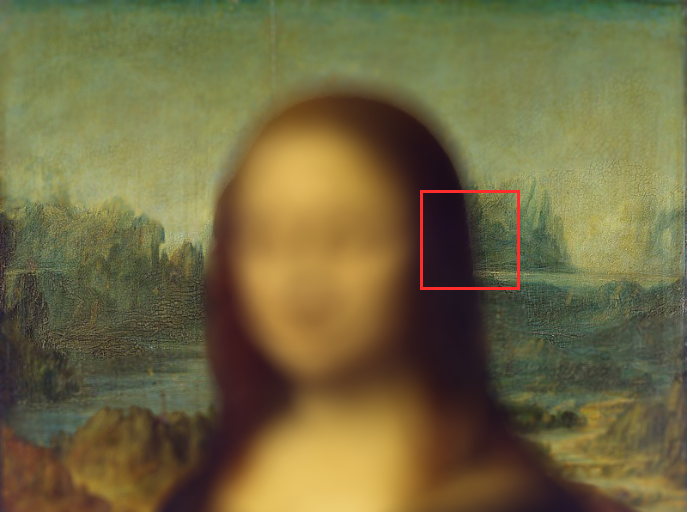} & 
    \includegraphics[width=0.24\linewidth]{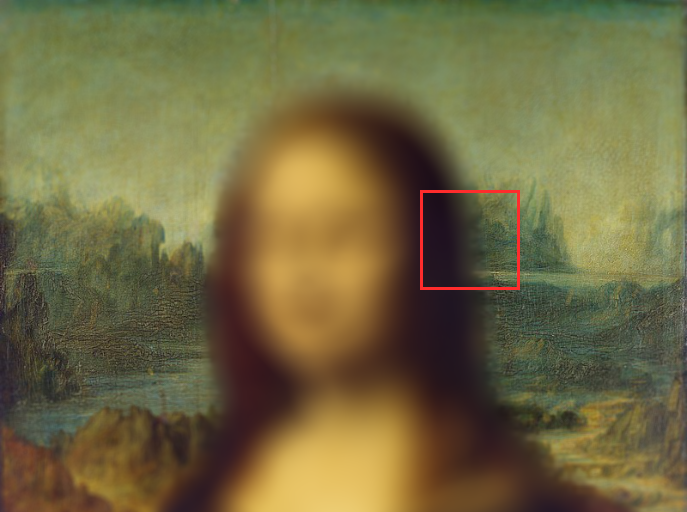} & 
    \includegraphics[width=0.24\linewidth]{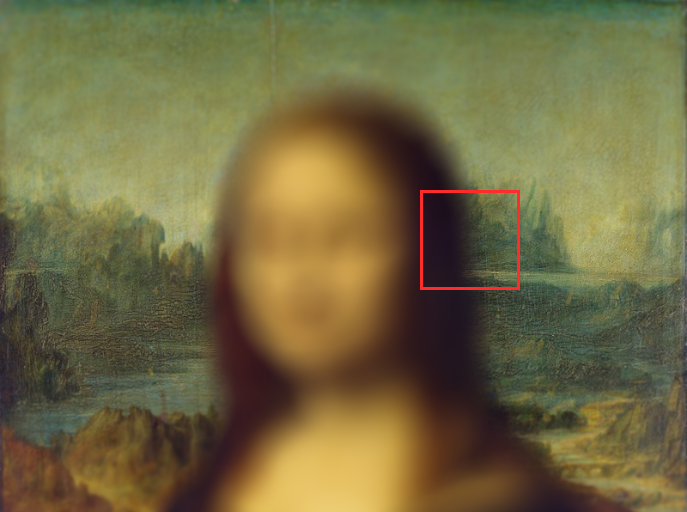} & 
    \includegraphics[width=0.24\linewidth]{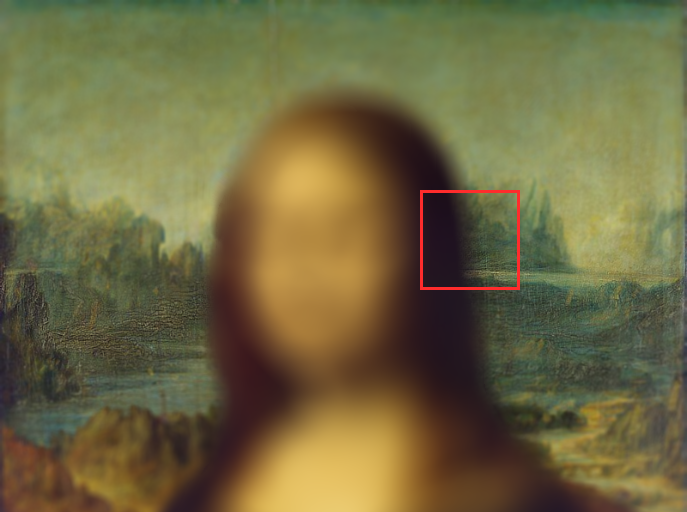} \\
    \includegraphics[width=0.24\linewidth]{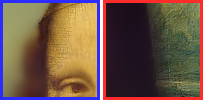} & 
    \includegraphics[width=0.24\linewidth]{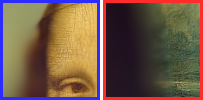} & 
    \includegraphics[width=0.24\linewidth]{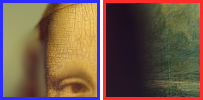} & 
    \includegraphics[width=0.24\linewidth]{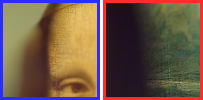} \\
    (a) SteReFo~\shortcite{busamSteReFoEfficientImage2019} & 
    (b) BokehMe~\shortcite{pengBokehMeWhenNeural2022} & 
    (c) MPIB~\shortcite{pengMPIBMPIBasedBokeh2022} & 
    (d) \Name{} (ours)\\
    \end{tabular}
    \caption{By being occlusion-aware, \Name{} renders realistic bokeh effects directly from the bokeh rendering process without any post-processing. Compared with the scattering/gathering-based method SteReFo or learning-based method BokehMe, \Name{} renders natural partial occlusion (see red parts). MPIB learns to render a partial occlusion effect but breaks on unseen data (see blue parts).  
    \Name{} is more robust than learning-based methods given the same inputs because the rendering process is not learned. \textbf{Best viewed by zooming in}.}
    \label{fig:teaser}
\end{teaserfigure}

\maketitle

\section{Introduction}
\begin{figure*}[t]
    \centering
    \includegraphics[width=\linewidth]{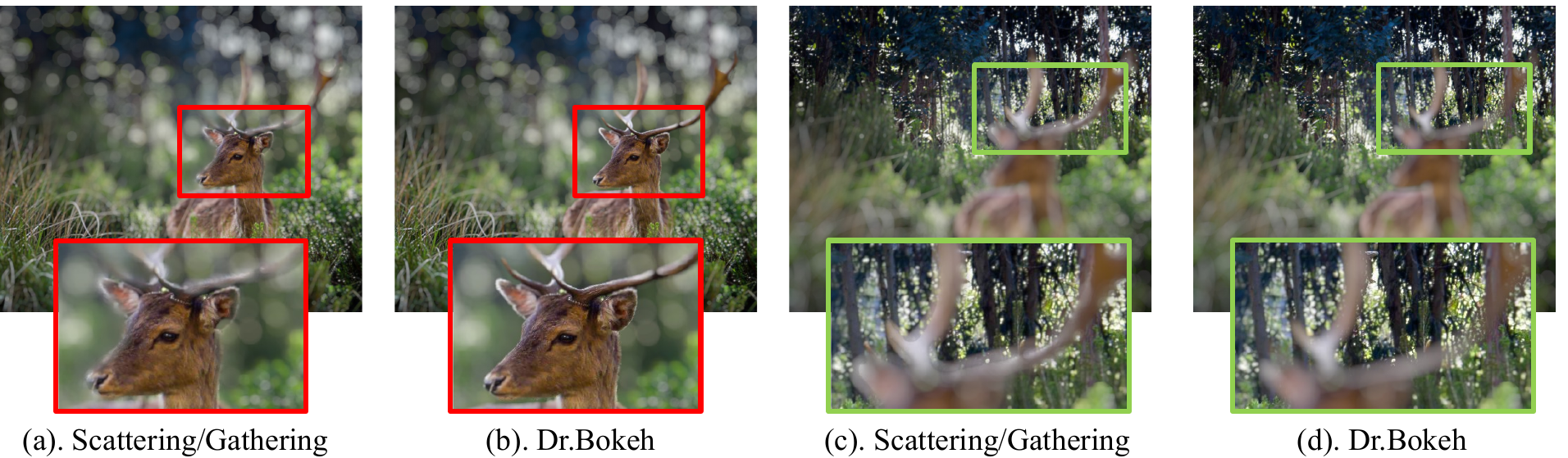}
    \caption{\textbf{Color bleeding and partial occlusion:} Color bleeding is a common artifact meaning the pixels in the out-of-focus scatter to in-focus regions. Partial occlusion is a semi-transparent effect on the out-of-focus boundary regions, mostly in the background in-focus case.  
    (a). na\"ive scattering/gathering suffers from the color bleeding problem. 
    (b). \Name{} solves this issue directly in the rendering process via correctly handling on-focal occlusion.  
    (c). Existing scattering/gathering method does not consider occluded pixels, leading to an unnatural partial occlusion effect. 
    (d). \Name{} addresses this challenge via inpainting occluded pixels and correctly handling non-focal occlusion.
    \textbf{Best viewed by zoom-in.}
    }
    \label{fig:occlusion_type_demo}
\end{figure*}

Bokeh is a physical effect produced by a camera lens, and it usually refers to the shape and quality of out-of-focus areas in an image. 
Such an effect can highlight the in-focus subject and enhance the aesthetic quality of the image. 
Various computational methods have been developed to create the bokeh effect from an all-in-focus photo, and they can be divided into classical and learning-based methods. Classical bokeh rendering methods take the all-in-focus RGB image and its depth map as inputs. 
Image filtering methods, like scattering and gathering operations, are commonly used to synthesize the bokeh effect. However, they often lead to the color bleeding problem (see Fig.~\ref{fig:occlusion_type_demo} (a)) on object boundaries because the filtering-based methods fail to handle the boundary correctly. 
Other methods~\cite{wadhwaSyntheticDepthoffieldSinglecamera2018,krausDepthofFieldRenderingPyramidal2007,zhangSyntheticDefocusLookahead2019} split the image into layers based on depth discontinuity, render bokeh effects on each layer and carefully blend the rendered layers to fix the color bleeding problem. Still, they cannot represent continuous lens blur effects due to the discretization of layers. The learning-based methods~\cite{pengBokehMeWhenNeural2022, wangDeepLensShallowDepth2018,xiaoDeepFocusLearnedImage2018f} compute bokeh by training on artifact-free data. 
However, all the aforementioned methods fail to render the partial occlusion effects naturally (see Fig.~\ref{fig:occlusion_type_demo} (c)) when the background is in-focus as the occluded background geometries contribute to the boundary partial occlusion regions. 

As bokeh is a physical phenomenon, we aim to fundamentally address the color bleeding and rendering partial occlusion challenges from the rendering process. 
We observe that the image filtering-based methods do not strictly follow the physics law, which leads to color bleeding and unnatural partial occlusion. 
First, from the light transportation perspective, the radiance from the background should be fully occluded by the foreground object when the foreground object is in focus (see Sec.~\ref{sec.occ_type}). 
Failure to model this in rendering leads to the color-bleeding artifact.  
Second, in the background in-focus case, the foreground object only partially occludes the radiance behind the boundaries. 
Missing this modeling leads to the unnatural partial occlusion problem. 
The first case shows the boundary occlusion \textit{"hard"} occlude the radiance behind, while the second case shows the boundary occlusion \textit{"soft"} occlude the radiance behind.
No matter which case, we find that the \textbf{occlusion} is the critical factor not addressed by image filtering methods. However, it is crucial in addressing color bleeding and rendering partial occlusion effects for realistic bokeh rendering.

We introduce \Name{}, a novel differentiable (Dr.) bokeh rendering algorithm that correctly handles boundary occlusion in the image-based filtering-based rendering process, is free of the color bleeding artifacts, and renders natural partial occlusion effects (see Fig.~\ref{fig:occlusion_type_demo} (b) and (d)). 
\Name{} does not need to be trained and can directly replace existing bokeh renderer in existing pipelines. 
Although the rendering process is not learning-based, the input information for \Name{}, e.g., depth and inpainted background, is needed and commonly acquired by learning-based methods. 
To support the data-driven framework, we soften the non-differentiable operations in \Name{} and make it fully differentiable so that \Name{} can be directly used in data-driven pipelines for end-to-end training. We validate the rendering quality and differentiability of \Name{} by extensive quantitative and qualitative evaluations on synthetic and real-world datasets. An example in Fig.~\ref{fig:occlusion_type_demo} shows how \Name{} correctly handles the color bleeding problem and renders boundary partial occlusion effects. By correctly handling the boundary occlusion, \Name{} can help the depth-from-defocus community by improving the depth quality as \Name{} follows the physics law. 

The main contributions of this work can be summarized as follows:  
(1) \Name{}, a novel occlusion-aware filtering-based bokeh renderer by introducing geometric occlusion terms. It addresses the color bleeding problem and renders the natural partial-occlusion effects directly in the rendering stage without training.
(2) A differentiable implementation allows plug-and-play bokeh rendering in data-driven pipelines. A carefully designed loss that helps the \Name{} address the depth ambiguity problem in depth-from-defocus problem.

\section{Related Work}
Our work is closely related to lens blur rendering, differentiable rendering, and image inpainting. 

\subsection{Lens blur}
Existing methods of modeling lens blur can be classified into 3D rendering and image space rendering.

\textit{Lens blur in 3D:} Classical graphics rendering~\cite{potmesil1981lens, pharr2016physically, rokita1996generating} creates the optical lens blur through ray tracing of the 3D scene from a configured camera setting with a lens (e.g., a thin lens model). The rendered images naturally contain optical bokeh that is physically accurate to how the camera lens is modeled. Real-time methods~\cite{scheuermann2004advanced, kass2006interactive, goransson2007practical, franke2018multi} for efficient DOF rendering have been explored. Different point spread functions~\cite{lee2008real,lee2009depth, leeRealtimeLensBlur2010, wu2013rendering, lei2013approximate, xu2014depth, yan2015fast} on layered representations are proposed to render the lens blur in hardware rendering pipelines efficiently.
Detailed effects, like lens blur on view-dependent surfaces~\cite{lee2009depth} or lens aberration effects~\cite{leeRealtimeLensBlur2010, wu2013rendering}, or challenging dynamic scenes~\cite{jeong2020real} can also be efficiently synthesized. Light field~\cite{vaidyanathan2015layered} or multiview images~\cite{liu2016depth} render natural partial occlusion effects. 

However, a 3D scene is not always available, and it is computationally expensive to render a fully-converged rendering as the camera sampling space increases.

\textit{Image-space lens blur:} Recently, lens blur has been more efficiently rendered in the image space by applying a depth-dependent blur to in-focus pixels. Classical methods~\cite{krausDepthofFieldRenderingPyramidal2007, yangVirtualDSLRHigh2016, wadhwaSyntheticDepthoffieldSinglecamera2018, zhangSyntheticDefocusLookahead2019} use an RGBD image and create shallow depth-of-field effects using kernel scattering or gathering operations. Na\"ive scattering or gathering-based methods result in color bleed artifacts due to inadequate occlusion handling.
\citeauthor{wadhwaSyntheticDepthoffieldSinglecamera2018}\shortcite{wadhwaSyntheticDepthoffieldSinglecamera2018}, \citeauthor{busamSteReFoEfficientImage2019}\shortcite{busamSteReFoEfficientImage2019} and \citeauthor{zhang2019depth}\shortcite{zhang2019depth} propose a fix to the boundary errors by carefully blending the blurred layers. 
Approached from a different angle, our method delves into the fundamentals of the light propagation process. It proposes a better light transport simulation that naturally avoids color bleeding and produces realistic partial occlusion effects. 

Existing learning-based methods primarily leverage light field rendering and neural rendering. 
Light-field based lens blur rendering~\cite{kalantariLearningbasedViewSynthesis2016, srinivasanApertureSupervisionMonocular2018} predicts scene depth and constructs the 4D light field by warping the all-in-focus image using the depth. 
The shallow depth-of-field image can be approximated by aggregating the light-field images. However, this approach is memory intensive as a high-quality lens blur effect requires a large set of light field images. 
Another class of methods leverages the differentiability of the gathering operation to learn a layered representation and render defocus blur~\cite{srinivasanApertureSupervisionMonocular2018,busamSteReFoEfficientImage2019,luoBokehRenderingDefocus2020}. These approaches assume each layer has a single depth value and apply a fixed blur kernel per layer. 
Other learning-based methods~\cite{ignatov2020rendering, nalbachDeepShadingConvolutional2017, xiaoDeepFocusLearnedImage2018f, wangDeepLensShallowDepth2018, pengMPIBMPIBasedBokeh2022} directly produce a defocused image using deep neural networks.
Recent work of \citeauthor{pengBokehMeWhenNeural2022}~\shortcite{pengBokehMeWhenNeural2022} first uses a classical method to render lens blur and then a neural network to fix the artifacts. Although the neural network corrects color bleeding, the artifacts and corrections are bounded by the training data. Our physically-based occlusion-aware rendering method can directly render realistically defocused images without any post-processing fix (see Fig.~\ref{fig:teaser}). 

Although these methods provide background-focused lens blur rendering, the lack of explicit prediction of the occluded backgrounds leads to unnatural partial occlusion effects (Fig.~\ref{fig:teaser}).
A recent work of~\citeauthor{pengMPIBMPIBasedBokeh2022}~\shortcite{pengMPIBMPIBasedBokeh2022} proposes to explicitly inpaint the occluded background and apply adaptive gathering operations on the multiplane image (MPI)~\cite{tuckerSingleViewViewSynthesis2020b} layers to make the network learn shallow depth-of-field rendering on multiple focal planes.
Due to the generalization limitation of the learning process and the discretization of layered depth, small inconsistent artifacts between two depth layers lead to a noticeable ``leaky'' artifact (see Fig.~\ref{fig:teaser}). 
Our method simulates smooth lens blur based on continuous depth. 
Our blur rendering process does not need to be trained, which makes \Name{} more robust than learning-based methods. Also, \Name{} is differentiable and can be directly plugged into classical lens blur or data-driven pipelines.

\subsection{Differentiable Rendering} 
Differentiable rendering makes the rendering process suitable for inverse problems~\cite{zhang2019differential, zhang2020path, li2018differentiable, zhao2020physics, jakob2022dr, liu2019soft}. 
Instead of handcrafting the rendering equations, others~\cite{eslami2018neural, lombardi2019neural,sheng2021ssn,sheng2022controllable,sheng2023pixht} propose to leverage the neural renderer directly. 
We refer readers for more details to~\cite{kato2020differentiable, tewari2020state}.

Existing methods for lens blur rendering~\cite{srinivasanApertureSupervisionMonocular2018, busamSteReFoEfficientImage2019,kanekoUnsupervisedLearningDepth2021,pengMPIBMPIBasedBokeh2022} rely on the light field or adaptive gathering operators on discrete depth layers to render blur results.
As the process is fully differentiable, depth maps can be obtained via blur supervision. The estimated depth is derived from discrete layers, leading to quantization errors and consequently lacking sufficient depth details. 

Our method can directly replace the blur rendering modules in those methods and output continuous depth.
\citeauthor{gurSingleImageDepth2019}~\shortcite{gurSingleImageDepth2019} propose a differentiable scattering-based bokeh rendering layer with a Gaussian blur kernel to learn continuous depth estimation.
The Gaussian blur kernel is not physically accurate and does not consider the geometry occlusion, leading to inaccurate depth and artifacts on boundaries.   
\Name{} follows physics laws and achieves better depth estimation results. 

\subsection{Image Inpainting} 
Our work is related to image inpainting in terms of how occluded pixels are handled.
Compared with traditional methods~\cite{criminisi2003object, hays2007scene}, CNN-based methods~\cite{iizuka2017globally, pathak2016context,yang2017high} supervised by a GAN loss~\cite{goodfellow2020generative} generate plausible contents using the spatial context. Conventional convolution and different network architectures~\cite{liu2020rethinking, nazeri2019edgeconnect, zhu2021image} have been extended by different convolutions~\cite{liu2018image, yu2019free, suvorov2022resolution} to deal with free-form inpainting mask. 

We directly utilize the off-the-shelf inpainting method~\cite{suvorov2022resolution} to fill in occluded contents. 
\section{Image Space Bokeh Rendering}
We introduce an occlusion-aware filtering-based bokeh rendering method, taking into account of \textit{occlusion} that is missing in point splatting-based methods (scattering or gathering) (Sec.~\ref{Sec. GBRE}). 
An accurate occlusion calculation requires a full 3D scene, either given or reconstructed, which is rarely available in consumer photography.
Instead, we approach the occlusion term from a 2D image space, specifically targeting bokeh rendering. 
We base our modeling on different types of occlusions (Sec.~\ref{sec.occ_type}).
Then we propose our layered bokeh rendering equation by exactly modeling the in-focal occlusion and approximating the non-focal occlusion, which is fundamentally difficult to retrieve (Sec.~\ref{sec.ocbre}).    

\subsection{Occlusion-aware Bokeh Rendering}\label{Sec. GBRE}
\begin{figure}[t]
    \centering
    \includegraphics[width=0.99\linewidth]{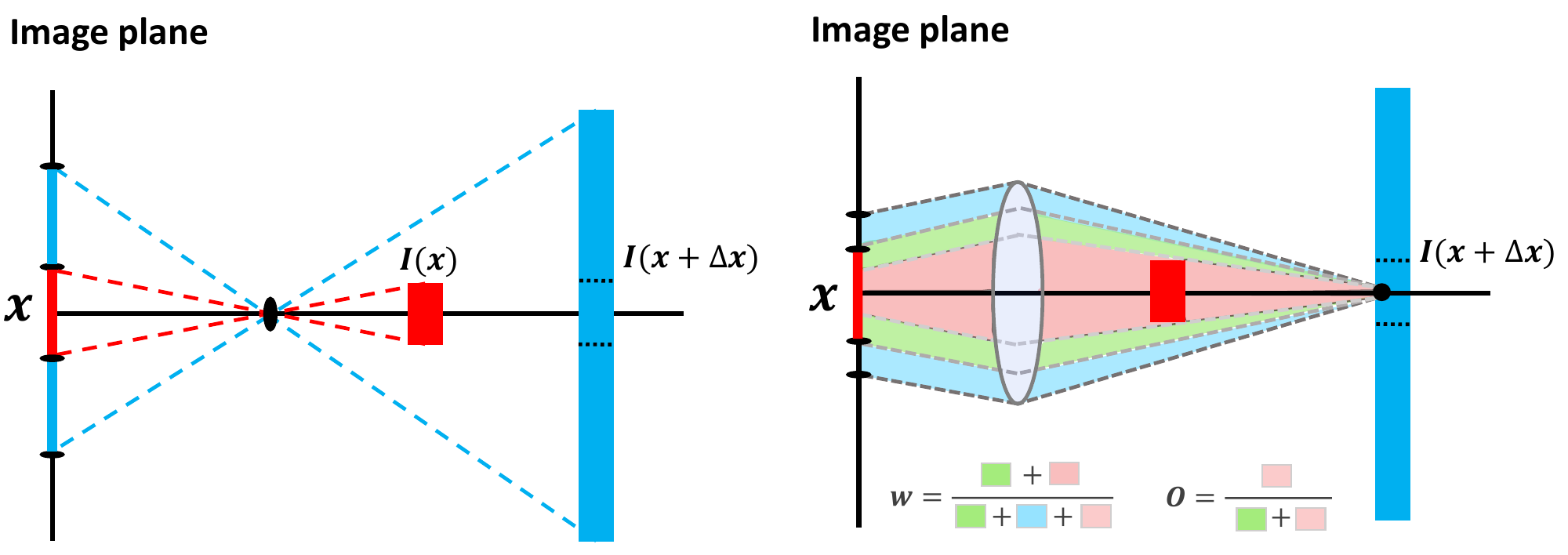}
    \caption{\textbf{Occlusion-based bokeh rendering:} The left image shows that the red and blue objects are projected to the image plane for a pinhole camera. 
    Each pixel in the image space has a corresponding 3D point under the pinhole camera assumption. 
    The right image shows that the point $x+\Delta x$ (originally occluded by $x$ in pinhole camera view) will be visible and contributes to the pixel where $x$ was projected from the scattering perspective. 
    The cone regions of all the colors visualize the regions that $x+\Delta x$ can cover. 
    $w$ is the percentage of energy for $x+\Delta x$ to be scattered to pixel $x$. 
    $O$ is the percentage of energy for $x+\Delta x$ to be scattered to pixel $x$ considering the occlusion by the front red object. 
    }
    \label{fig:occlusion_eqn}
\end{figure}

The filtering-based bokeh rendering method with the occlusion term (see Fig.~\ref{fig:occlusion_eqn}) has the form:   
\begin{equation}
B(x) = \sum\limits_{y \in \Omega(x)} I(y) w(y, x) O(y, x), \label{Eq. GBRE}
\end{equation}
where $B(x)$ is the pixel value for rendered bokeh image, $\Omega(x)$ is the set of all neighborhood pixels of $x$ contributing to the defocus blur, $I(y)$ denotes the in-focus pixel values of $y$, $w(y, x)$ is the energy weight term for pixel $y$ to scatter to $x$, and $O(y, x)$ is the geometry occlusion term for pixel $y$ to scatter to $x$.
 
Existing methods~\cite{pengBokehMeWhenNeural2022, yangVirtualDSLRHigh2016, busamSteReFoEfficientImage2019} often denote $I$ as the all-in-focus image. 
In the generalized rendering equation that we propose, $I(y)$ includes not only the visible neighborhood pixels but also the occluded pixels that contribute to $B(x)$ (see Fig.~\ref{fig:occlusion_eqn}). 

The energy term $0\leq{w}\leq{1}$ is a unitless value that represents the fraction of energy $I(y)$ scattered to $B(x)$. 
The $w$ term is determined jointly by the energy distribution of the scattering region $S$
and the lens shape $K$ as described below. 
To compute the energy distribution of the scattering region $S$, we first calculate the blur kernel radius~$k$ at depth $z_p$ using Lensmaker's formula:
\begin{equation}
k = \alpha \  L\  f\left|\frac{1}{z_p} - \frac{1}{z_f}\right|, \label{Eq. Coc}
\end{equation} 
where $\alpha$ is a configurable scaling factor, $L$ is the lens size, $f$ is the focal length, and $z_f$ is the scene depth at focal length.
Existing gathering-based methods \cite{wadhwaSyntheticDepthoffieldSinglecamera2018, busamSteReFoEfficientImage2019} often discretize the scene into layers and apply convolution with blur kernels whose radius is computed with Eqn.~(\ref{Eq. Coc}). 
If $||x-y|| \geq k$, $w(y)=0$, meaning there is no contribution from the pixel $I(y)$.
These methods make different assumptions on the distribution for pixels with $||x-y|| \leq k$: uniform distribution~\cite{yangVirtualDSLRHigh2016,wadhwaSyntheticDepthoffieldSinglecamera2018}, distance-based~\cite{wadhwaSyntheticDepthoffieldSinglecamera2018}, radial-based~\cite{busamSteReFoEfficientImage2019} have been used to compute the energy term $w$.
The lens shape is another factor that affects~$w$.
Instead of using a circular disk kernel~\cite{wadhwaSyntheticDepthoffieldSinglecamera2018}, stylized lens shapes like a star-, heart-, or triangular-shaped kernels have been used for this term~\cite{yangVirtualDSLRHigh2016}. 
\begin{figure}[t]
    \centering
    \includegraphics[width=\linewidth]{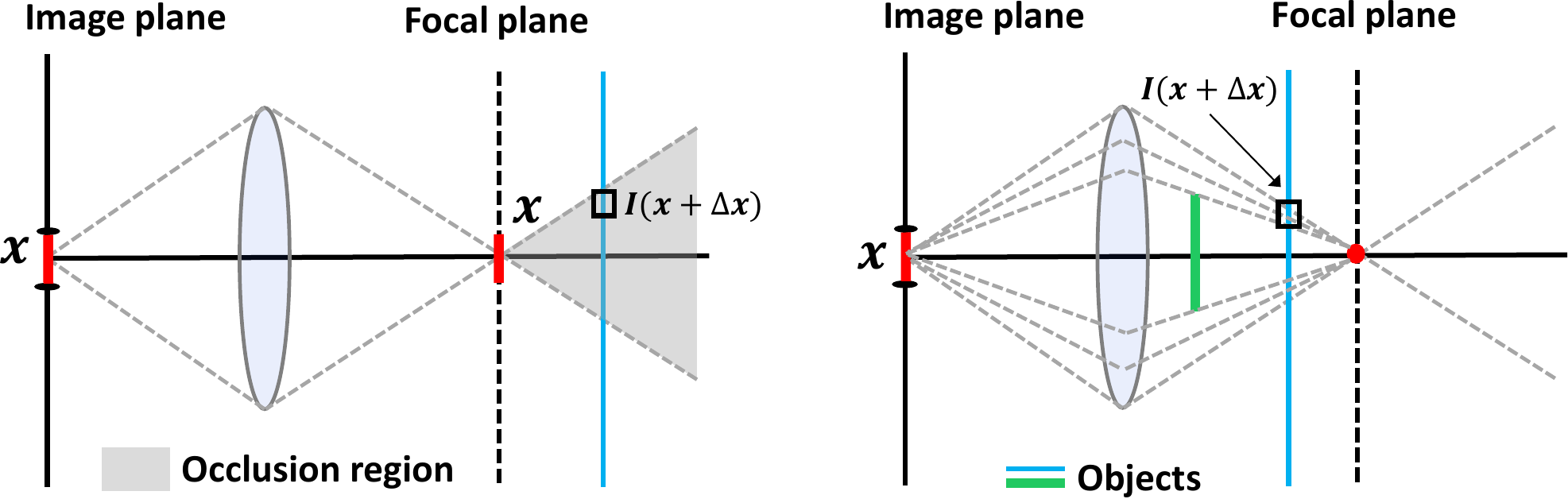}
    \caption{\textbf{Two types of occlusion:} In the occlusion by the focal point (left), the parts behind the red visible object (although as small as a pixel) should not scatter to the position $x$ in the image plane during the bokeh rendering process. In the occlusion by non-focal objects (right) like the green stick, some objects originally occluded by the green objects under the pinhole camera view can contribute to the position $x$ in the image plane.}
    \label{fig:occlusion_type}
\end{figure}

Finally, the newly introduced occlusion term $0\leq{O(x,y)}\leq{1}$ describes whether $I(y)$ is occluded when it scatters to pixel~$x$. 
The occlusion term is critical in bokeh rendering, especially when handling boundary regions, which are sensitive to color bleeding artifacts caused by blending pixels from incorrect depth layers.
As shown in Fig.~\ref{fig:occlusion_type}, ignoring the occlusion term leads to boundary artifacts when focusing on the foreground. 
It also leads to unnatural partial occlusion artifacts when focusing on the background. 
Existing methods~\cite{wadhwaSyntheticDepthoffieldSinglecamera2018, busamSteReFoEfficientImage2019} address the color bleeding problem by pasting back the segmented in-focus object to the final rendered image to override the error regions or using a neural network~\cite{pengBokehMeWhenNeural2022} to correct the errors. 
These methods, however, still fail to model natural partial occlusion effects in the challenging cases of focusing on the background. 
\citeauthor{pengMPIBMPIBasedBokeh2022}~\shortcite{pengMPIBMPIBasedBokeh2022} uses a data-driven method to learn to render realistic partial occlusion. 
Similar to any deep learning methods, \citeauthor{pengMPIBMPIBasedBokeh2022}~\shortcite{pengMPIBMPIBasedBokeh2022} is not robust on unseen data and may break due to generalization issues.  
We propose to address both challenges fundamentally from the rendering process without any post-processing or training of a neural network.

\subsection{Occlusion Types}\label{sec.occ_type}
We classify the occlusion into two types (see Fig.~\ref{fig:occlusion_type}): 
(1)~\textbf{on-focal occlusion}: the occluder is on the focal plane when it blocks all the contributions from behind, and (2)~\textbf{non-focal occlusion}: the occluder is not on the focal plane, and only block a portion of rays. 
Note that the objects invisible to the camera in the all-in-focus image may still contribute to the final lens blur result  (Fig.~\ref{fig:occlusion_type} in the thin-lens camera cases). 
As shown in Fig.~\ref{fig:occlusion_type_demo}, correctly handling \textbf{on-focal occlusion} makes \Name{} free of color bleeding artifacts, and correctly handling the \textbf{non-focal occlusion} helps \Name{} render natural partial occlusion effects. 
\begin{figure}[t]
    \centering
    \includegraphics[width=0.99\linewidth]{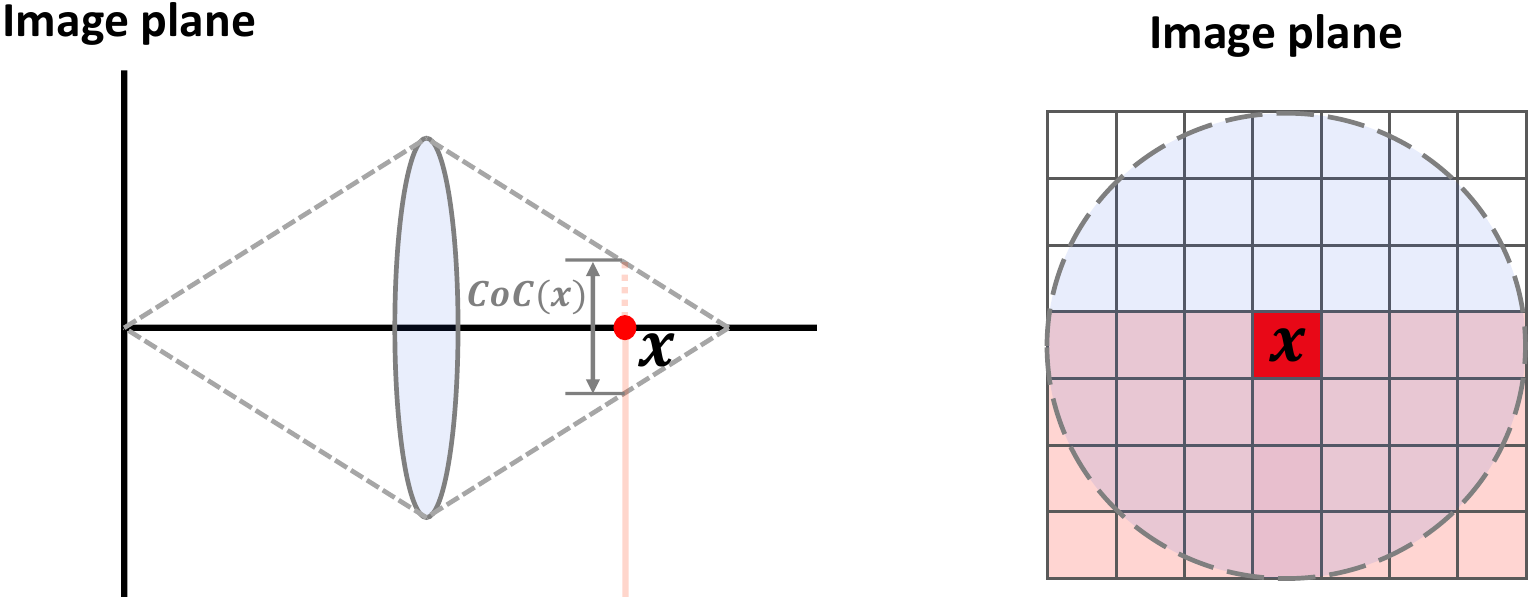}
    \caption{\textbf{Non-focal occlusion:} Given an image plane, a thin lens and a rectangular-shaped pink object in the scene, the left image shows its side view. 
    The right image is the local region for the projected pixel position for $x$ in the image plane.
    The region covered by the blue circle visualizes the CoC region.
    The pink regions are the rectangular-shaped object.
    The inter-layer occlusion is computed by integrating the occluders over the CoC regions for pixel $x$ and is used to occlude the contributions of layers behind.}
    \label{fig:inter-layer-occlusion}
\end{figure}

The first type of occlusion is straightforward and explains why some pixels within the scattering region should not scatter to the foreground when the foreground is in-focus. 
The second type of occlusion needs to be modeled with more care.
When the occluder is not at the focal plane (Fig.~\ref{fig:occlusion_type}), the occluded points, which are invisible in the all-in-focus image, become partially visible, while the occluder becomes semitransparent around its blurry boundary. 
This type of occlusion requires correctly modeling the scene geometries, the lens parameters, and the relative distances between two pixels in the image plane, which is a complex function to model in the image space. 
However, under the planar surface assumption, we can approximate the occlusion in the way shown in Fig.~\ref{fig:inter-layer-occlusion}.
If we assume the scene is composed of planes aligned with the camera view, similar to the pink rectangle in Fig.~\ref{fig:inter-layer-occlusion}, all the points on the pink rectangle are on the same focal plane. They block the rays coming from behind. 
In this special case, the percentage of blocked rays by point $x$ and its neighbors equals the percentage of pixels the object covers in the CoC region, as shown in Fig.~\ref{fig:inter-layer-occlusion}.
Therefore, by searching around the circle of confusion (CoC) region of $x$, we can approximate the percentage of rays coming from behind that will be occluded by the pink object.
In practice, \Name{} applies the aforementioned method to approximate non-focal occlusion computation for non-planar layers.    
Experiments show that our non-focal occlusion approximation still renders realistic bokeh effects.

\subsection{Layered Occlusion-aware Bokeh Rendering Equation}\label{sec.ocbre}
Given the observations in Sec.~\ref{sec.occ_type}, we propose a layered occlusion-aware bokeh rendering equation:

\begin{equation}
    B_l(x) = \sum^n_{l=1} V_l(x) \Pi^{l-1}_{k=1}(1-V_k(x)) \frac{\sum_{y \in \Omega} I_l(y) w_l(y, x) O_l(y, x)}{\sum_{y \in \Omega} w_l(y, x) O_l(y, x)}, \label{Eq. oabre}
\end{equation}    
where $O_l(y)$ and $V_l(x)$ are the in-layer occlusion and inter-layer occlusion, $n$ is the total amount of layers. 
The layered occlusion-aware bokeh rendering equation Eqn.~(\ref{Eq. oabre}) computes the scattering results with in-layer occlusion for each layer, then blends the layers with inter-layer occlusion terms from front to back according to the inter-layer occlusion. 
All the blending weights sum up to one to ensure energy conservation. 
This new rendering method corrects the classical scattering-/gathering-based methods by enabling spatially varying defocus blur with continuous change of blur radius and handles correctly both in-layer occlusion and inter-layer occlusion at the discontinuity boundaries.

Without entire 3D scene geometries, it is challenging to reconstruct all occluded layers given only single image input.  
We propose to approximate the layer representation by peeling the scene at an object level, i.e., one layer is defined by pixels that belong to the same object.

The in-layer occlusion term $O_l(y)$ ensures each layer has the correct on-focal occlusion to avoid any color bleeding problem. 
The inter-layer occlusion term $V_l(x)$ ensures the correct handling of non-focal occlusion and blending of different layers, resulting in natural partial occlusion effects.
In detail, the $V_l(x)$ term decides the occlusion percentage of the radiance behind and also the amount of energy coming from layer $l$.
In Fig.~\ref{fig:inter-layer-occlusion}, $V_l(x)$ for the pink rectangle layer reweights the energy from the rectangle layer scattering and the energy coming from layers behind such that their coefficients sum up to be one, which ensures energy conservation.   

The \textbf{on-focal occlusion} is denoted by $O_l(y, x) \in \{0,1\}$, a unitless binary value that describes the in-layer occlusion between the neighborhood pixels $y$ and $x$: 
\begin{equation}
    O_l(y, x) = 
    \begin{cases}
    0, & \text{if } d_{x}=0 \text{ and }  d_{x} > d_{y} \\
    1, & \text{otherwise}
    \end{cases} \label{Eq. in-layer-occ}
\end{equation}
where $d_x$ is the relative disparity (inverse of depth): $d_x = 1/{z_x} - 1/z_f$.
As shown in Sec.~\ref{sec.occ_type}, the in-layer occlusion only happens when the point corresponding to pixel~$x$ in 3D is on the focal plane and the corresponding 3D point of neighborhood pixel $x+dx$ is behind the 3D point of the pixel $x$. 
In practice, we model $O_l(y, x)$ as a probability instead of a binary value. There are several advantages: 1) The continuous value $O_l(y, x)$ better models the real physics as points not exactly on the focal plane but near the focal plane should partially occlude some amount of radiance from behind; 2) The probability is a "soft" value that models a smooth boundary occlusion. 3) the softened $O_l(y, x)$ is differentiable. The equation details are discussed in Sec.~\ref{sec.soft}  

The \textbf{non-focal occlusion} is denoted by $V_l(x)\in[0, 1]$ is a unitless value that describes the inter-layer \textit{visibility} between layer $l$ and all the layers behind: 
\begin{equation}
V_l(x) = \frac{1}{A_{\Omega'}} \sum_{y \in \Omega'} a_l(y),  \label{Eq. Visible}
\end{equation}
where $\Omega'$ is the set of all the in-layer neighborhood pixels within the circle of confusion (CoC) region for~$x$, $A_{\Omega'}$ is the area of $\Omega'$ and~$a_l$ is the alpha value for the layer $l$. The CoC region can be computed by Eqn.~(\ref{Eq. Coc}). 

The energy term $w_l(y, x)$ for the pixel $y$ in layer $l$ is: 
\begin{equation}
    w_l(y, x) = \frac{S_l(y, x)K(y)a(y)}{A_l(y)} = \frac{\mathds{1}(\|y-x\|<r)\mathds{1}(\|y-x\|<k) a(y)}{\pi r^2},
    \label{Eqn. w}
\end{equation} 
where $a(y)$ is the alpha value of $y$, $r$ is the scatter radius, $k$ is the lens size, and $A_l(y)$ is the area of the CoC of $y$. 
The norm is L2-norm measuring the Euclidean distance, and $w$ is similar to~\cite{lee2008real} as it considers the energy attenuation for different scatter radius $r$. But we further model the lens shape term $K$ to support stylized lens shape. By default, $K$ is a perfect circle as described in Eqn.~(\ref{Eqn. w}).

\section{Differentiable Bokeh Rendering}
Differentiability would allow the bokeh renderer to fit into any neural network for end-to-end training. 
However, the occlusion-aware bokeh rendering method in Sec.~\ref{sec.ocbre} is not differentiable. 
We introduce a fully differentiable bokeh rendering method in Sec.~\ref{sec.soft}, discuss the derivative details in Sec.~\ref{sec.Diff} and show an application that benefits from it: depth from defocus cues in Sec.~\ref{sec.depth-from-defocus}. 

\subsection{Soften the Non-differentiable Operations}\label{sec.soft}
\begin{figure}
    \centering
    \includegraphics[width=0.99\linewidth]{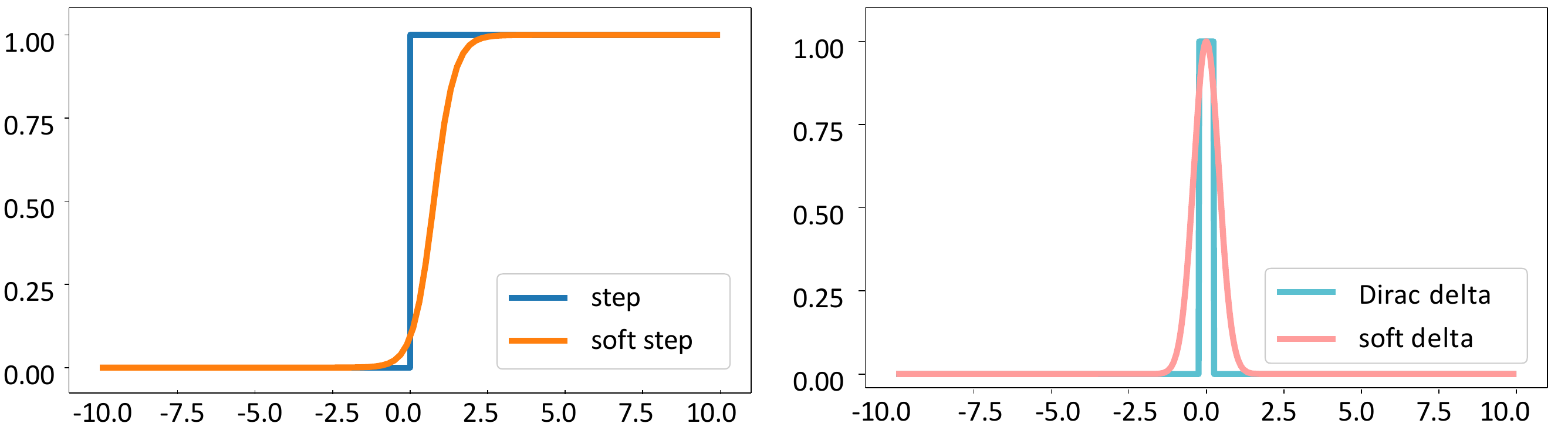}
    \caption{\textbf{Softening of the two non-differentiable operations:} Softening of the non-differentiable step function (left). Note we shift the soft-step function to the right to make sure that for $x=0$, $y$ is close to zero. The right image shows softening of the non-differentiable delta function.}
    \label{fig:diff_operations}
\end{figure}

The occlusion-aware bokeh equation Eqn.~(\ref{Eq. GBRE}) includes two non-differentiable terms: the occlusion term $O$ and the scattering term~$S_l$. The terms are non-differentiable as it involves non-differentiable operations similar to step function or Dirac delta function as shown in Fig.~\ref{fig:diff_operations}. 
We approximate those non-differentiable terms with differentiable operations, e.g., a step function can be approximated with a soft-step function. The occlusion term $O$ is the Dirac delta function $\delta_x$ with a value infinity at zero and zero everywhere else.  
We replace the $O_l$ by:
\begin{equation}
    O_l(y,x) = 1-exp(-3d_x^2)\left(\frac12\tanh(10(d_{y}-d_x - 0.1))-\frac12\right).
\end{equation} \label{Eq. soft O}
The scattering term $S_l$ is a step function(if the neighborhood $x+\Delta x$ can scatter to $x$ then is one, otherwise zero) we replace it by a differentiable function:
\begin{equation}
    S_l(y, x) = 1/\left(1+10\exp\left(-3(\alpha|d_y|+1-\|d_y-d_x\|_2^2)\right)\right),
\end{equation} \label{Eq. soft S}
where $\alpha$ is a camera parameter controlling the blur radius. 
The coefficients in Eqns.~(\ref{Eq. soft O},~\ref{Eq. soft S}) are empirically selected to fit the original function and are reasonable to the bokeh rendering process.

\subsection{Derivatives of the Bokeh Rendering Method}\label{sec.Diff}
The current machine learning frameworks like Pytorch~\cite{paszke2019pytorch} provide automatic differentiation mechanisms for basic mathematical operations.  
\Name{} cannot directly be implemented using provided auto-differentiable layers and need custom forward/backward calculation in the CUDA layer.  
Also, the derivatives, especially w.r.t. disparity, are complicated as most of the terms in the $B(x)$ involve disparity. 
So we derive the derivatives in this section. 

For simplicity, we only derive the partial derivative for a single layer, which is enough for the implementation.  
Multiple layers can be easily derived based on the per-layer partial derivatives. 
The partial derivatives of each RGB channel are similar. 
We use $I$ to denote the three channels. 
According to the chain rule, the partial derivatives with respect to RGB $I$, depth $d$, and alpha $a$ are:
\begin{equation}
    \frac{\partial L}{\partial B(x)} \frac{\partial B(x)}{\partial I(x)} =
    a(x)\sum_{y\in \Omega(x)} \frac{\frac{\partial L}{\partial B(y)} w(x, y) O(x, y)} {\sum_{y' \in \Omega(y) w(y', y) O(y',y)}}.
    \label{Eq. diff_I}
\end{equation}

The partial derivative for $d$ is:  

\begin{multline}
    \frac{\partial L}{\partial B(x)} \frac{\partial B(x)}{\partial d(x)} = 
        \sum_{y\in \Omega}  \frac{\partial L}{\partial B(y)} \frac{I(y) (W(y, x)- w(y, x)O(y, x))}{W(y, x)^2}  \\ 
       \cdot \left( \frac{\partial w(y, x)}{\partial d(x)} O(y, x) + \frac{\partial O(y, x)}{\partial d(x)} w(y, x) \right). \label{Eq. diff_d}
\end{multline}

The partial derivative for $a$ is: 

\begin{multline}
    \frac{\partial L}{\partial B(x)} \frac{\partial B(x)}{\partial a(x)} 
        = \sum_{y \in \Omega(x)}\frac{\partial L}{\partial B(y)}  I(y) O(x, y) \frac{\partial w(x, y)}{\partial a(x)} \\ 
                \cdot  \frac{W(y) - w(x,y) O(x, y)}{W(y)^2},
                \label{Eq. diff_a}
\end{multline}
where the $W(y)$ is: 
\begin{equation}
    W(y) = \sum_{y' \in \Omega(y)} w(y', y) O(y', y) .
\end{equation}
Note that $w(y, x)$ is the energy term for pixel $y$ to scatter to $x$, and $w(x, y)$ is the energy term for $x$ to scatter to $y$. 
So $w(y, x)$ is not equivalent to $w(x, y)$. 
$O$ is similar to $w$ that $O(y, x)$ is not equivalent to $O(x, y)$.   
More details of related terms can be found in Appendix.

\subsection{Depth from Defocus}\label{sec.depth-from-defocus}
\begin{figure}[t]
    \centering
    \includegraphics[width=\linewidth]{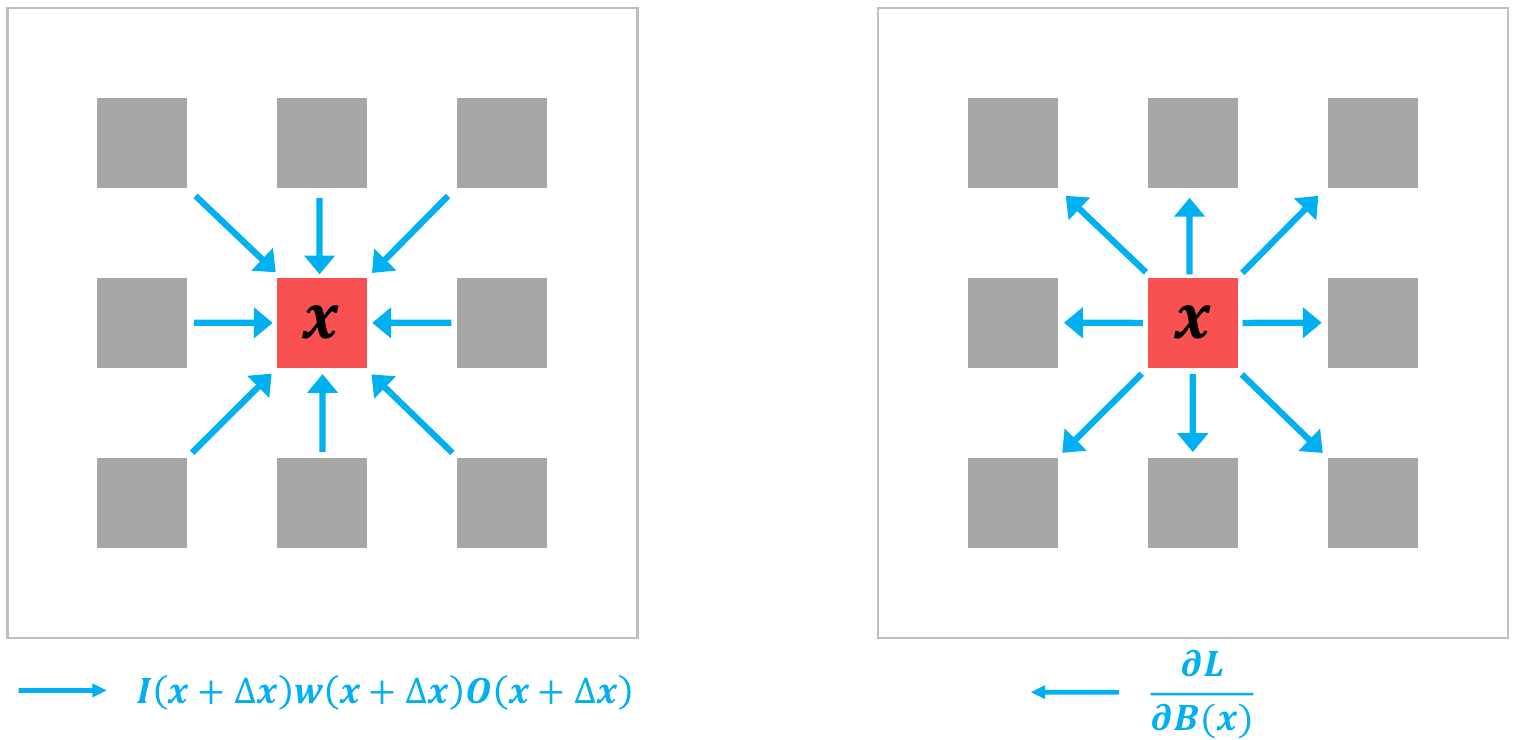}
    \caption{\textbf{Per-pixel loss is not enough:} In the left image, neighborhood pixels jointly contribute to~$x$. There exists a case that the $x$ is correct, i.e. the sum of the contributions is correct, but all the neighborhood values are totally wrong, e.g., neighborhood values are shuffled. In this case, there is no loss for pixel~$x$. Then there is no backward gradient for the neighborhood pixels as shown in the right image, even though the neighborhood pixels are wrong.}
    \label{fig:hierarchy_loss}
\end{figure}
Depth from defocus is a depth estimation method that utilizes the correlation between depth and defocuses blur to train a neural network to estimate depth supervised by blur data~\cite{srinivasanApertureSupervisionMonocular2018, gurSingleImageDepth2019}. 
\Name{} can replace the bokeh rendering module in the previous depth from defocus methods and achieves better depth quality than the state-of-the-art methods.

Depth from defocus needs a special loss function design, and we propose the loss in the following form: 
\begin{equation}
L(y, \hat{y}) = \lambda_1 L_1(y, \hat{y}) + \lambda_2 G(\hat{y}) + \lambda_3 H_{SSIM}(y, \hat{y}), \label{eq.loss}
\end{equation} 
where $\lambda_i$ are coefficients, $L_1$ is the norm, $G$ is a regularization term in gradient space for smoothness, and $H_{SSIM}$ is the hierarchy SSIM loss to supervise \Name{} to learn a better depth. 
We follow existing works~\cite{srinivasanApertureSupervisionMonocular2018, gurSingleImageDepth2019} to have the regularization term $G$ as:  
\begin{equation}
    G = \frac{1}{N}\sum_{i=1}^N |\partial_xD_i|e^{-|\partial_xI_i|} + |\partial_yD_i|e^{-|\partial_yI_i|},
\end{equation}
where $N$ is the $N$ layers of the pyramid of the image, and $G$ is a smooth and regularization term in monocular-depth estimation.

Except for the gradient loss, we noticed that per-pixel loss, such as $L1$ or $L2$ norm cannot supervise the neural network efficiently due to the ambiguity introduced by the bokeh computation process. 
As the example shown in Fig.~\ref{fig:hierarchy_loss}, the per-pixel loss fails to supervise the network to optimize the neighborhood values. 
The reason is that pixel scattering or gathering is a patch-level operation, which means that the per-pixel loss signal is not enough in describing the patch-level error.  
To guide the network not only care about per-pixel results but also regional results, we propose adding a hierarchy SSIM term to learn a better depth.  
The default SSIM has a patch size of 11. 
As the maximum scattering range is highly likely to be larger than 11, we propose to use a hierarchy SSIM loss: a set of SSIM loss with different patch sizes to give the pixel the regional feedback instead of per-pixel feedback.

\section{Implementation and Results}
Since implementing the backward differential computation is non-trivial, we will release our code to allow for the reproducibility of our work. Here we introduce our forward bokeh rendering pipeline and relevant implementation details in Sec.~\ref{sec.imp}. Quantitative and qualitative evaluations of bokeh rendering and differentiability are discussed in Sec.~\ref{sec.lens_blur_eval} and~\ref{sec.diff_eval}. 

\subsection{Implementation}\label{sec.imp}
\begin{figure}[t]
    \centering
    \includegraphics[width=\linewidth]{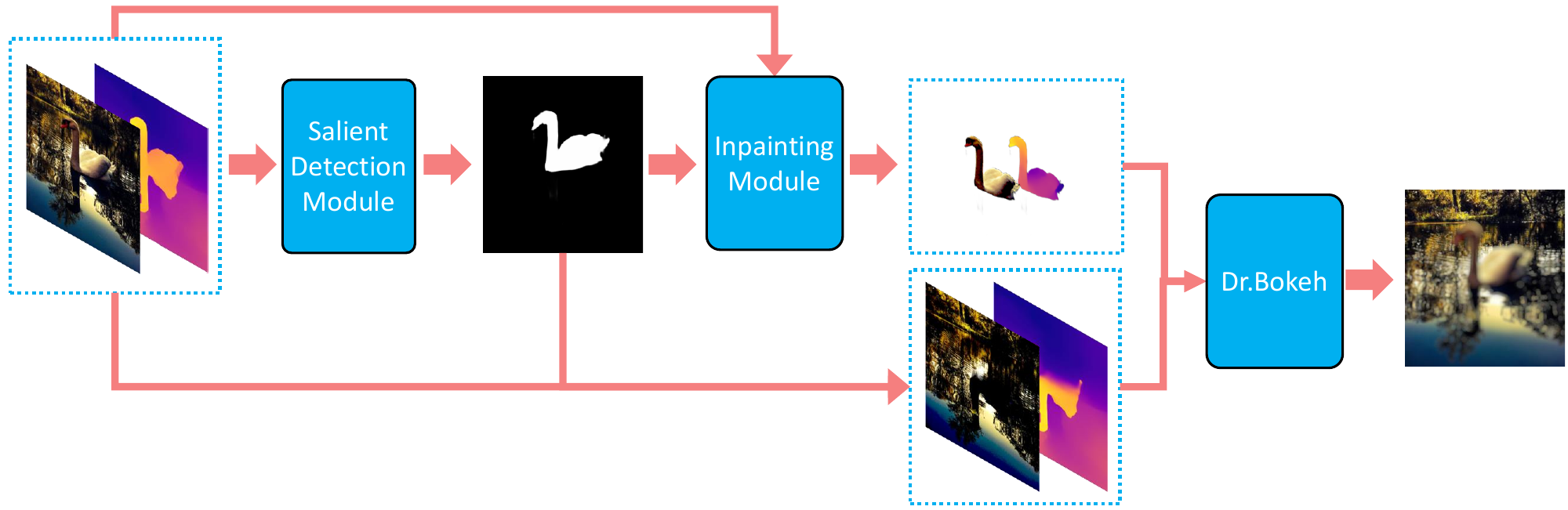}
    \caption{\textbf{\Name{} rendering pipeline:} The rendering pipeline takes the RGBD as input. A salient object detection module extracts the salient object. Then the pipeline computes the occluded RGBD values behind the salient objects. Then given the foreground RGBAD and background RGBD, \Name{} renders a realistic bokeh image.}
    \label{fig:pipeline}
\end{figure}

\textbf{Forward-Bokeh Rendering:}  
For a $W \times H$ pixel image with $L$ layers and searching neighborhood range $R$, the computation complexity for \Name{} is $\mathcal{O}(W \times H \times L \times R^2)$. 
We implement \Name{} with CUDA acceleration and integrate it in Pytorch as a new computation layer. 
In practice, for a $256\times256$ image with $L=2$ and $R=21$, it takes 19 ms on average. 
A larger $L$ improves the bokeh rendering quality, but qualitative results in Sec.~\ref{sec.eval_real} and supplementary materials show that two layers are enough for many real-world cases. 
Our bokeh rendering pipeline (Fig.~\ref{fig:pipeline}) includes a salient object detection module, an inpainting module, and the \Name{} renderer.  
An off-the-shelf salient object detection model~\cite{wei2020label} provides the layered scene representation for all examples in this paper. 
Please note the salient object detection module can be replaced with any other segmentation or matting network depending on the input category for good performance, e.g., an image matting network that predicts a detailed matting layer for portrait images. 
The salient object mask is then used to guide background RGBD inpainting, and we use LaMa~\cite{suvorov2022resolution} for high-resolution inpainting to generate all the results. 
\Name{} takes user-defined camera parameters, including the focal plane distance, blur radius, and lens shape, to synthesize the bokeh image.

\textbf{Backward-Derivatives:} To make \Name{} fit into data-driven pipelines, we implement the differentiable operation in Pytorch~\cite{paszke2019pytorch}. 
Although Pytorch is auto-differentiable, the computation in \Name{} involves many operations not supported by the auto-differentiable computation layers in Pytorch. 
We implement the derivative computation from scratch in CUDA and integrate the computation layer as a new layer in Pytorch.  

The vanilla implementation suffers from the vanishing gradient problem for large depth inputs as the two sides of the soft-step function in Fig.~\ref{fig:diff_operations} are flat. 
The optimization results heavily depend on the initialization and may stay at the local minimum without further improvement. 
To address this issue, we adaptively ``leak'' the gradient of two sides of the soft step functions similar to leaky ReLU. 
In practice, we notice this slight change significantly improves the optimization results. 
As shown in Fig.~\ref{fig:depth_quality}, there are two large blob-like areas in the second-row example for GaussPSF~\cite{gurSingleImageDepth2019} because GaussPSF applies a Gaussian kernel for blur computation. 
Although it provides easy gradient computations, GaussPSF also suffers from the vanishing gradient problem.  Using our ``leaky'' soft step, \Name{} can be robust to bad network initializations and learn a better depth. 

\textbf{Depth-From-Defocus Implementation:} Following the Aperture~\cite{srinivasanApertureSupervisionMonocular2018} and GaussPSF~\cite{gurSingleImageDepth2019}, we train a CNN to predict the depth. 
The only difference is that we replace the bokeh rendering module with \Name{}.

\subsection{Lens Blur Rendering Evaluation}\label{sec.lens_blur_eval}
Quantitatively evaluating lens blur quality is still challenging given a single RGB or RGBD image as input and no benchmark exists yet.   
So we follow existing works~\cite{wangDeepLensShallowDepth2018, pengBokehMeWhenNeural2022, pengMPIBMPIBasedBokeh2022} to create a synthetic benchmark (Sec.~\ref{sec.eval_synthetic}) except that we render high-quality lens blur results by ray tracing through a real thin-lens. 
In addition, quantitative evaluations on lens blur are not always reliable~\cite{ignatov2020aim, pengMPIBMPIBasedBokeh2022}, so we collected real-world images and applied a user study to evaluate our method qualitatively (Sec.~\ref{sec.eval_real}).
We also provide more qualitative results in supplementary materials.

\begin{figure}[t]
    \centering
    \includegraphics[width=\linewidth]{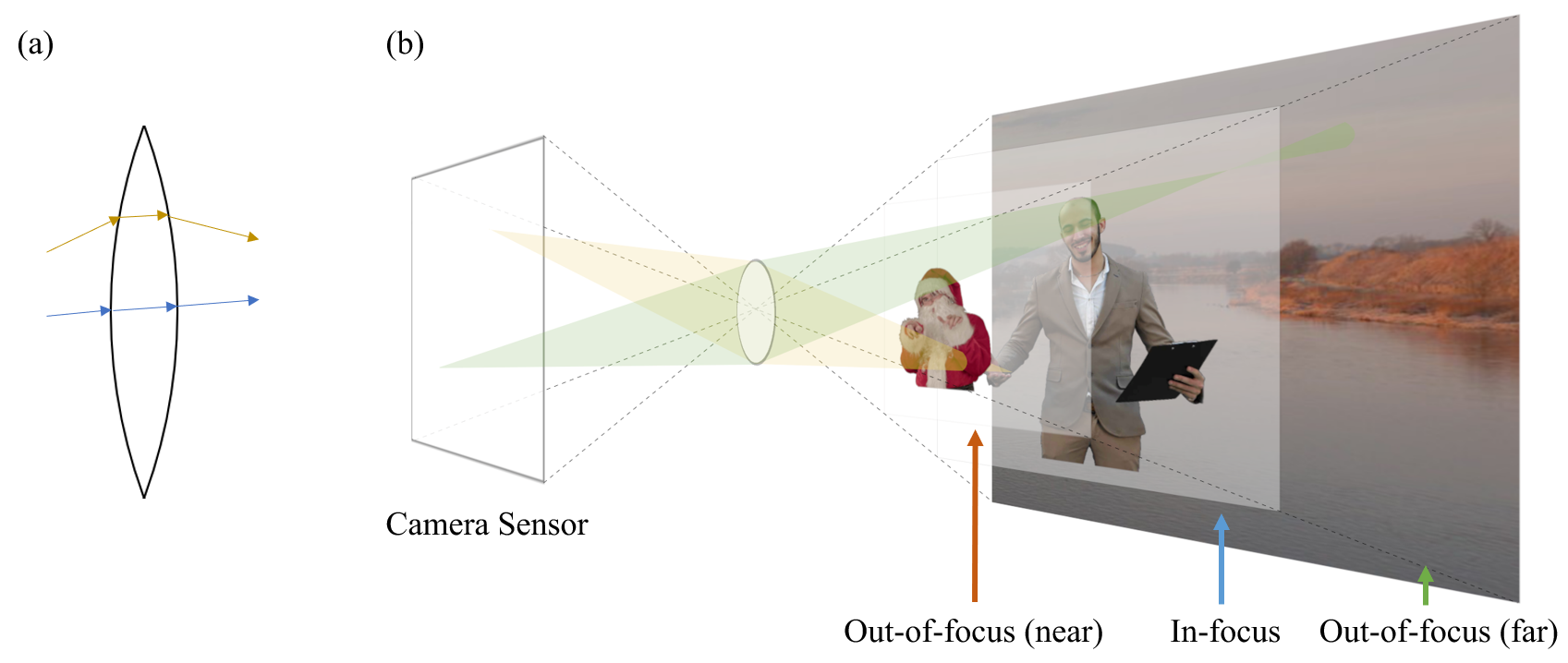}
    \caption{We simulate bokeh by computing how rays scatter and focus through a spherical lens system. (a) Side-view of how rays get refracted into and out of the lens. 
    Rays hitting somewhere near the center of the lens are barely distorted (blue), while the ones hitting the peripheral get distorted more (brown). 
    (b) Setup of the rendering scene. The billboards are resized to cover the exact FOV of the camera sensor. 
    A point on the image plane shoots many rays that go through the lens aperture and get refracted.
    As a result, in-focus billboards will be rendered sharply, while ones that are closer/farther will be blurry.
    }
    \label{fig:ray-lens-refraction}
\end{figure}

\begin{table*}[t]
\centering
\caption{Result on the synthetic benchmark. Comparing with SteReFo~\cite{busamSteReFoEfficientImage2019}, DeepLens~\cite{wangDeepLensShallowDepth2018}, BokehMe~\cite{pengBokehMeWhenNeural2022}, and MPIB~\cite{pengMPIBMPIBasedBokeh2022}. \Name{} outperforms state-of-the-art methods in all the metrics.} \label{tab:eval_quantitatve}
\small
\begin{tabular}{l|ccccc}
\shline
\textbf{Method}  & \textbf{RMSE} $\downarrow$ & \textbf{RMSE-s} $\downarrow$  & \textbf{SSIM} $\uparrow$  & \textbf{PSNR} $\uparrow$  & \textbf{ZNCC} $\uparrow$ \\
\hline
SteroFo & 0.0179 $\pm$ 0.0092 & 0.0178 $\pm$ 0.0092 & 0.9753 $\pm$ 0.0168 & 35.9201 $\pm$ 4.2145 & 0.9966 $\pm$ 0.0028\\
DeepLens & 0.0461 $\pm$ 0.0188 & 0.0403 $\pm$ 0.0147 & 0.9476 $\pm$ 0.0263 & 27.3505 $\pm$ 3.6262 & 0.9827 $\pm$ 0.0125\\
BokehMe & 0.0144 $\pm$ 0.0077 & 0.0143 $\pm$ 0.0077 & 0.9708 $\pm$ 0.0248 & 37.7721 $\pm$ 4.1209 & 0.9976 $\pm$ 0.0027\\
MPIB & 0.0152 $\pm$ 0.0075 & 0.0151 $\pm$ 0.0074 & 0.9702 $\pm$ 0.0256 & 37.2024 $\pm$ 3.8708 & 0.9974 $\pm$ 0.0027\\
\textbf{\Name{}} & \textbf{0.0133 $\pm$ 0.0077} & \textbf{0.0133 $\pm$  0.0076} & \textbf{ 0.9757 $\pm$ 0.0211} & \textbf{38.7288 $\pm$ 4.7439} & \textbf{0.9979 $\pm$ 0.0024} \\
\shline
\end{tabular}
\end{table*}

\begin{figure*}[t]
    \centering
    \centering
    \setlength{\tabcolsep}{1pt}
    \begin{tabular}{cccccc}
    \includegraphics[width=0.16\linewidth]{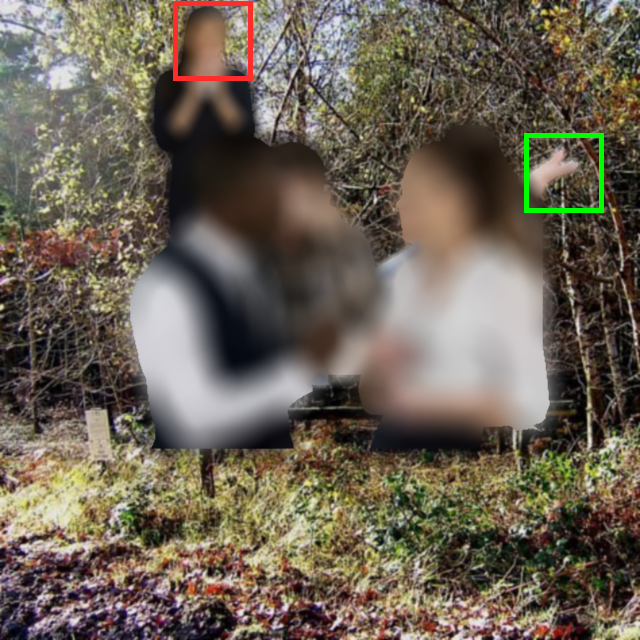} & 
    \includegraphics[width=0.16\linewidth]{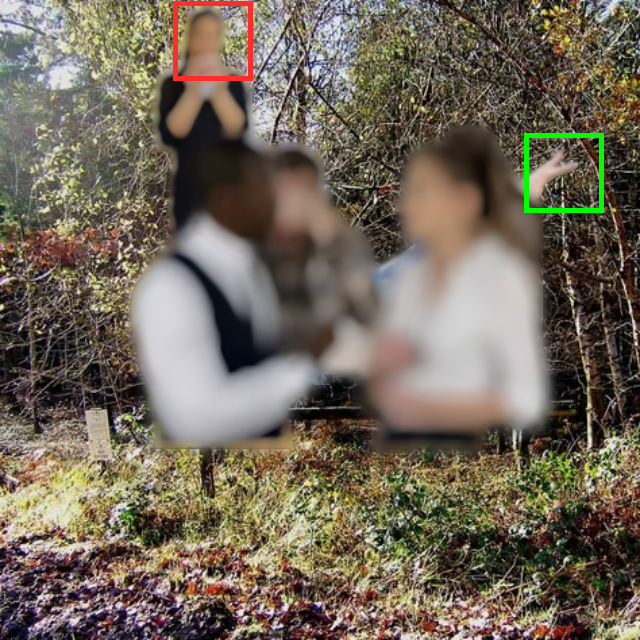} & 
    \includegraphics[width=0.16\linewidth]{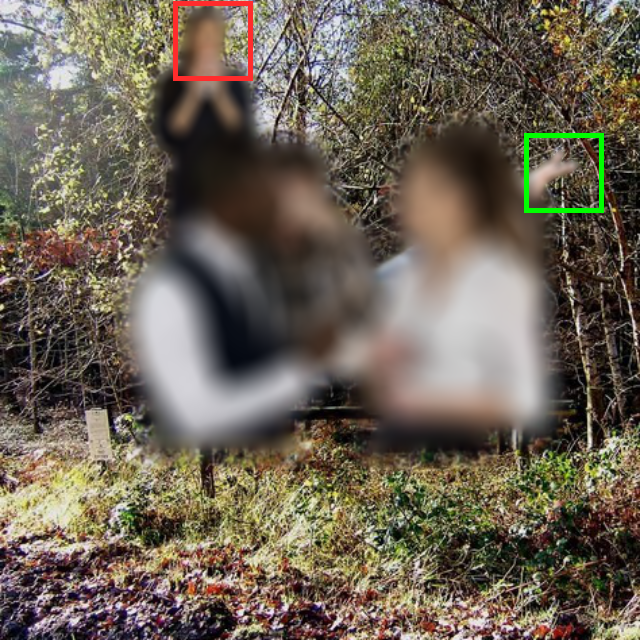} & 
    \includegraphics[width=0.16\linewidth]{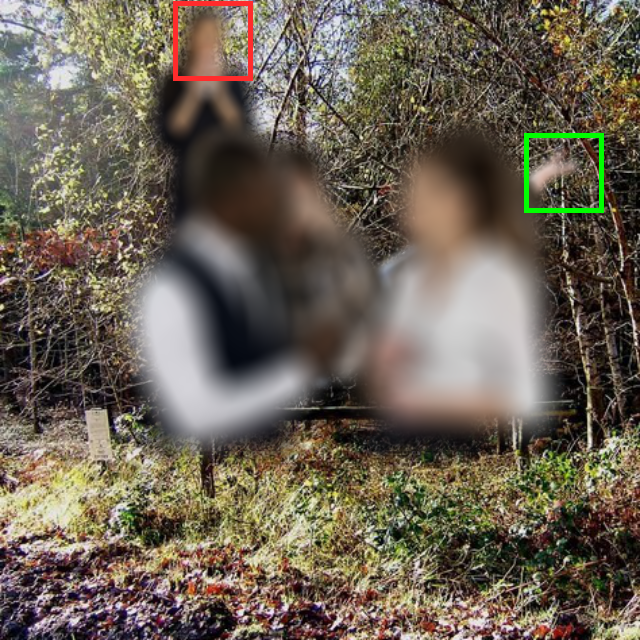} & 
    \includegraphics[width=0.16\linewidth]{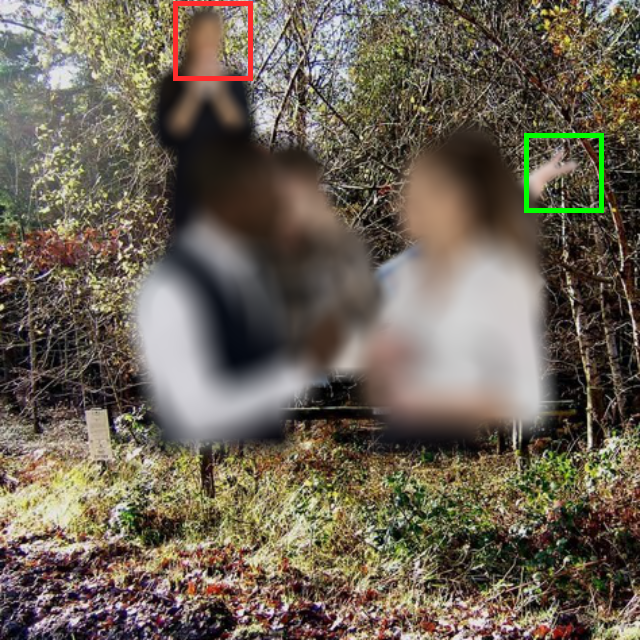} & 
    \includegraphics[width=0.16\linewidth]{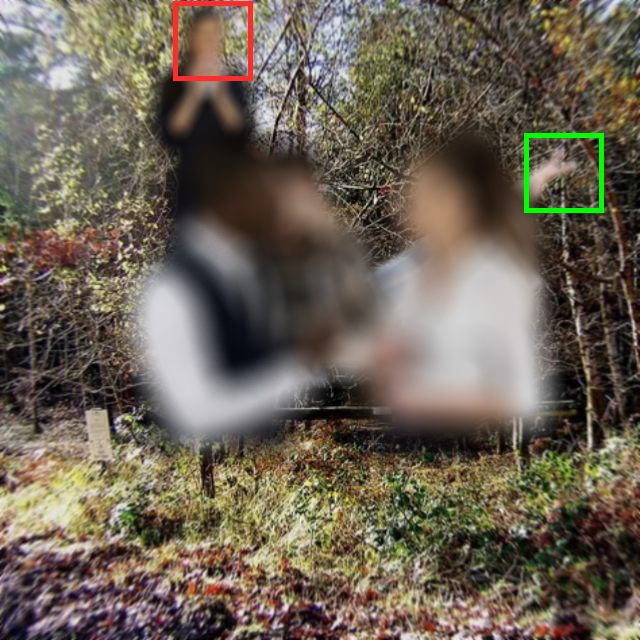} \\

    \includegraphics[width=0.16\linewidth]{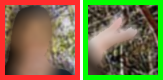} & 
    \includegraphics[width=0.16\linewidth]{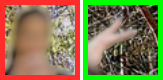} & 
    \includegraphics[width=0.16\linewidth]{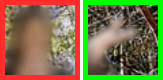} & 
    \includegraphics[width=0.16\linewidth]{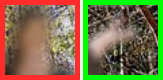} & 
    \includegraphics[width=0.16\linewidth]{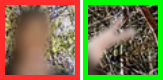} & 
    \includegraphics[width=0.16\linewidth]{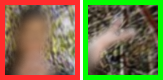} \\
    (a) SteReFo & (b) DeepLens & (c) BokehMe & (d) MPIB & (e) \Name{} (ours) & (f) Ground Truth \\
    \end{tabular}
    \caption{\textbf{Qualitative Comparison of Depth of Field (DoF) Results on Synthetic Benchmarks:} The scattering/gathering-based method (SteRoFo) exhibits unnatural partial occlusion. Learning-based methods (DeepLens and BokehMe) struggle to render natural partial occlusion in the absence of explicit modeling. Although the state-of-the-art method MPIB was trained to address the partial occlusion challenge, our method \Name{} achieves the best DoF quality without necessitating training. Best viewed by zooming in.}
    \label{fig:eval_syn_example}
\end{figure*}

\subsubsection{Synthetic Benchmark for Quantitative Evaluation} \label{sec.eval_synthetic} 
\paragraph{Dataset:} Existing works~\cite{wangDeepLensShallowDepth2018, pengMPIBMPIBasedBokeh2022} 
setup the scene by compositing multiple layered images and utilizing an approximated pseudo ray tracer to render the lens blur ground truth. Instead, we implemented a renderer that ray traces through a real thin lens to generate the lens blur ground truth in order to evaluate the effectiveness of \Name{} (see Fig. ~\ref{fig:ray-lens-refraction}). 
The lens is modeled as the intersection of two identical spheres of radius $R_c$, such that the radius of the intersection circle is the aperture radius $R_a = L/2$. 
The thickness of the lens is computed as $d = 2\sqrt{R_c^2-R_a^2}$, which gives the lens' focal length~$f$ together with the lens' refractive index $\eta$, using the lensmaker's equation~\cite{greivenkamp2004field}:
$$
\frac{1}{f} = (\eta - 1)\left(\frac{2}{R_c} + \frac{(\eta-1)d}{\eta R_c^2}\right).
$$

The camera is set forth with a chosen FOV, and the image plane is placed at distance~$D_I > f$.  The color of a pixel is computed by tracing rays from the pixel through various random points on the lens. 
More details can be found in the appendix.

The scene (5-layer billboards) setup is similar to the dataset by DeepLens~\cite{wangDeepLensShallowDepth2018} and MPIB~\cite{pengMPIBMPIBasedBokeh2022}. 
The foreground objects are randomly sampled from Adobe Matting Dataset~\cite{xu2017deep} and AIM-500~\cite{li2021deep}. 
The background scenes are randomly sampled from the landmark dataset~\cite{weyand2020GLDv2}. 
The benchmark includes 100 scenes with different blur radiuses and focal planes. 
Each scene has an all-in-focus image, a ground truth depth, a layered ground truth scene representation, and a bokeh ground truth. 

\textbf{Metric:}
We apply the RMSE metric and a scale-invariant RMSE (RMSE-s)~\cite{sun2019single} as we noticed that different methods have different gamma correction implementations. 
We also apply the SSIM and ZNCC for perception evaluation.

\textbf{Comparison to related work:}
We compare \Name{} to a gathering-based method SteReFo~\cite{busamSteReFoEfficientImage2019}, a fully learning-based method DeepLens~\cite{wangDeepLensShallowDepth2018} and a hybrid of classical and learning-based method BokehMe~\cite{pengBokehMeWhenNeural2022} and MPIB~\cite{pengMPIBMPIBasedBokeh2022}.  
Different methods take different kernel parameters. 
So we search all the blur kernels and pick the best result from each method. 
All methods take the same depth as input. 

Each step in the lens blur rendering pipeline affects the rendering quality. 
But different methods have different pipelines, which makes the quantitative evaluation easy to be unfair.
For example, DeepLens predicts its own depth and then predicts the lens blur.    
MPIB and \Name{} involves background inpainting, which looks reasonable perceptually but is easy to have large quantitative errors.
To be fair to all the methods, the quantitative evaluation only measures the rendering step quality in all the pipelines instead of measuring the whole pipeline.  
We use the ground truth depth for all the methods. 
For DeepLens, we replace the predicted depth with the ground truth depth. 
For fairness, we replace the predicted occluded pixels with the ground truth pixels for MPIB and \Name{}.

\begin{figure*}[hbt]
    \centering
    \setlength{\tabcolsep}{1pt}
    \begin{tabular}{ccccc}
    \includegraphics[width=0.19\linewidth]{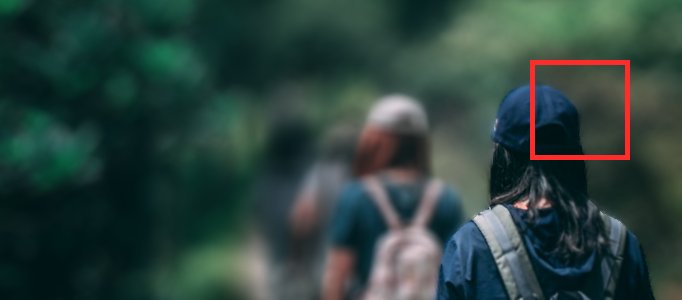} & 
    \includegraphics[width=0.19\linewidth]{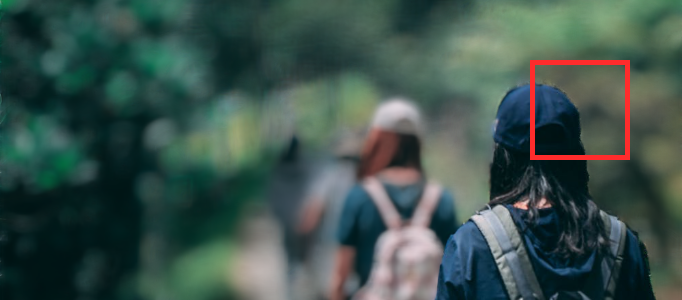} & 
    \includegraphics[width=0.19\linewidth]{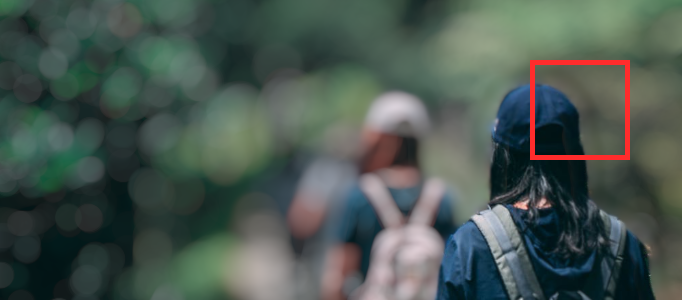} & 
    \includegraphics[width=0.19\linewidth]{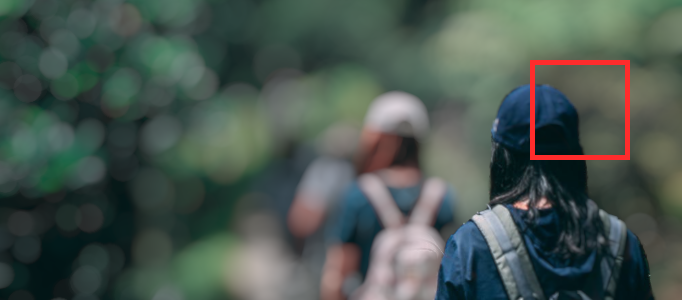} & 
    \includegraphics[width=0.19\linewidth]{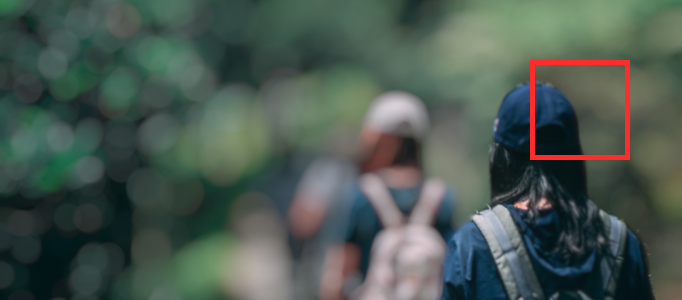} \\ 
    \includegraphics[width=0.19\linewidth]{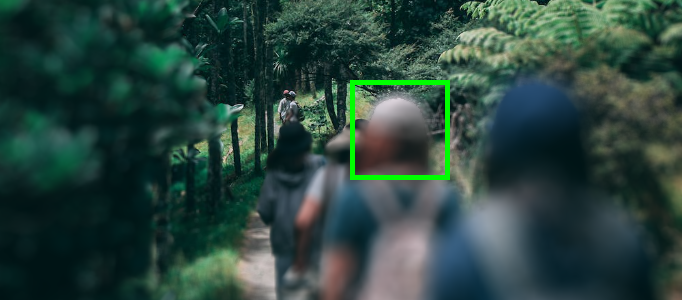} & 
    \includegraphics[width=0.19\linewidth]{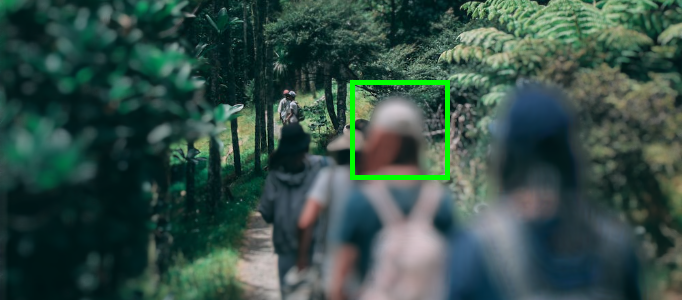} & 
    \includegraphics[width=0.19\linewidth]{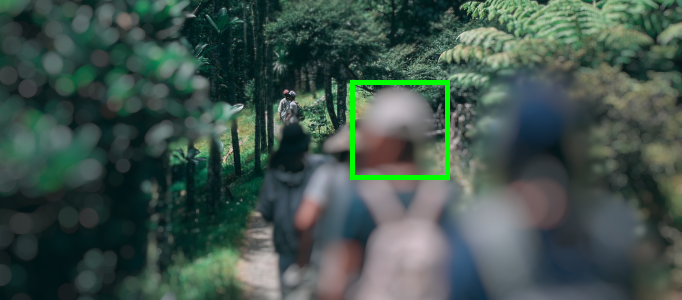} & 
    \includegraphics[width=0.19\linewidth]{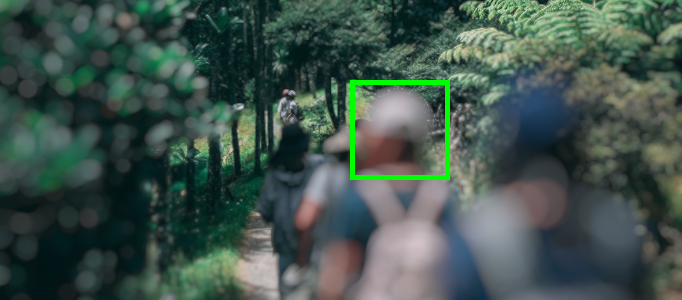} & 
    \includegraphics[width=0.19\linewidth]{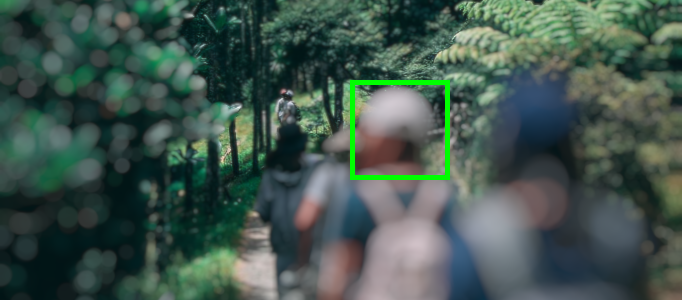} \\ 
    \includegraphics[width=0.19\linewidth]{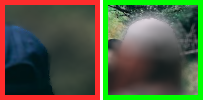} & 
    \includegraphics[width=0.19\linewidth]{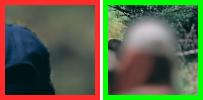} & 
    \includegraphics[width=0.19\linewidth]{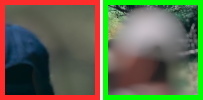} & 
    \includegraphics[width=0.19\linewidth]{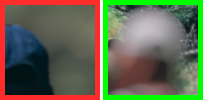} & 
    \includegraphics[width=0.19\linewidth]{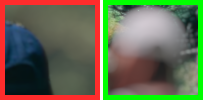} \\ 

    \includegraphics[width=0.19\linewidth]{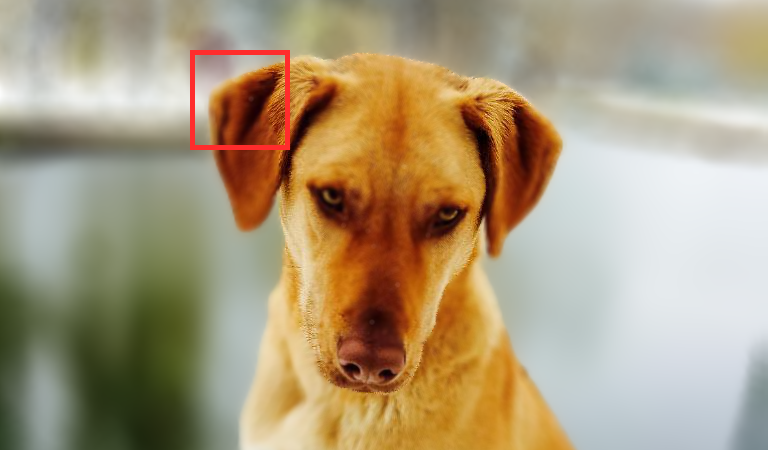} & 
    \includegraphics[width=0.19\linewidth]{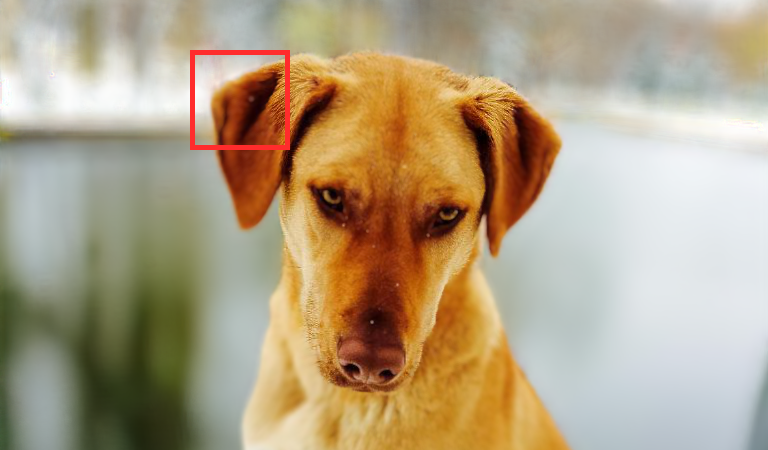} & 
    \includegraphics[width=0.19\linewidth]{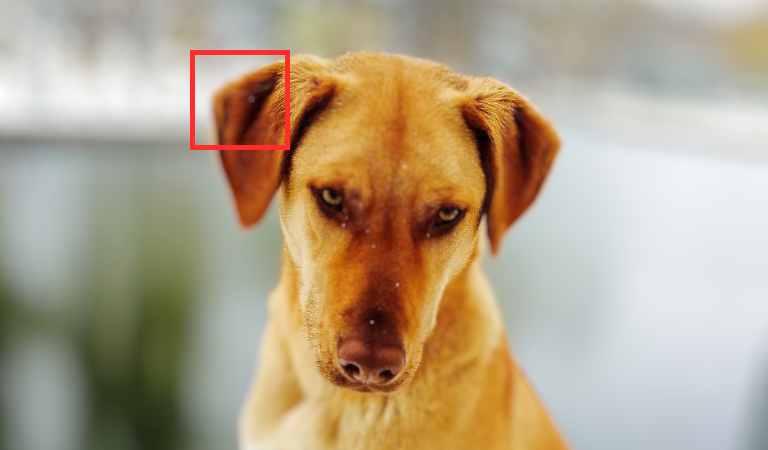} & 
    \includegraphics[width=0.19\linewidth]{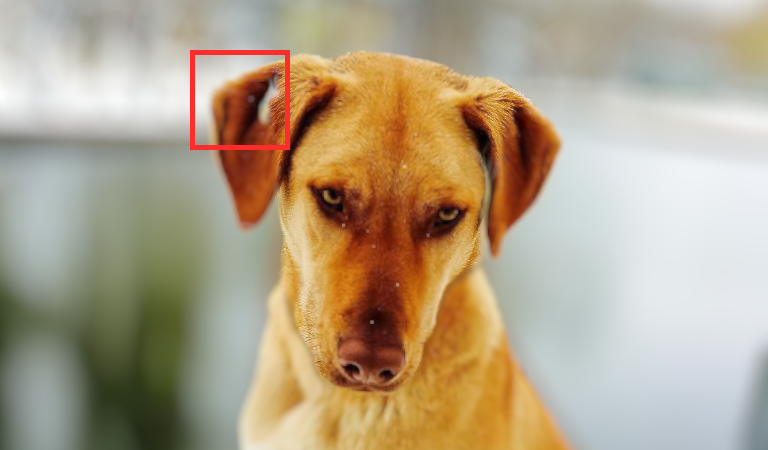} & 
    \includegraphics[width=0.19\linewidth]{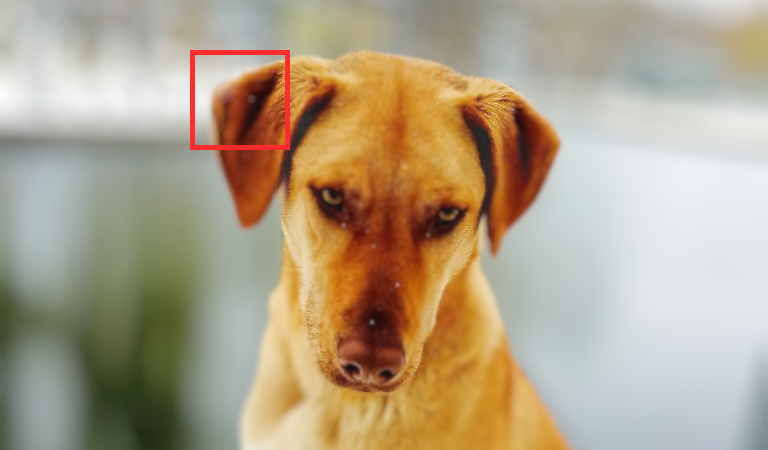} \\ 
    \includegraphics[width=0.19\linewidth]{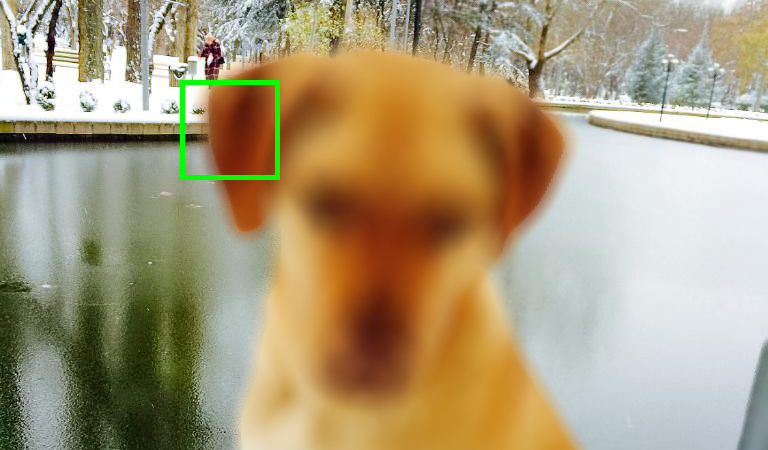} & 
    \includegraphics[width=0.19\linewidth]{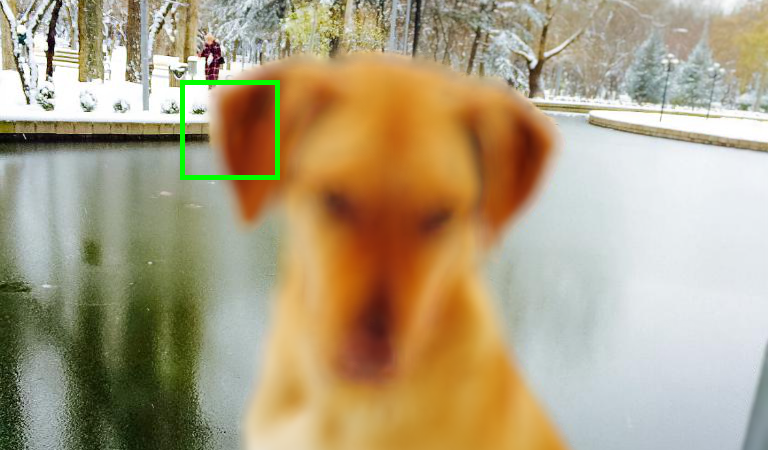} & 
    \includegraphics[width=0.19\linewidth]{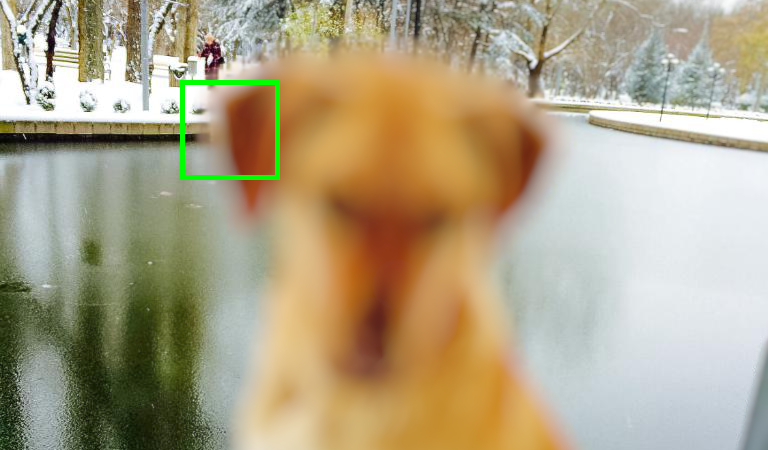} & 
    \includegraphics[width=0.19\linewidth]{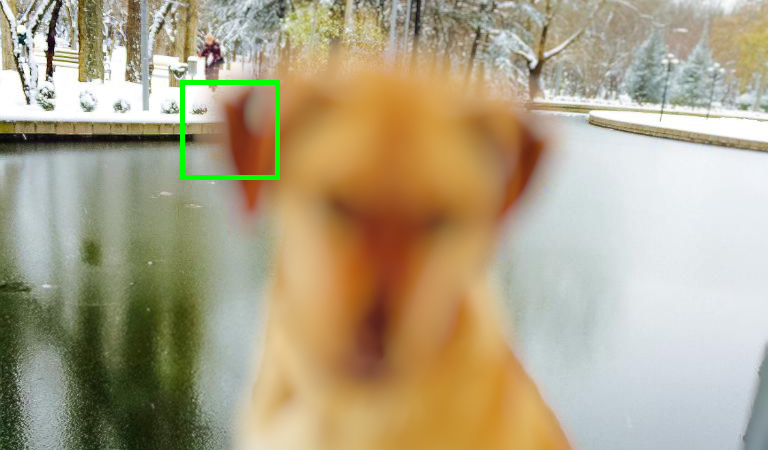} & 
    \includegraphics[width=0.19\linewidth]{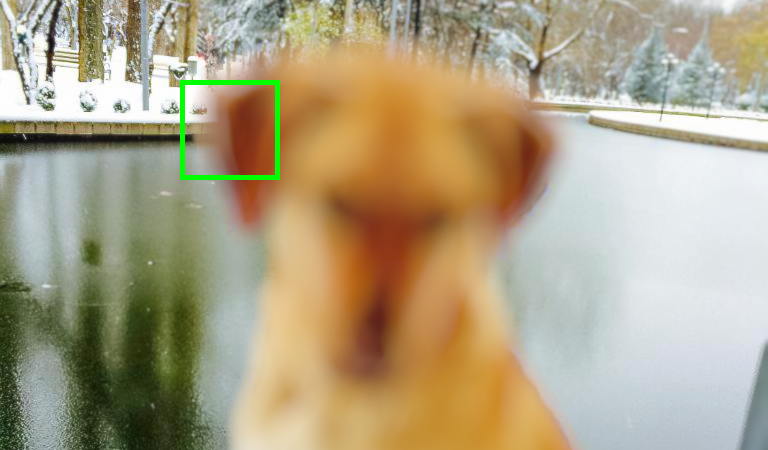} \\ 
    \includegraphics[width=0.19\linewidth]{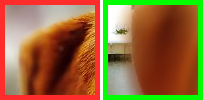} & 
    \includegraphics[width=0.19\linewidth]{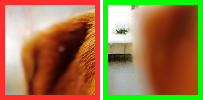} & 
    \includegraphics[width=0.19\linewidth]{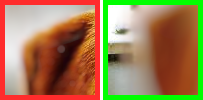} & 
    \includegraphics[width=0.19\linewidth]{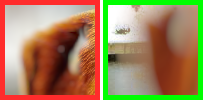} & 
    \includegraphics[width=0.19\linewidth]{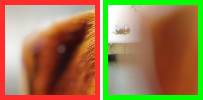} \\ 

    \includegraphics[width=0.19\linewidth]{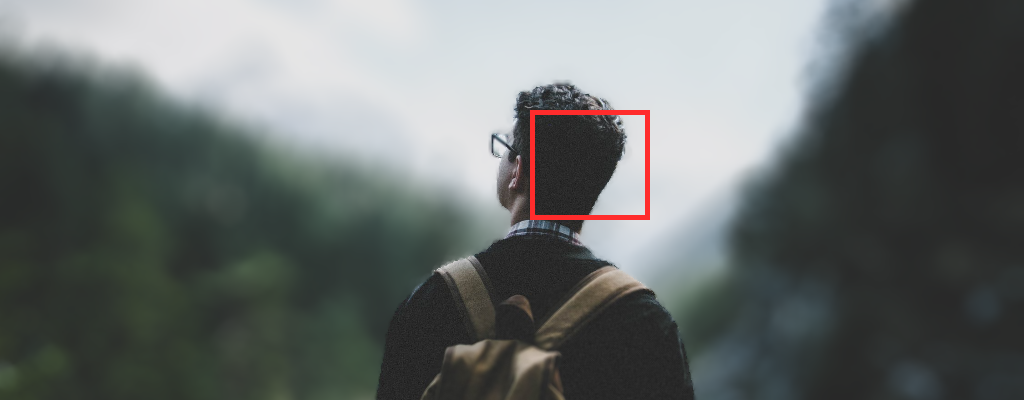} & 
    \includegraphics[width=0.19\linewidth]{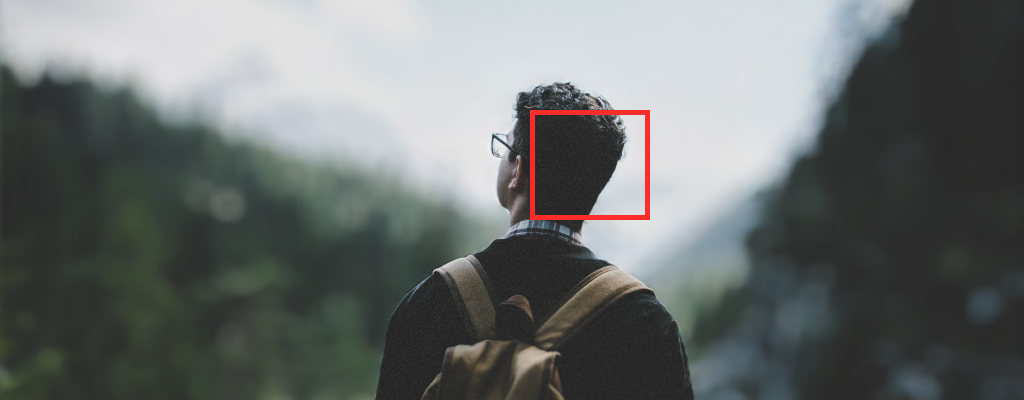} & 
    \includegraphics[width=0.19\linewidth]{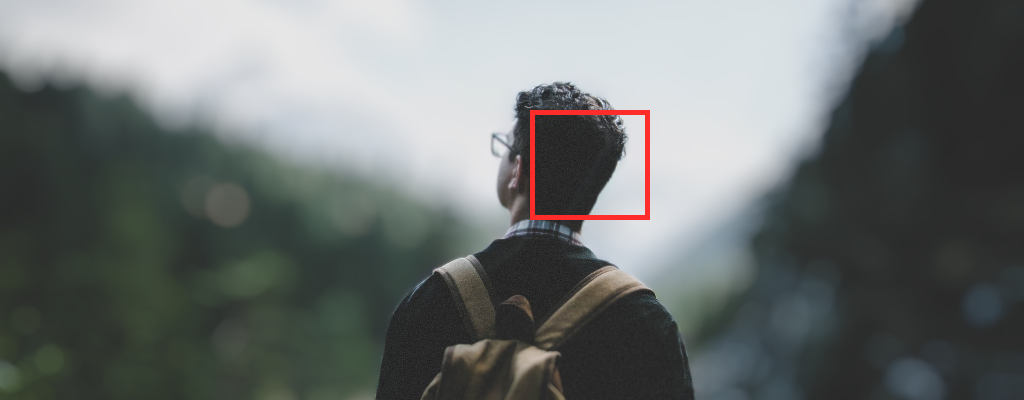} & 
    \includegraphics[width=0.19\linewidth]{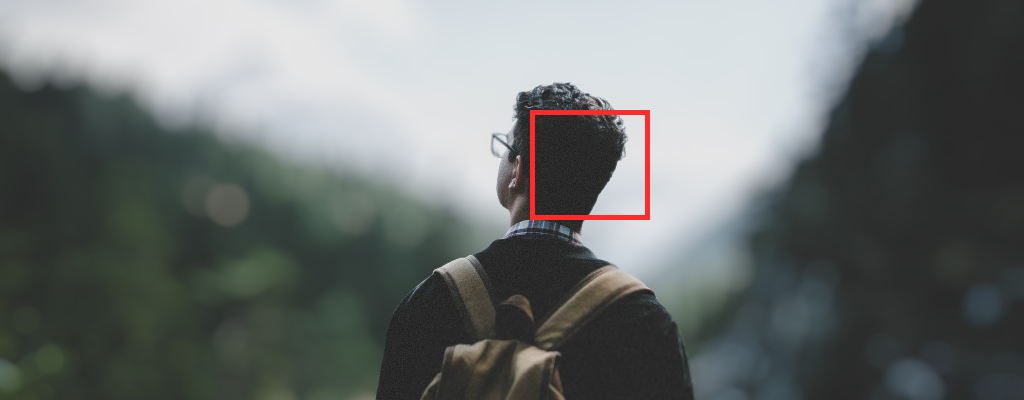} & 
    \includegraphics[width=0.19\linewidth]{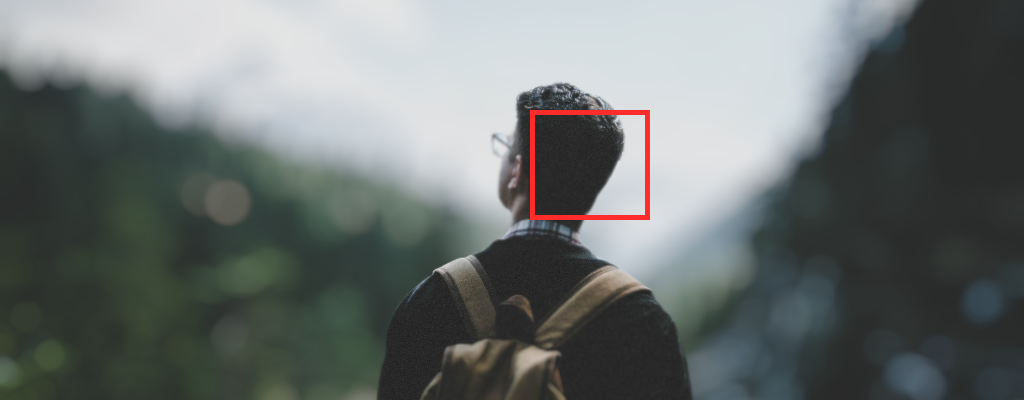} \\ 
    \includegraphics[width=0.19\linewidth]{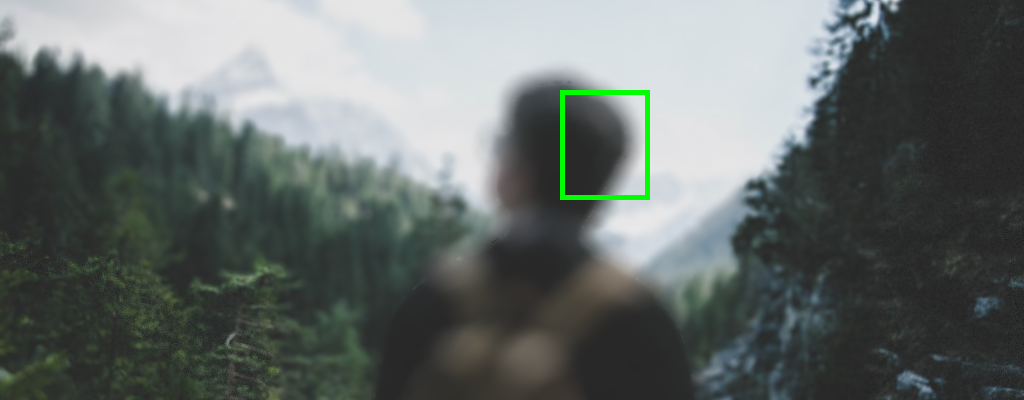} & 
    \includegraphics[width=0.19\linewidth]{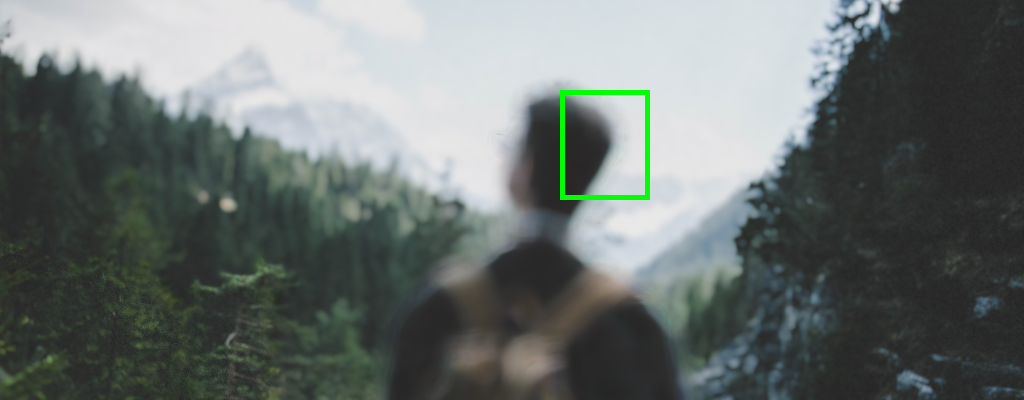} & 
    \includegraphics[width=0.19\linewidth]{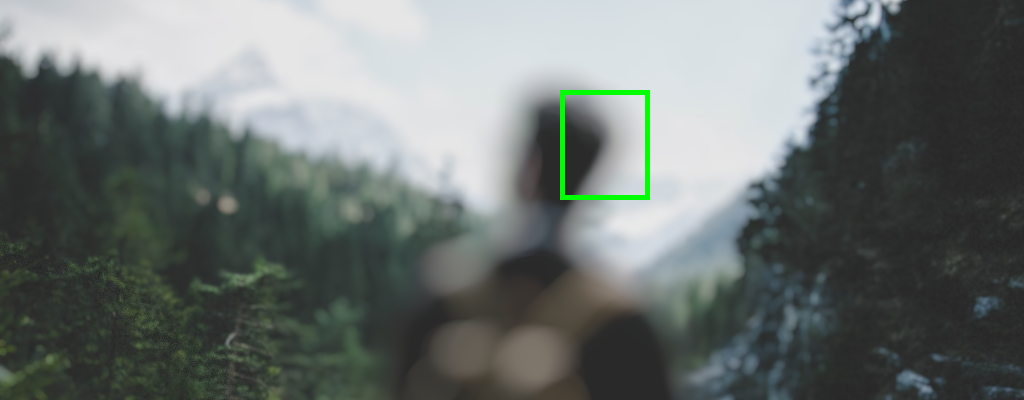} & 
    \includegraphics[width=0.19\linewidth]{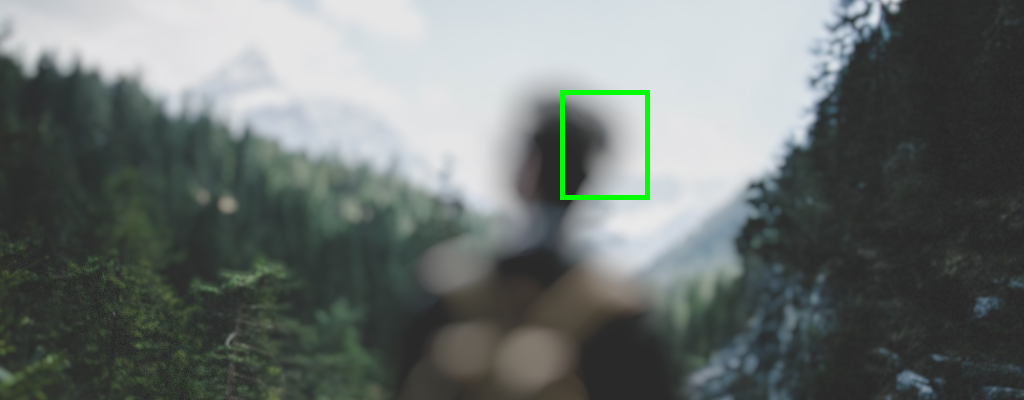} & 
    \includegraphics[width=0.19\linewidth]{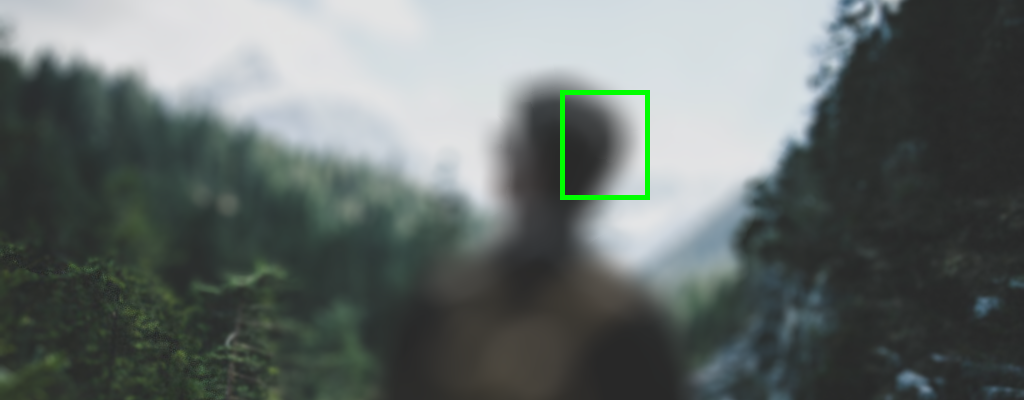} \\ 
    \includegraphics[width=0.19\linewidth]{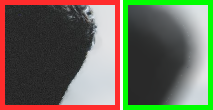} & 
    \includegraphics[width=0.19\linewidth]{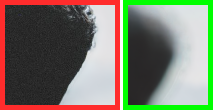} & 
    \includegraphics[width=0.19\linewidth]{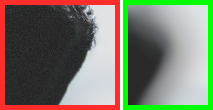} & 
    \includegraphics[width=0.19\linewidth]{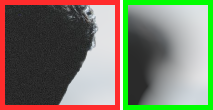} & 
    \includegraphics[width=0.19\linewidth]{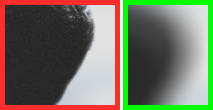} \\

    (a) SteReFo~\shortcite{busamSteReFoEfficientImage2019} & 
    (b) DeepLens~\shortcite{wangDeepLensShallowDepth2018} & 
    (c) BokehMe~\shortcite{pengBokehMeWhenNeural2022} & 
    (d) MPIB~\shortcite{pengMPIBMPIBasedBokeh2022} & 
    (e) \Name{} (ours)\\
    \end{tabular}
    \caption{\textbf{Qualitative comparisons on real-world images:} Classical methods (SteReFo) are competitive for foreground in-focus but fail to render natural boundary partial occlusion. Learning-based methods (DeepLens and BokehMe) suffer from unnatural partial occlusion effects. MPIB learns to render the partial occlusion effect but has leaking artifacts due to generalization issues (see the second and the third-row examples). \Name{} renders natural partial occlusion effects and is more robust for either foreground or background in-focus cases given the same inputs.} \label{fig:qualitaive}
\end{figure*}

We show quantitative and qualitative results in Tab.~\ref{tab:eval_quantitatve} and Fig.~\ref{fig:eval_syn_example}.
Although the gathering-based method SteroFo achieves a high SSIM value, it is easy to observe its unnature partial occlusion results in Fig.~\ref{fig:eval_syn_example} (a). 
DeepLens does not perform well in metrics. 
The potential reason is that the synthesized training data for DeepLens fails to have realistic foreground blurs. 
Although MPIB was designed to learn better partial occlusion effects, BokehMe still performs slightly better than MPIB in the metrics. 
The reason may be that MPIB is a fully learning-based method and does not generalize well to unseen datasets compared to the hybrid method BokehMe. 
However, MPIB qualitatively renders more natural partial occlusion effects, as shown in Fig.~\ref{fig:eval_syn_example} (d).  
Our method \Name{} performs the best in the quantitative metrics and can render realistic partial occlusion effects.

\subsubsection{Real-data for Qualitative Evaluation}\label{sec.eval_real}
\paragraph{Dataset:}
Existing works~\cite{ignatov2020aim, pengMPIBMPIBasedBokeh2022} suggest that the numerical metric for evaluating lens blur is unreliable, which also shows up in our quantitative evaluations. 
Thus, we further apply a qualitative evaluation to evaluate our method. We collect real-world images with different subjects, background scenarios, and lighting as testing data for the user study. 
The user study contained 40 questions of two-alternative forced-choice (2AFC). 
Each question is a pair of lens blur results generated by \Name{} and a comparison method (SteReFo, DeepLens, BokehMe, and MPIB).

\paragraph{Comparison to related works:}
The metric and the comparing works are the same as the quantitative evaluation: 41 people ($75\%$ male, $25\%$ female, $46\%$ no photography experience, $26\%$ some experience, $28\%$ experienced) participated in our user study. We discard all replies that were too short (under three minutes) or always clicked on the same side. 
Our results show that $81\%, 74\%, 68\%, 61\%$ of participants support that the image generated by \Name{} is more realistic than SteReFo, DeepLens, BokehMe, and MPIB. 
In particular, the T-test value in the more realistic lens blur effect for an image generated by \Name{} and BokehMe, DeepLens, and SteReFo are 5.75, 8.12, and 11.63 and are significant at 0.001 levels, which indicates the \Name{} is significantly better than those reference methods from a user perception perspective.

Fig.~\ref{fig:qualitaive} demonstrates the qualitative comparisons with all the previous works.
For foreground in-focus cases, \Name{} can preserve the sharp boundary consistently, while the learning-based methods still have generalization issues such as unnatural partial occlusion effects and leaky background artifacts (see Fig.~\ref{fig:qualitaive} the second and third examples). 
For background in-focus cases, \Name{} has natural partial occlusion effects and is more robust than SOTA learning-based method MPIB. 
More qualitative comparison results can be found in the supplementary materials.

\subsection{Differentiable Evaluation}\label{sec.diff_eval}

\textbf{Comparison to Related Work:} We relate our differentiable lens blur rendering to two previous works: Aperture Supervision (Aperture)~\cite{srinivasanApertureSupervisionMonocular2018} and Guassian-based PSF (GaussPSF)\cite{gurSingleImageDepth2019}. 
Aperture trains a neural network to predict depth layers by blur image supervision. 
GaussPSF replaces the bokeh rendering module in Aperture with differentiable Gaussian kernels and trains the neural network in a similar routine. 
Compared to GaussPSF, our occlusion-aware \Name{} is a more accurate differentiable bokeh rendering module in terms of lens blur physics. 
For a fair comparison in the following benchmarks, we use the same depth estimation network~\cite{Xian_2020_CVPR} for all the comparison methods and the same loss functions Eqn.~(\ref{eq.loss}) for all methods.

\textbf{Benchmark:}  We evaluate the differentiability of \Name{} on the real-world benchmark: \textit{Light Field Dataset}~\cite{srinivasan2017learning}. 
There is no depth ground truth for the real datasets, so we only quantitatively evaluate the final rendered bokeh images and qualitatively show depth qualities from all the methods. 
The bokeh images rendered from the light-field camera are good bokeh approximations. 
There are 3,343 images. 
Similar to previous works, we split the dataset into 3,006 training images and 337 testing images. 

\begin{table}[hbt]
\centering
\caption{Result on the light field benchmark. Comparing with aperture supervision~\cite{srinivasanApertureSupervisionMonocular2018} and GaussPSF~\cite{gurSingleImageDepth2019}. 
Note the metrics are measured on the DoF image due to the lack of depth ground truth.} \label{tab:diff_LF}
\small
\begin{tabular}{l|ccc}
\shline
\textbf{Method}  & \textbf{RMSE} $\downarrow$ & \textbf{SSIM} $\uparrow$  & \textbf{PSNR} $\uparrow$  \\
\hline
Aperture & $0.0133 \pm 0.0038$  &  $0.9774 \pm 0.0104$ & $37.8396 \pm 2.2216$ \\
GaussPSF & $0.0132 \pm 0.0032$  &  $0.9767 \pm 0.0101$ & $37.8391 \pm 1.9961$ \\
\textbf{\Name{}} & \textbf{0.0123 $\pm$ 0.0029}  & \textbf{0.9807 $\pm$ 0.0086} & \textbf{38.4422 $\pm$ 1.9447}\\ 
\shline
\end{tabular}
\end{table}

\begin{figure}[t]
    \centering 
    \setlength{\tabcolsep}{1pt}
    \begin{tabular}{cccc}
    \includegraphics[width=0.24\linewidth]{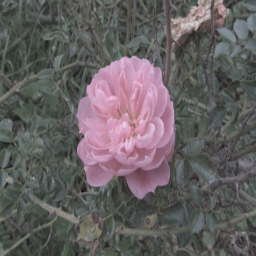} & 
    \includegraphics[width=0.24\linewidth]{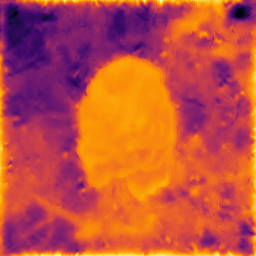} & 
    \includegraphics[width=0.24\linewidth]{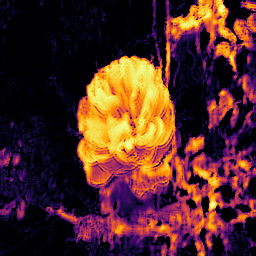} & 
    \includegraphics[width=0.24\linewidth]{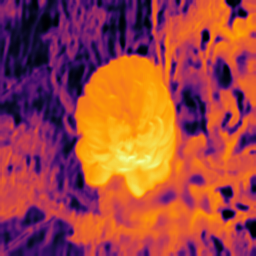} \\ 
    \includegraphics[width=0.24\linewidth]{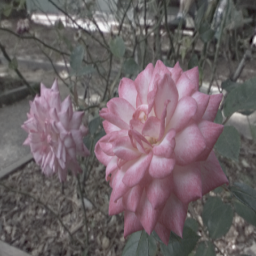} & 
    \includegraphics[width=0.24\linewidth]{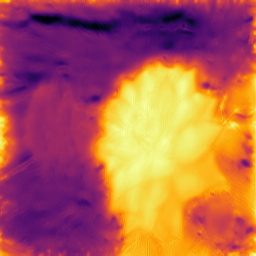} & 
    \includegraphics[width=0.24\linewidth]{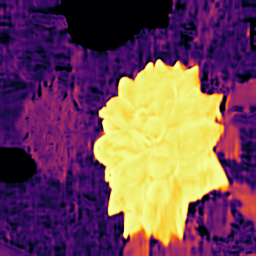} & 
    \includegraphics[width=0.24\linewidth]{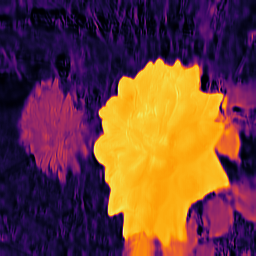} \\
    \includegraphics[width=0.24\linewidth]{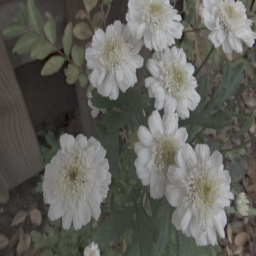} & 
    \includegraphics[width=0.24\linewidth]{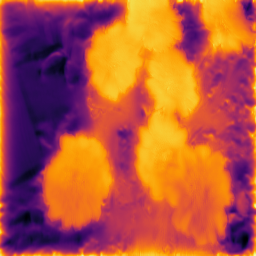} & 
    \includegraphics[width=0.24\linewidth]{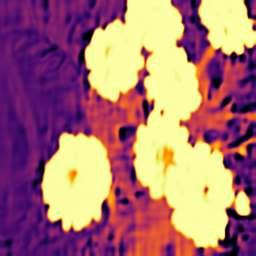} & 
    \includegraphics[width=0.24\linewidth]{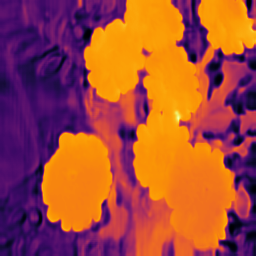} \\
    Input & Aperture & GaussPSF & \Name{} 
    \end{tabular}
    
    \caption{\textbf{Depth optimization for one pair data:} The first column is the all-in-focus input image. The second column shows results by Aperture~\shortcite{srinivasanApertureSupervisionMonocular2018}; the third column by GaussPSF~\shortcite{gurSingleImageDepth2019}, and the last column results by \Name{}. The depth map optimized by \Name{} has more details and is more accurate.}    
    \label{fig:depth_quality}
\end{figure}

\begin{figure}[t]
    \centering
    \setlength{\tabcolsep}{1pt}
    \begin{tabular}{cccc}
    \includegraphics[width=0.24\linewidth]{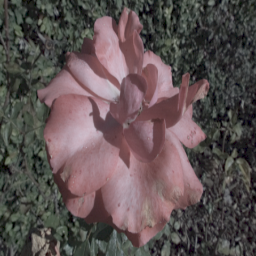} & 
    \includegraphics[width=0.24\linewidth]{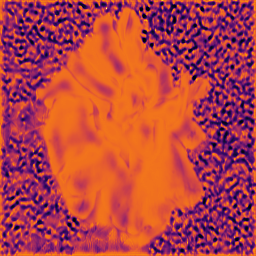} & 
    \includegraphics[width=0.24\linewidth]{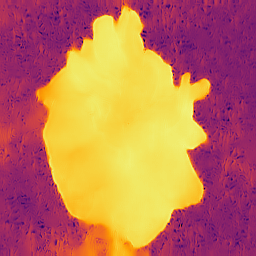} & 
    \includegraphics[width=0.24\linewidth]{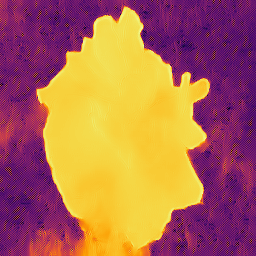} \\
    \includegraphics[width=0.24\linewidth]{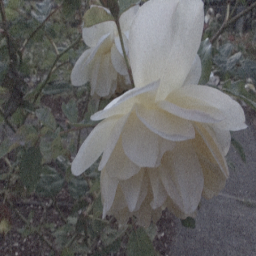} & 
    \includegraphics[width=0.24\linewidth]{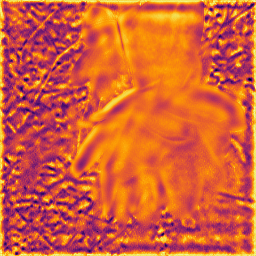} & 
    \includegraphics[width=0.24\linewidth]{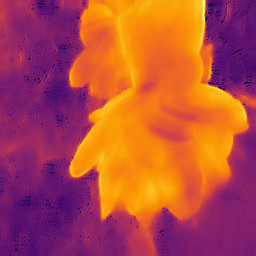} & 
    \includegraphics[width=0.24\linewidth]{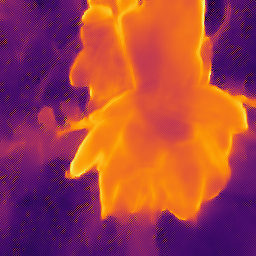} \\
    \includegraphics[width=0.24\linewidth]{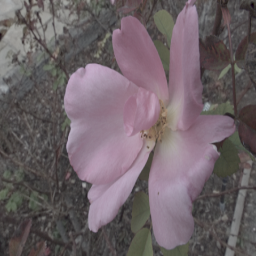} & 
    \includegraphics[width=0.24\linewidth]{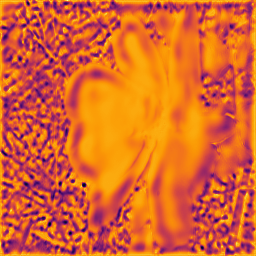} & 
    \includegraphics[width=0.24\linewidth]{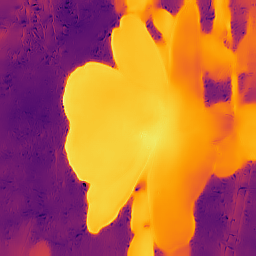} & 
    \includegraphics[width=0.24\linewidth]{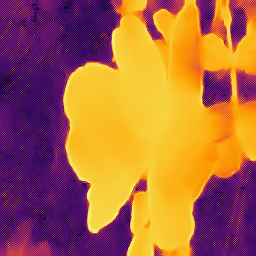} \\
    Input & Aperture & GaussPSF & \Name{}
    \end{tabular}
    \caption{\textbf{Depth from defocus dataset}. The first column is the all-in-focus input image. The second column shows results by Aperture~\shortcite{srinivasanApertureSupervisionMonocular2018}; the third column by GaussPSF~\shortcite{gurSingleImageDepth2019}, and the last column by \Name{}. The depth map by Aperture is noisy. GaussPSF predicts smoother depth.\Name{} predicts smoother depth and keeps more boundary details.}
    \label{fig:depth-from-defocus}
\end{figure}

\textbf{Depth Quality:} Tab.~\ref{tab:diff_LF} shows the quantitative evaluation results.
\Name{} outperforms all the previous works in all metrics, which shows that the more accurate blur renderer improves the learning process. 
The quantitative evaluation is measured on bokeh images and we show the qualitative results of the generated depth map in Fig.~\ref{fig:depth_quality} and ~\ref{fig:depth-from-defocus}. 
The depth map can either be obtained by direct optimization over an all-in-focus image and a bokeh image pair or by training a neural network to predict the depth based on a large-scale defocus dataset. 
The direct optimization over one-pair data can clearly show the depth quality supervised by the differentiable rendering layer, while the depth predicted by the trained neural network can illustrate the overall performance of the differentiable layer in the data-driven pipeline. 
  
As shown in Fig.~\ref{fig:depth_quality}, \Name{} can obtain the best quality depth image supervised by the defocus image as \Name{} is more accurate in terms of the lens blur physics. 
Fig.~\ref{fig:depth-from-defocus} further shows that \Name{} helps the neural networks learn to predict the best quality depth compared with related works. 

\begin{table}[t]
\centering
\caption{\textbf{Ablation study.} 
Correct handling of boundary occlusion improves the existing gathering-based depth form defocus performance.  
The proposed hierarchy SSIM loss further helps \Name{} learns better depth.
} \label{tab:diff_ablation}
\small
\begin{tabular}{l|ccc}
\shline
\textbf{Method}  & \textbf{RMSE} $\downarrow$  & \textbf{SSIM} $\uparrow$  & \textbf{PSNR} $\uparrow$  \\
\hline
w.o. occlusion & $0.0139 \pm 0.0034$  & $0.9729 \pm 0.0115$ & $37.4016 \pm 2.0212$ \\
w. occlusion  & $0.0136 \pm 0.0033$  &  $0.9740 \pm 0.0113$ & $37.5888 \pm 2.0473$ \\
\hline
L1 + Grad            & $0.0146 \pm 0.0036$  &  $0.9673 \pm 0.0136$ & $36.9398 \pm 2.0497$ \\ 
L1 + Grad+ SSIM      & $0.0136 \pm 0.0033$  &  $0.9740 \pm 0.0113$ & $37.5888 \pm 2.0473$  \\
L1 + Grad+ HSSIM     & $0.0123 \pm 0.0031$  &  $0.9770 \pm 0.0107$ & $38.4487 \pm 2.0902$ \\
\shline
\end{tabular}
\end{table}

\begin{figure}[t]
    \centering
    \includegraphics[width=0.24\linewidth]{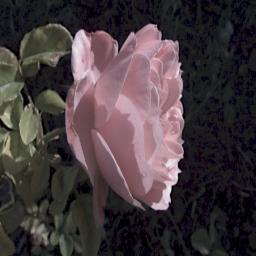}
    \includegraphics[width=0.24\linewidth]{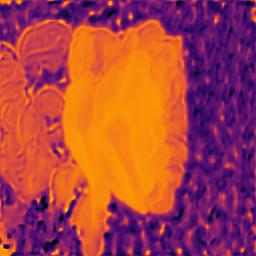}
    \includegraphics[width=0.24\linewidth]{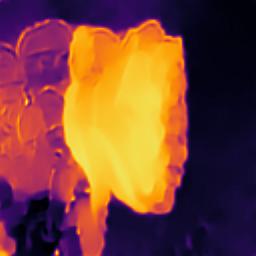}
    \includegraphics[width=0.24\linewidth]{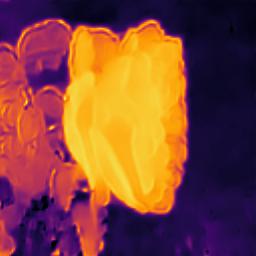}
    \caption{\textbf{Qualitative comparisons of different loss:} The first image is the RGB input. The second image is the result of L1 + Grad loss. The third image is the result of L1 + Grad + SSIM loss. The last image is the result of L1 + Grad + HSSIM loss.}
    \label{fig:eval_ablation_loss}
\end{figure}

\textbf{Ablation Study:} We conduct two experiments to understand the contribution of the occlusion term and the proposed hierarchy SSIM (HSSIM) loss (Sec.~\ref{sec.Diff}). 
We first compare \Name{} with a similar differentiable rendering layer but without the occlusion term by training on the \textit{light field} benchmark.
Evaluations (see Tab.~\ref{tab:diff_ablation}) on the benchmark show that the occlusion term helps the neural network training. 
Second, in the loss function experiment, we compare our loss function (Eqn.~\ref{eq.loss}) with two similar versions: one is just a $L1$ loss with the gradient loss, and the other is the $L1$ loss with the gradient loss and the SSIM loss. 
Note the gradient loss is a regularization loss. 
Tab.~\ref{tab:diff_ablation} shows our loss function outperforms other baselines. 
As shown in Fig.~\ref{fig:eval_ablation_loss}, the L1 + Grad loss makes the depth map relatively noisy. 
The L1 + Grad + SSIM makes the results smoother but loses some details. 
Our L1 + Grad + HSSIM gets a smooth depth map while preserving the boundary details.

\section{Conclusions}
\begin{figure}[t]
    \centering
    \includegraphics[width=0.49\linewidth]{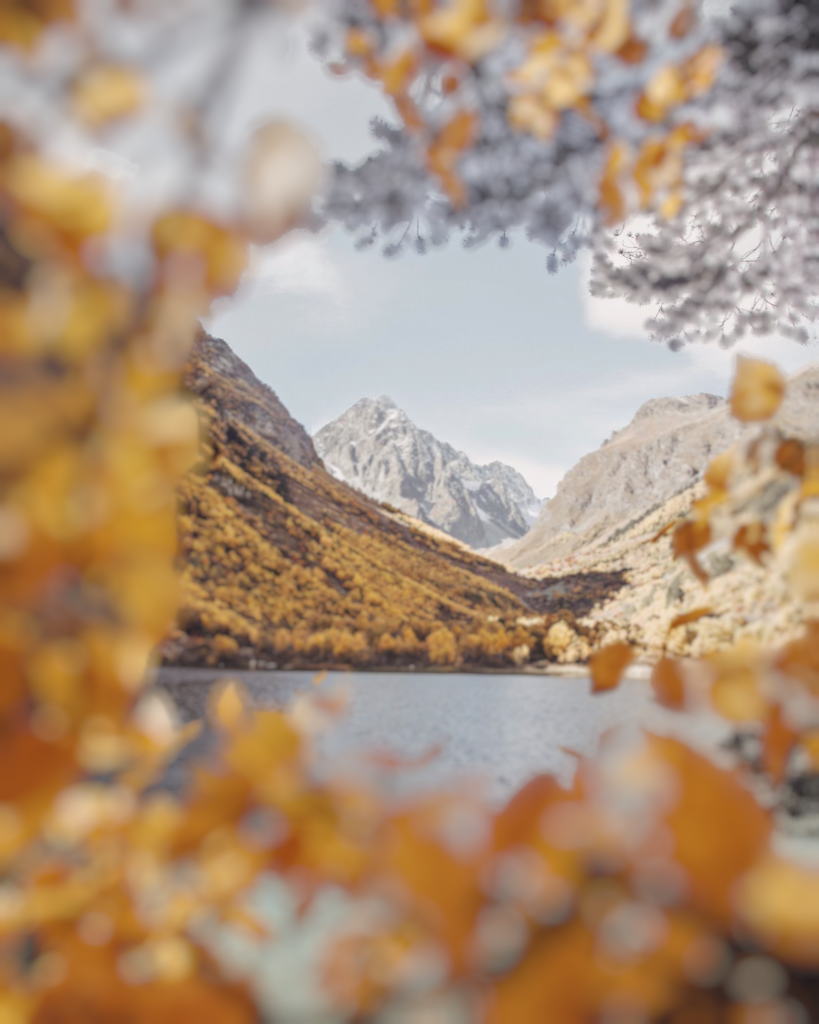}
    \includegraphics[width=0.49\linewidth, trim={0 0 0 4.3cm}, clip]{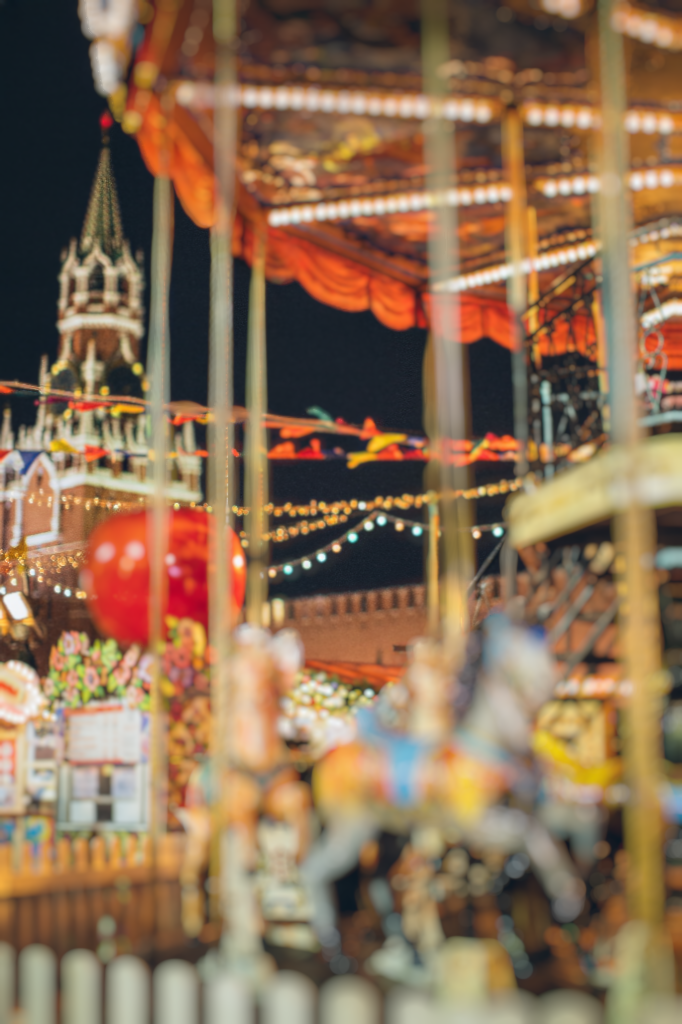}
    \caption{\textbf{Failure cases:} the salient object detection cannot find a good target for complex scenes, which leads to unnatural partial occlusions.}
    \label{fig:failure}
\end{figure}
We have introduced \Name{}, a novel physically-based differentiable and occlusion-aware DoF rendering algorithm. 
\Name{} addresses the color bleeding problem and renders realistic partial occlusion for DoF effect synthesis directly in the rendering stage without any training on the rendering. 
Moreover, \Name{} is a fully differentiable DoF rendering module that allows it to be plug-and-play in data-driven pipelines. Qualitative and quantitative comparisons validate that \Name{} achieves the state-of-the-art lens blur quality in different focus settings and the state-of-the-art depth quality in the depth-from-defocus community. 

\textit{Limitation and Future Work:} Our method has several limitations. 
First, \Name{} relies on the layered image inputs generated by other methods. 
Therefore, the layered image quality affects the final bokeh quality, as shown in Fig.~\ref{fig:failure}. 
Second, although the two-layer scene approximation works well in many cases shown in the qualitative evaluations, it fails when the two layers are not enough to represent a complex scene, as shown in Fig.~\ref{fig:failure}.
Third, lack of the synthesis of noise, which can often be observed in images captured by real-world devices like DSLR, the depth quality learned by \Name{} may be affected. 
Therefore, one of the future works is to propose a better scene understanding and robust layered scene representation. 
The differentiability of \Name{} may help this process. 
In addition, extending \Name{} with noise synthesis may further help the depth-from-defocus task.

\bibliographystyle{ACM-Reference-Format}
\bibliography{main}


\begin{thebibliography}{70}


\ifx \showCODEN    \undefined \def \showCODEN     #1{\unskip}     \fi
\ifx \showDOI      \undefined \def \showDOI       #1{#1}\fi
\ifx \showISBNx    \undefined \def \showISBNx     #1{\unskip}     \fi
\ifx \showISBNxiii \undefined \def \showISBNxiii  #1{\unskip}     \fi
\ifx \showISSN     \undefined \def \showISSN      #1{\unskip}     \fi
\ifx \showLCCN     \undefined \def \showLCCN      #1{\unskip}     \fi
\ifx \shownote     \undefined \def \shownote      #1{#1}          \fi
\ifx \showarticletitle \undefined \def \showarticletitle #1{#1}   \fi
\ifx \showURL      \undefined \def \showURL       {\relax}        \fi
\providecommand\bibfield[2]{#2}
\providecommand\bibinfo[2]{#2}
\providecommand\natexlab[1]{#1}
\providecommand\showeprint[2][]{arXiv:#2}

\bibitem[Busam et~al\mbox{.}(2019)]%
        {busamSteReFoEfficientImage2019}
\bibfield{author}{\bibinfo{person}{Benjamin Busam}, \bibinfo{person}{Matthieu
  Hog}, \bibinfo{person}{Steven McDonagh}, {and} \bibinfo{person}{Gregory
  Slabaugh}.} \bibinfo{year}{2019}\natexlab{}.
\newblock \showarticletitle{{SteReFo}: {Efficient} {Image} {Refocusing} with
  {Stereo} {Vision}}. In \bibinfo{booktitle}{\emph{2019 {IEEE}/{CVF}
  {International} {Conference} on {Computer} {Vision} {Workshop} ({ICCVW})}}.
  \bibinfo{publisher}{IEEE}, \bibinfo{address}{Seoul, Korea (South)},
  \bibinfo{pages}{3295--3304}.
\newblock
\showISBNx{978-1-72815-023-9}
\urldef\tempurl%
\url{https://doi.org/10.1109/ICCVW.2019.00411}
\showDOI{\tempurl}


\bibitem[Criminisi et~al\mbox{.}(2003)]%
        {criminisi2003object}
\bibfield{author}{\bibinfo{person}{Antonio Criminisi}, \bibinfo{person}{Patrick
  Perez}, {and} \bibinfo{person}{Kentaro Toyama}.}
  \bibinfo{year}{2003}\natexlab{}.
\newblock \showarticletitle{Object removal by exemplar-based inpainting}. In
  \bibinfo{booktitle}{\emph{2003 IEEE Computer Society Conference on Computer
  Vision and Pattern Recognition, 2003. Proceedings.}},
  Vol.~\bibinfo{volume}{2}. IEEE, \bibinfo{pages}{II--II}.
\newblock


\bibitem[Eslami et~al\mbox{.}(2018)]%
        {eslami2018neural}
\bibfield{author}{\bibinfo{person}{SM~Ali Eslami}, \bibinfo{person}{Danilo
  Jimenez~Rezende}, \bibinfo{person}{Frederic Besse}, \bibinfo{person}{Fabio
  Viola}, \bibinfo{person}{Ari~S Morcos}, \bibinfo{person}{Marta Garnelo},
  \bibinfo{person}{Avraham Ruderman}, \bibinfo{person}{Andrei~A Rusu},
  \bibinfo{person}{Ivo Danihelka}, \bibinfo{person}{Karol Gregor},
  {et~al\mbox{.}}} \bibinfo{year}{2018}\natexlab{}.
\newblock \showarticletitle{Neural scene representation and rendering}.
\newblock \bibinfo{journal}{\emph{Science}} \bibinfo{volume}{360},
  \bibinfo{number}{6394} (\bibinfo{year}{2018}), \bibinfo{pages}{1204--1210}.
\newblock


\bibitem[Franke et~al\mbox{.}(2018)]%
        {franke2018multi}
\bibfield{author}{\bibinfo{person}{Linus Franke}, \bibinfo{person}{Nikolai
  Hofmann}, \bibinfo{person}{Marc Stamminger}, {and} \bibinfo{person}{Kai
  Selgrad}.} \bibinfo{year}{2018}\natexlab{}.
\newblock \showarticletitle{Multi-layer depth of field rendering with tiled
  splatting}.
\newblock \bibinfo{journal}{\emph{Proceedings of the ACM on Computer Graphics
  and Interactive Techniques}} \bibinfo{volume}{1}, \bibinfo{number}{1}
  (\bibinfo{year}{2018}), \bibinfo{pages}{1--17}.
\newblock


\bibitem[Goodfellow et~al\mbox{.}(2020)]%
        {goodfellow2020generative}
\bibfield{author}{\bibinfo{person}{Ian Goodfellow}, \bibinfo{person}{Jean
  Pouget-Abadie}, \bibinfo{person}{Mehdi Mirza}, \bibinfo{person}{Bing Xu},
  \bibinfo{person}{David Warde-Farley}, \bibinfo{person}{Sherjil Ozair},
  \bibinfo{person}{Aaron Courville}, {and} \bibinfo{person}{Yoshua Bengio}.}
  \bibinfo{year}{2020}\natexlab{}.
\newblock \showarticletitle{Generative adversarial networks}.
\newblock \bibinfo{journal}{\emph{Commun. ACM}} \bibinfo{volume}{63},
  \bibinfo{number}{11} (\bibinfo{year}{2020}), \bibinfo{pages}{139--144}.
\newblock


\bibitem[G{\"o}ransson and Karlsson(2007)]%
        {goransson2007practical}
\bibfield{author}{\bibinfo{person}{Jhonny G{\"o}ransson} {and}
  \bibinfo{person}{Andreas Karlsson}.} \bibinfo{year}{2007}\natexlab{}.
\newblock \showarticletitle{Practical post-process depth of field}.
\newblock \bibinfo{journal}{\emph{GPU Gems}} \bibinfo{volume}{3},
  \bibinfo{number}{583-606} (\bibinfo{year}{2007}), \bibinfo{pages}{2}.
\newblock


\bibitem[Greivenkamp(2004)]%
        {greivenkamp2004field}
\bibfield{author}{\bibinfo{person}{John~E Greivenkamp}.}
  \bibinfo{year}{2004}\natexlab{}.
\newblock \bibinfo{booktitle}{\emph{Field guide to geometrical optics}}.
  Vol.~\bibinfo{volume}{1}.
\newblock \bibinfo{publisher}{SPIE press Bellingham, Washington}.
\newblock


\bibitem[Gur and Wolf(2019)]%
        {gurSingleImageDepth2019}
\bibfield{author}{\bibinfo{person}{Shir Gur} {and} \bibinfo{person}{Lior
  Wolf}.} \bibinfo{year}{2019}\natexlab{}.
\newblock \showarticletitle{Single {Image} {Depth} {Estimation} {Trained} via
  {Depth} {From} {Defocus} {Cues}}. In \bibinfo{booktitle}{\emph{2019
  {IEEE}/{CVF} {Conference} on {Computer} {Vision} and {Pattern} {Recognition}
  ({CVPR})}}. \bibinfo{publisher}{IEEE}, \bibinfo{address}{Long Beach, CA,
  USA}, \bibinfo{pages}{7675--7684}.
\newblock
\showISBNx{978-1-72813-293-8}
\urldef\tempurl%
\url{https://doi.org/10.1109/CVPR.2019.00787}
\showDOI{\tempurl}


\bibitem[Hays and Efros(2007)]%
        {hays2007scene}
\bibfield{author}{\bibinfo{person}{James Hays} {and} \bibinfo{person}{Alexei~A
  Efros}.} \bibinfo{year}{2007}\natexlab{}.
\newblock \showarticletitle{Scene completion using millions of photographs}.
\newblock \bibinfo{journal}{\emph{ACM Transactions on Graphics (ToG)}}
  \bibinfo{volume}{26}, \bibinfo{number}{3} (\bibinfo{year}{2007}),
  \bibinfo{pages}{4--es}.
\newblock


\bibitem[Ignatov et~al\mbox{.}(2020a)]%
        {ignatov2020rendering}
\bibfield{author}{\bibinfo{person}{Andrey Ignatov}, \bibinfo{person}{Jagruti
  Patel}, {and} \bibinfo{person}{Radu Timofte}.}
  \bibinfo{year}{2020}\natexlab{a}.
\newblock \showarticletitle{Rendering natural camera bokeh effect with deep
  learning}. In \bibinfo{booktitle}{\emph{Proceedings of the IEEE/CVF
  Conference on Computer Vision and Pattern Recognition Workshops}}.
  \bibinfo{pages}{418--419}.
\newblock


\bibitem[Ignatov et~al\mbox{.}(2020b)]%
        {ignatov2020aim}
\bibfield{author}{\bibinfo{person}{Andrey Ignatov}, \bibinfo{person}{Radu
  Timofte}, \bibinfo{person}{Ming Qian}, \bibinfo{person}{Congyu Qiao},
  \bibinfo{person}{Jiamin Lin}, \bibinfo{person}{Zhenyu Guo},
  \bibinfo{person}{Chenghua Li}, \bibinfo{person}{Cong Leng},
  \bibinfo{person}{Jian Cheng}, \bibinfo{person}{Juewen Peng}, {et~al\mbox{.}}}
  \bibinfo{year}{2020}\natexlab{b}.
\newblock \showarticletitle{AIM 2020 challenge on rendering realistic bokeh}.
  In \bibinfo{booktitle}{\emph{European Conference on Computer Vision}}.
  Springer, \bibinfo{pages}{213--228}.
\newblock


\bibitem[Iizuka et~al\mbox{.}(2017)]%
        {iizuka2017globally}
\bibfield{author}{\bibinfo{person}{Satoshi Iizuka}, \bibinfo{person}{Edgar
  Simo-Serra}, {and} \bibinfo{person}{Hiroshi Ishikawa}.}
  \bibinfo{year}{2017}\natexlab{}.
\newblock \showarticletitle{Globally and locally consistent image completion}.
\newblock \bibinfo{journal}{\emph{ACM Transactions on Graphics (ToG)}}
  \bibinfo{volume}{36}, \bibinfo{number}{4} (\bibinfo{year}{2017}),
  \bibinfo{pages}{1--14}.
\newblock


\bibitem[Jakob et~al\mbox{.}(2022)]%
        {jakob2022dr}
\bibfield{author}{\bibinfo{person}{Wenzel Jakob},
  \bibinfo{person}{S{\'e}bastien Speierer}, \bibinfo{person}{Nicolas Roussel},
  {and} \bibinfo{person}{Delio Vicini}.} \bibinfo{year}{2022}\natexlab{}.
\newblock \showarticletitle{DR. JIT: a just-in-time compiler for differentiable
  rendering}.
\newblock \bibinfo{journal}{\emph{ACM Transactions on Graphics (TOG)}}
  \bibinfo{volume}{41}, \bibinfo{number}{4} (\bibinfo{year}{2022}),
  \bibinfo{pages}{1--19}.
\newblock


\bibitem[Jeong et~al\mbox{.}(2020)]%
        {jeong2020real}
\bibfield{author}{\bibinfo{person}{Yuna Jeong}, \bibinfo{person}{Seung~Youp
  Baek}, \bibinfo{person}{Yechan Seok}, \bibinfo{person}{Gi~Beom Lee}, {and}
  \bibinfo{person}{Sungkil Lee}.} \bibinfo{year}{2020}\natexlab{}.
\newblock \showarticletitle{Real-time dynamic bokeh rendering with efficient
  look-up table sampling}.
\newblock \bibinfo{journal}{\emph{IEEE Transactions on Visualization and
  Computer Graphics}} \bibinfo{volume}{28}, \bibinfo{number}{2}
  (\bibinfo{year}{2020}), \bibinfo{pages}{1373--1384}.
\newblock


\bibitem[Kalantari et~al\mbox{.}(2016)]%
        {kalantariLearningbasedViewSynthesis2016}
\bibfield{author}{\bibinfo{person}{Nima~Khademi Kalantari},
  \bibinfo{person}{Ting-Chun Wang}, {and} \bibinfo{person}{Ravi Ramamoorthi}.}
  \bibinfo{year}{2016}\natexlab{}.
\newblock \showarticletitle{Learning-based view synthesis for light field
  cameras}.
\newblock \bibinfo{journal}{\emph{ACM Trans. Graph.}} \bibinfo{volume}{35},
  \bibinfo{number}{6} (\bibinfo{date}{Nov.} \bibinfo{year}{2016}),
  \bibinfo{pages}{1--10}.
\newblock
\showISSN{0730-0301, 1557-7368}
\urldef\tempurl%
\url{https://doi.org/10.1145/2980179.2980251}
\showDOI{\tempurl}


\bibitem[Kaneko(2021)]%
        {kanekoUnsupervisedLearningDepth2021}
\bibfield{author}{\bibinfo{person}{Takuhiro Kaneko}.}
  \bibinfo{year}{2021}\natexlab{}.
\newblock \showarticletitle{Unsupervised {Learning} of {Depth} and
  {Depth}-of-{Field} {Effect} from {Natural} {Images} with {Aperture}
  {Rendering} {Generative} {Adversarial} {Networks}}. In
  \bibinfo{booktitle}{\emph{2021 {IEEE}/{CVF} {Conference} on {Computer}
  {Vision} and {Pattern} {Recognition} ({CVPR})}}. \bibinfo{publisher}{IEEE},
  \bibinfo{address}{Nashville, TN, USA}, \bibinfo{pages}{15674--15683}.
\newblock
\showISBNx{978-1-66544-509-2}
\urldef\tempurl%
\url{https://doi.org/10.1109/CVPR46437.2021.01542}
\showDOI{\tempurl}


\bibitem[Kass et~al\mbox{.}(2006)]%
        {kass2006interactive}
\bibfield{author}{\bibinfo{person}{Michael Kass}, \bibinfo{person}{Aaron
  Lefohn}, {and} \bibinfo{person}{John~D Owens}.}
  \bibinfo{year}{2006}\natexlab{}.
\newblock \showarticletitle{Interactive depth of field using simulated
  diffusion on a GPU}.
\newblock  (\bibinfo{year}{2006}).
\newblock


\bibitem[Kato et~al\mbox{.}(2020)]%
        {kato2020differentiable}
\bibfield{author}{\bibinfo{person}{Hiroharu Kato}, \bibinfo{person}{Deniz
  Beker}, \bibinfo{person}{Mihai Morariu}, \bibinfo{person}{Takahiro Ando},
  \bibinfo{person}{Toru Matsuoka}, \bibinfo{person}{Wadim Kehl}, {and}
  \bibinfo{person}{Adrien Gaidon}.} \bibinfo{year}{2020}\natexlab{}.
\newblock \showarticletitle{Differentiable rendering: A survey}.
\newblock \bibinfo{journal}{\emph{arXiv preprint arXiv:2006.12057}}
  (\bibinfo{year}{2020}).
\newblock


\bibitem[Kraus and Strengert(2007)]%
        {krausDepthofFieldRenderingPyramidal2007}
\bibfield{author}{\bibinfo{person}{M. Kraus} {and} \bibinfo{person}{M.
  Strengert}.} \bibinfo{year}{2007}\natexlab{}.
\newblock \showarticletitle{Depth-of-{Field} {Rendering} by {Pyramidal} {Image}
  {Processing}}.
\newblock \bibinfo{journal}{\emph{Computer Graphics Forum}}
  \bibinfo{volume}{26}, \bibinfo{number}{3} (\bibinfo{year}{2007}),
  \bibinfo{pages}{645--654}.
\newblock
\showISSN{1467-8659}
\urldef\tempurl%
\url{https://doi.org/10.1111/j.1467-8659.2007.01088.x}
\showDOI{\tempurl}
\newblock
\shownote{\_eprint:
  https://onlinelibrary.wiley.com/doi/pdf/10.1111/j.1467-8659.2007.01088.x}.


\bibitem[Lee et~al\mbox{.}(2009)]%
        {lee2009depth}
\bibfield{author}{\bibinfo{person}{Sungkil Lee}, \bibinfo{person}{Elmar
  Eisemann}, {and} \bibinfo{person}{Hans-Peter Seidel}.}
  \bibinfo{year}{2009}\natexlab{}.
\newblock \showarticletitle{Depth-of-field rendering with multiview synthesis}.
\newblock \bibinfo{journal}{\emph{ACM Transactions on Graphics (TOG)}}
  \bibinfo{volume}{28}, \bibinfo{number}{5} (\bibinfo{year}{2009}),
  \bibinfo{pages}{1--6}.
\newblock


\bibitem[Lee et~al\mbox{.}(2010)]%
        {leeRealtimeLensBlur2010}
\bibfield{author}{\bibinfo{person}{Sungkil Lee}, \bibinfo{person}{Elmar
  Eisemann}, {and} \bibinfo{person}{Hans-Peter Seidel}.}
  \bibinfo{year}{2010}\natexlab{}.
\newblock \showarticletitle{Real-time lens blur effects and focus control}.
\newblock \bibinfo{journal}{\emph{ACM Trans. Graph.}} \bibinfo{volume}{29},
  \bibinfo{number}{4} (\bibinfo{date}{July} \bibinfo{year}{2010}),
  \bibinfo{pages}{1--7}.
\newblock
\showISSN{0730-0301, 1557-7368}
\urldef\tempurl%
\url{https://doi.org/10.1145/1778765.1778802}
\showDOI{\tempurl}


\bibitem[Lee et~al\mbox{.}(2008)]%
        {lee2008real}
\bibfield{author}{\bibinfo{person}{Sungkil Lee},
  \bibinfo{person}{Gerard~Jounghyun Kim}, {and} \bibinfo{person}{Seungmoon
  Choi}.} \bibinfo{year}{2008}\natexlab{}.
\newblock \showarticletitle{Real-time depth-of-field rendering using point
  splatting on per-pixel layers}. In \bibinfo{booktitle}{\emph{Computer
  Graphics Forum}}, Vol.~\bibinfo{volume}{27}. Wiley Online Library,
  \bibinfo{pages}{1955--1962}.
\newblock


\bibitem[Lei and Hughes(2013)]%
        {lei2013approximate}
\bibfield{author}{\bibinfo{person}{Kefei Lei} {and} \bibinfo{person}{John~F
  Hughes}.} \bibinfo{year}{2013}\natexlab{}.
\newblock \showarticletitle{Approximate depth of field effects using few
  samples per pixel}. In \bibinfo{booktitle}{\emph{Proceedings of the ACM
  SIGGRAPH Symposium on Interactive 3D Graphics and Games}}.
  \bibinfo{pages}{119--128}.
\newblock


\bibitem[Li et~al\mbox{.}(2021)]%
        {li2021deep}
\bibfield{author}{\bibinfo{person}{Jizhizi Li}, \bibinfo{person}{Jing Zhang},
  {and} \bibinfo{person}{Dacheng Tao}.} \bibinfo{year}{2021}\natexlab{}.
\newblock \showarticletitle{Deep automatic natural image matting}.
\newblock \bibinfo{journal}{\emph{arXiv preprint arXiv:2107.07235}}
  (\bibinfo{year}{2021}).
\newblock


\bibitem[Li et~al\mbox{.}(2018)]%
        {li2018differentiable}
\bibfield{author}{\bibinfo{person}{Tzu-Mao Li}, \bibinfo{person}{Miika
  Aittala}, \bibinfo{person}{Fr{\'e}do Durand}, {and} \bibinfo{person}{Jaakko
  Lehtinen}.} \bibinfo{year}{2018}\natexlab{}.
\newblock \showarticletitle{Differentiable monte carlo ray tracing through edge
  sampling}.
\newblock \bibinfo{journal}{\emph{ACM Transactions on Graphics (TOG)}}
  \bibinfo{volume}{37}, \bibinfo{number}{6} (\bibinfo{year}{2018}),
  \bibinfo{pages}{1--11}.
\newblock


\bibitem[Liu et~al\mbox{.}(2018)]%
        {liu2018image}
\bibfield{author}{\bibinfo{person}{Guilin Liu}, \bibinfo{person}{Fitsum~A
  Reda}, \bibinfo{person}{Kevin~J Shih}, \bibinfo{person}{Ting-Chun Wang},
  \bibinfo{person}{Andrew Tao}, {and} \bibinfo{person}{Bryan Catanzaro}.}
  \bibinfo{year}{2018}\natexlab{}.
\newblock \showarticletitle{Image inpainting for irregular holes using partial
  convolutions}. In \bibinfo{booktitle}{\emph{Proceedings of the European
  conference on computer vision (ECCV)}}. \bibinfo{pages}{85--100}.
\newblock


\bibitem[Liu et~al\mbox{.}(2020)]%
        {liu2020rethinking}
\bibfield{author}{\bibinfo{person}{Hongyu Liu}, \bibinfo{person}{Bin Jiang},
  \bibinfo{person}{Yibing Song}, \bibinfo{person}{Wei Huang}, {and}
  \bibinfo{person}{Chao Yang}.} \bibinfo{year}{2020}\natexlab{}.
\newblock \showarticletitle{Rethinking image inpainting via a mutual
  encoder-decoder with feature equalizations}. In
  \bibinfo{booktitle}{\emph{European Conference on Computer Vision}}. Springer,
  \bibinfo{pages}{725--741}.
\newblock


\bibitem[Liu et~al\mbox{.}(2019)]%
        {liu2019soft}
\bibfield{author}{\bibinfo{person}{Shichen Liu}, \bibinfo{person}{Tianye Li},
  \bibinfo{person}{Weikai Chen}, {and} \bibinfo{person}{Hao Li}.}
  \bibinfo{year}{2019}\natexlab{}.
\newblock \showarticletitle{Soft rasterizer: A differentiable renderer for
  image-based 3d reasoning}. In \bibinfo{booktitle}{\emph{Proceedings of the
  IEEE/CVF International Conference on Computer Vision}}.
  \bibinfo{pages}{7708--7717}.
\newblock


\bibitem[Liu and Rokne(2016)]%
        {liu2016depth}
\bibfield{author}{\bibinfo{person}{Xin Liu} {and} \bibinfo{person}{Jon~G
  Rokne}.} \bibinfo{year}{2016}\natexlab{}.
\newblock \showarticletitle{Depth of field synthesis from sparse views}.
\newblock \bibinfo{journal}{\emph{Computers \& Graphics}}  \bibinfo{volume}{55}
  (\bibinfo{year}{2016}), \bibinfo{pages}{21--32}.
\newblock


\bibitem[Lombardi et~al\mbox{.}(2019)]%
        {lombardi2019neural}
\bibfield{author}{\bibinfo{person}{Stephen Lombardi}, \bibinfo{person}{Tomas
  Simon}, \bibinfo{person}{Jason Saragih}, \bibinfo{person}{Gabriel Schwartz},
  \bibinfo{person}{Andreas Lehrmann}, {and} \bibinfo{person}{Yaser Sheikh}.}
  \bibinfo{year}{2019}\natexlab{}.
\newblock \showarticletitle{Neural volumes: Learning dynamic renderable volumes
  from images}.
\newblock \bibinfo{journal}{\emph{arXiv preprint arXiv:1906.07751}}
  (\bibinfo{year}{2019}).
\newblock


\bibitem[Luo et~al\mbox{.}(2020)]%
        {luoBokehRenderingDefocus2020}
\bibfield{author}{\bibinfo{person}{Xianrui Luo}, \bibinfo{person}{Juewen Peng},
  \bibinfo{person}{Ke Xian}, \bibinfo{person}{Zijin Wu}, {and}
  \bibinfo{person}{Zhiguo Cao}.} \bibinfo{year}{2020}\natexlab{}.
\newblock \showarticletitle{Bokeh {Rendering} from {Defocus} {Estimation}}.
\newblock In \bibinfo{booktitle}{\emph{Computer {Vision} – {ECCV} 2020
  {Workshops}}}, \bibfield{editor}{\bibinfo{person}{Adrien Bartoli} {and}
  \bibinfo{person}{Andrea Fusiello}} (Eds.). Vol.~\bibinfo{volume}{12537}.
  \bibinfo{publisher}{Springer International Publishing},
  \bibinfo{address}{Cham}, \bibinfo{pages}{245--261}.
\newblock
\showISBNx{978-3-030-67069-6 978-3-030-67070-2}
\urldef\tempurl%
\url{https://doi.org/10.1007/978-3-030-67070-2_15}
\showDOI{\tempurl}
\newblock
\shownote{Series Title: Lecture Notes in Computer Science}.


\bibitem[Nalbach et~al\mbox{.}(2017)]%
        {nalbachDeepShadingConvolutional2017}
\bibfield{author}{\bibinfo{person}{Oliver Nalbach}, \bibinfo{person}{Elena
  Arabadzhiyska}, \bibinfo{person}{Dushyant Mehta}, \bibinfo{person}{Hans-Peter
  Seidel}, {and} \bibinfo{person}{Tobias Ritschel}.}
  \bibinfo{year}{2017}\natexlab{}.
\newblock \showarticletitle{Deep {Shading}: {Convolutional} {Neural} {Networks}
  for {Screen}-{Space} {Shading}}.
\newblock \bibinfo{journal}{\emph{Computer Graphics Forum}}
  \bibinfo{volume}{36}, \bibinfo{number}{4} (\bibinfo{date}{July}
  \bibinfo{year}{2017}), \bibinfo{pages}{65--78}.
\newblock
\showISSN{01677055}
\urldef\tempurl%
\url{https://doi.org/10.1111/cgf.13225}
\showDOI{\tempurl}
\newblock
\shownote{arXiv:1603.06078 [cs]}.


\bibitem[Nazeri et~al\mbox{.}(2019)]%
        {nazeri2019edgeconnect}
\bibfield{author}{\bibinfo{person}{Kamyar Nazeri}, \bibinfo{person}{Eric Ng},
  \bibinfo{person}{Tony Joseph}, \bibinfo{person}{Faisal Qureshi}, {and}
  \bibinfo{person}{Mehran Ebrahimi}.} \bibinfo{year}{2019}\natexlab{}.
\newblock \showarticletitle{Edgeconnect: Structure guided image inpainting
  using edge prediction}. In \bibinfo{booktitle}{\emph{Proceedings of the
  IEEE/CVF International Conference on Computer Vision Workshops}}.
  \bibinfo{pages}{0--0}.
\newblock


\bibitem[Paszke et~al\mbox{.}(2019)]%
        {paszke2019pytorch}
\bibfield{author}{\bibinfo{person}{Adam Paszke}, \bibinfo{person}{Sam Gross},
  \bibinfo{person}{Francisco Massa}, \bibinfo{person}{Adam Lerer},
  \bibinfo{person}{James Bradbury}, \bibinfo{person}{Gregory Chanan},
  \bibinfo{person}{Trevor Killeen}, \bibinfo{person}{Zeming Lin},
  \bibinfo{person}{Natalia Gimelshein}, \bibinfo{person}{Luca Antiga},
  {et~al\mbox{.}}} \bibinfo{year}{2019}\natexlab{}.
\newblock \showarticletitle{Pytorch: An imperative style, high-performance deep
  learning library}.
\newblock \bibinfo{journal}{\emph{Advances in neural information processing
  systems}}  \bibinfo{volume}{32} (\bibinfo{year}{2019}).
\newblock


\bibitem[Pathak et~al\mbox{.}(2016)]%
        {pathak2016context}
\bibfield{author}{\bibinfo{person}{Deepak Pathak}, \bibinfo{person}{Philipp
  Krahenbuhl}, \bibinfo{person}{Jeff Donahue}, \bibinfo{person}{Trevor
  Darrell}, {and} \bibinfo{person}{Alexei~A Efros}.}
  \bibinfo{year}{2016}\natexlab{}.
\newblock \showarticletitle{Context encoders: Feature learning by inpainting}.
  In \bibinfo{booktitle}{\emph{Proceedings of the IEEE conference on computer
  vision and pattern recognition}}. \bibinfo{pages}{2536--2544}.
\newblock


\bibitem[Peng et~al\mbox{.}(2022a)]%
        {pengBokehMeWhenNeural2022}
\bibfield{author}{\bibinfo{person}{Juewen Peng}, \bibinfo{person}{Zhiguo Cao},
  \bibinfo{person}{Xianrui Luo}, \bibinfo{person}{Hao Lu}, \bibinfo{person}{Ke
  Xian}, {and} \bibinfo{person}{Jianming Zhang}.}
  \bibinfo{year}{2022}\natexlab{a}.
\newblock \showarticletitle{{BokehMe}: {When} {Neural} {Rendering} {Meets}
  {Classical} {Rendering}}. In \bibinfo{booktitle}{\emph{2022 {IEEE}/{CVF}
  {Conference} on {Computer} {Vision} and {Pattern} {Recognition} ({CVPR})}}.
  \bibinfo{publisher}{IEEE}, \bibinfo{address}{New Orleans, LA, USA},
  \bibinfo{pages}{16262--16271}.
\newblock
\showISBNx{978-1-66546-946-3}
\urldef\tempurl%
\url{https://doi.org/10.1109/CVPR52688.2022.01580}
\showDOI{\tempurl}


\bibitem[Peng et~al\mbox{.}(2022b)]%
        {pengMPIBMPIBasedBokeh2022}
\bibfield{author}{\bibinfo{person}{Juewen Peng}, \bibinfo{person}{Jianming
  Zhang}, \bibinfo{person}{Xianrui Luo}, \bibinfo{person}{Hao Lu},
  \bibinfo{person}{Ke Xian}, {and} \bibinfo{person}{Zhiguo Cao}.}
  \bibinfo{year}{2022}\natexlab{b}.
\newblock \showarticletitle{{MPIB}: {An} {MPI}-{Based} {Bokeh} {Rendering}
  {Framework} for {Realistic} {Partial} {Occlusion} {Effects}}. In
  \bibinfo{booktitle}{\emph{Computer {Vision} – {ECCV} 2022}}
  \emph{(\bibinfo{series}{Lecture {Notes} in {Computer} {Science}})},
  \bibfield{editor}{\bibinfo{person}{Shai Avidan}, \bibinfo{person}{Gabriel
  Brostow}, \bibinfo{person}{Moustapha Cissé}, \bibinfo{person}{Giovanni~Maria
  Farinella}, {and} \bibinfo{person}{Tal Hassner}} (Eds.).
  \bibinfo{publisher}{Springer Nature Switzerland}, \bibinfo{address}{Cham},
  \bibinfo{pages}{590--607}.
\newblock
\showISBNx{978-3-031-20068-7}
\urldef\tempurl%
\url{https://doi.org/10.1007/978-3-031-20068-7_34}
\showDOI{\tempurl}


\bibitem[Pharr et~al\mbox{.}(2016)]%
        {pharr2016physically}
\bibfield{author}{\bibinfo{person}{Matt Pharr}, \bibinfo{person}{Wenzel Jakob},
  {and} \bibinfo{person}{Greg Humphreys}.} \bibinfo{year}{2016}\natexlab{}.
\newblock \bibinfo{booktitle}{\emph{Physically based rendering: From theory to
  implementation}}.
\newblock \bibinfo{publisher}{Morgan Kaufmann}.
\newblock


\bibitem[Potmesil and Chakravarty(1981)]%
        {potmesil1981lens}
\bibfield{author}{\bibinfo{person}{Michael Potmesil} {and}
  \bibinfo{person}{Indranil Chakravarty}.} \bibinfo{year}{1981}\natexlab{}.
\newblock \showarticletitle{A lens and aperture camera model for synthetic
  image generation}.
\newblock \bibinfo{journal}{\emph{ACM SIGGRAPH Computer Graphics}}
  \bibinfo{volume}{15}, \bibinfo{number}{3} (\bibinfo{year}{1981}),
  \bibinfo{pages}{297--305}.
\newblock


\bibitem[Rokita(1996)]%
        {rokita1996generating}
\bibfield{author}{\bibinfo{person}{Przemyslaw Rokita}.}
  \bibinfo{year}{1996}\natexlab{}.
\newblock \showarticletitle{Generating depth of-field effects in virtual
  reality applications}.
\newblock \bibinfo{journal}{\emph{IEEE Computer Graphics and Applications}}
  \bibinfo{volume}{16}, \bibinfo{number}{2} (\bibinfo{year}{1996}),
  \bibinfo{pages}{18--21}.
\newblock


\bibitem[Scheuermann et~al\mbox{.}(2004)]%
        {scheuermann2004advanced}
\bibfield{author}{\bibinfo{person}{Thorsten Scheuermann} {et~al\mbox{.}}}
  \bibinfo{year}{2004}\natexlab{}.
\newblock \showarticletitle{Advanced depth of field}.
\newblock \bibinfo{journal}{\emph{GDC 2004}}  \bibinfo{volume}{8}
  (\bibinfo{year}{2004}).
\newblock


\bibitem[Sheng et~al\mbox{.}(2022)]%
        {sheng2022controllable}
\bibfield{author}{\bibinfo{person}{Yichen Sheng}, \bibinfo{person}{Yifan Liu},
  \bibinfo{person}{Jianming Zhang}, \bibinfo{person}{Wei Yin},
  \bibinfo{person}{A~Cengiz Oztireli}, \bibinfo{person}{He Zhang},
  \bibinfo{person}{Zhe Lin}, \bibinfo{person}{Eli Shechtman}, {and}
  \bibinfo{person}{Bedrich Benes}.} \bibinfo{year}{2022}\natexlab{}.
\newblock \showarticletitle{Controllable shadow generation using pixel height
  maps}. In \bibinfo{booktitle}{\emph{European Conference on Computer Vision}}.
  Springer, \bibinfo{pages}{240--256}.
\newblock


\bibitem[Sheng et~al\mbox{.}(2021)]%
        {sheng2021ssn}
\bibfield{author}{\bibinfo{person}{Yichen Sheng}, \bibinfo{person}{Jianming
  Zhang}, {and} \bibinfo{person}{Bedrich Benes}.}
  \bibinfo{year}{2021}\natexlab{}.
\newblock \showarticletitle{SSN: Soft shadow network for image compositing}. In
  \bibinfo{booktitle}{\emph{Proceedings of the IEEE/CVF Conference on Computer
  Vision and Pattern Recognition}}. \bibinfo{pages}{4380--4390}.
\newblock


\bibitem[Sheng et~al\mbox{.}(2023)]%
        {sheng2023pixht}
\bibfield{author}{\bibinfo{person}{Yichen Sheng}, \bibinfo{person}{Jianming
  Zhang}, \bibinfo{person}{Julien Philip}, \bibinfo{person}{Yannick
  Hold-Geoffroy}, \bibinfo{person}{Xin Sun}, \bibinfo{person}{He Zhang},
  \bibinfo{person}{Lu Ling}, {and} \bibinfo{person}{Bedrich Benes}.}
  \bibinfo{year}{2023}\natexlab{}.
\newblock \showarticletitle{PixHt-Lab: Pixel Height Based Light Effect
  Generation for Image Compositing}. In \bibinfo{booktitle}{\emph{Proceedings
  of the IEEE/CVF Conference on Computer Vision and Pattern Recognition}}.
  \bibinfo{pages}{16643--16653}.
\newblock


\bibitem[Srinivasan et~al\mbox{.}(2018)]%
        {srinivasanApertureSupervisionMonocular2018}
\bibfield{author}{\bibinfo{person}{Pratul~P. Srinivasan},
  \bibinfo{person}{Rahul Garg}, \bibinfo{person}{Neal Wadhwa},
  \bibinfo{person}{Ren Ng}, {and} \bibinfo{person}{Jonathan~T. Barron}.}
  \bibinfo{year}{2018}\natexlab{}.
\newblock \showarticletitle{Aperture {Supervision} for {Monocular} {Depth}
  {Estimation}}. In \bibinfo{booktitle}{\emph{2018 {IEEE}/{CVF} {Conference} on
  {Computer} {Vision} and {Pattern} {Recognition}}}. \bibinfo{publisher}{IEEE},
  \bibinfo{address}{Salt Lake City, UT}, \bibinfo{pages}{6393--6401}.
\newblock
\showISBNx{978-1-5386-6420-9}
\urldef\tempurl%
\url{https://doi.org/10.1109/CVPR.2018.00669}
\showDOI{\tempurl}


\bibitem[Srinivasan et~al\mbox{.}(2017)]%
        {srinivasan2017learning}
\bibfield{author}{\bibinfo{person}{Pratul~P Srinivasan},
  \bibinfo{person}{Tongzhou Wang}, \bibinfo{person}{Ashwin Sreelal},
  \bibinfo{person}{Ravi Ramamoorthi}, {and} \bibinfo{person}{Ren Ng}.}
  \bibinfo{year}{2017}\natexlab{}.
\newblock \showarticletitle{Learning to synthesize a 4D RGBD light field from a
  single image}. In \bibinfo{booktitle}{\emph{Proceedings of the IEEE
  International Conference on Computer Vision}}. \bibinfo{pages}{2243--2251}.
\newblock


\bibitem[Sun et~al\mbox{.}(2019)]%
        {sun2019single}
\bibfield{author}{\bibinfo{person}{Tiancheng Sun}, \bibinfo{person}{Jonathan~T
  Barron}, \bibinfo{person}{Yun-Ta Tsai}, \bibinfo{person}{Zexiang Xu},
  \bibinfo{person}{Xueming Yu}, \bibinfo{person}{Graham Fyffe},
  \bibinfo{person}{Christoph Rhemann}, \bibinfo{person}{Jay Busch},
  \bibinfo{person}{Paul~E Debevec}, {and} \bibinfo{person}{Ravi Ramamoorthi}.}
  \bibinfo{year}{2019}\natexlab{}.
\newblock \showarticletitle{Single image portrait relighting.}
\newblock \bibinfo{journal}{\emph{ACM Transactions on Graphics (TOG)}}
  \bibinfo{volume}{38}, \bibinfo{number}{4} (\bibinfo{year}{2019}),
  \bibinfo{pages}{79--1}.
\newblock


\bibitem[Suvorov et~al\mbox{.}(2022)]%
        {suvorov2022resolution}
\bibfield{author}{\bibinfo{person}{Roman Suvorov}, \bibinfo{person}{Elizaveta
  Logacheva}, \bibinfo{person}{Anton Mashikhin}, \bibinfo{person}{Anastasia
  Remizova}, \bibinfo{person}{Arsenii Ashukha}, \bibinfo{person}{Aleksei
  Silvestrov}, \bibinfo{person}{Naejin Kong}, \bibinfo{person}{Harshith Goka},
  \bibinfo{person}{Kiwoong Park}, {and} \bibinfo{person}{Victor Lempitsky}.}
  \bibinfo{year}{2022}\natexlab{}.
\newblock \showarticletitle{Resolution-robust large mask inpainting with
  fourier convolutions}. In \bibinfo{booktitle}{\emph{Proceedings of the
  IEEE/CVF Winter Conference on Applications of Computer Vision}}.
  \bibinfo{pages}{2149--2159}.
\newblock


\bibitem[Tewari et~al\mbox{.}(2020)]%
        {tewari2020state}
\bibfield{author}{\bibinfo{person}{Ayush Tewari}, \bibinfo{person}{Ohad Fried},
  \bibinfo{person}{Justus Thies}, \bibinfo{person}{Vincent Sitzmann},
  \bibinfo{person}{Stephen Lombardi}, \bibinfo{person}{Kalyan Sunkavalli},
  \bibinfo{person}{Ricardo Martin-Brualla}, \bibinfo{person}{Tomas Simon},
  \bibinfo{person}{Jason Saragih}, \bibinfo{person}{Matthias Nie{\ss}ner},
  {et~al\mbox{.}}} \bibinfo{year}{2020}\natexlab{}.
\newblock \showarticletitle{State of the art on neural rendering}. In
  \bibinfo{booktitle}{\emph{Computer Graphics Forum}},
  Vol.~\bibinfo{volume}{39}. Wiley Online Library, \bibinfo{pages}{701--727}.
\newblock


\bibitem[Tucker and Snavely(2020)]%
        {tuckerSingleViewViewSynthesis2020b}
\bibfield{author}{\bibinfo{person}{Richard Tucker} {and} \bibinfo{person}{Noah
  Snavely}.} \bibinfo{year}{2020}\natexlab{}.
\newblock \showarticletitle{Single-{View} {View} {Synthesis} {With}
  {Multiplane} {Images}}. In \bibinfo{booktitle}{\emph{2020 {IEEE}/{CVF}
  {Conference} on {Computer} {Vision} and {Pattern} {Recognition} ({CVPR})}}.
  \bibinfo{publisher}{IEEE}, \bibinfo{address}{Seattle, WA, USA},
  \bibinfo{pages}{548--557}.
\newblock
\showISBNx{978-1-72817-168-5}
\urldef\tempurl%
\url{https://doi.org/10.1109/CVPR42600.2020.00063}
\showDOI{\tempurl}


\bibitem[Vaidyanathan et~al\mbox{.}(2015)]%
        {vaidyanathan2015layered}
\bibfield{author}{\bibinfo{person}{Karthik Vaidyanathan},
  \bibinfo{person}{Jacob Munkberg}, \bibinfo{person}{Petrik Clarberg}, {and}
  \bibinfo{person}{Marco Salvi}.} \bibinfo{year}{2015}\natexlab{}.
\newblock \showarticletitle{Layered light field reconstruction for defocus
  blur}.
\newblock \bibinfo{journal}{\emph{ACM Transactions on Graphics (TOG)}}
  \bibinfo{volume}{34}, \bibinfo{number}{2} (\bibinfo{year}{2015}),
  \bibinfo{pages}{1--12}.
\newblock


\bibitem[Wadhwa et~al\mbox{.}(2018)]%
        {wadhwaSyntheticDepthoffieldSinglecamera2018}
\bibfield{author}{\bibinfo{person}{Neal Wadhwa}, \bibinfo{person}{Rahul Garg},
  \bibinfo{person}{David~E. Jacobs}, \bibinfo{person}{Bryan~E. Feldman},
  \bibinfo{person}{Nori Kanazawa}, \bibinfo{person}{Robert Carroll},
  \bibinfo{person}{Yair Movshovitz-Attias}, \bibinfo{person}{Jonathan~T.
  Barron}, \bibinfo{person}{Yael Pritch}, {and} \bibinfo{person}{Marc Levoy}.}
  \bibinfo{year}{2018}\natexlab{}.
\newblock \showarticletitle{Synthetic depth-of-field with a single-camera
  mobile phone}.
\newblock \bibinfo{journal}{\emph{ACM Trans. Graph.}} \bibinfo{volume}{37},
  \bibinfo{number}{4} (\bibinfo{date}{Aug.} \bibinfo{year}{2018}),
  \bibinfo{pages}{1--13}.
\newblock
\showISSN{0730-0301, 1557-7368}
\urldef\tempurl%
\url{https://doi.org/10.1145/3197517.3201329}
\showDOI{\tempurl}


\bibitem[Wang et~al\mbox{.}(2018)]%
        {wangDeepLensShallowDepth2018}
\bibfield{author}{\bibinfo{person}{Lijun Wang}, \bibinfo{person}{Xiaohui Shen},
  \bibinfo{person}{Jianming Zhang}, \bibinfo{person}{Oliver Wang},
  \bibinfo{person}{Zhe Lin}, \bibinfo{person}{Chih-Yao Hsieh},
  \bibinfo{person}{Sarah Kong}, {and} \bibinfo{person}{Huchuan Lu}.}
  \bibinfo{year}{2018}\natexlab{}.
\newblock \showarticletitle{{DeepLens}: {Shallow} {Depth} {Of} {Field} {From}
  {A} {Single} {Image}}.
\newblock \bibinfo{journal}{\emph{arXiv:1810.08100 [cs]}} (\bibinfo{date}{Oct.}
  \bibinfo{year}{2018}).
\newblock
\urldef\tempurl%
\url{http://arxiv.org/abs/1810.08100}
\showURL{%
\tempurl}
\newblock
\shownote{arXiv: 1810.08100}.


\bibitem[Wei et~al\mbox{.}(2020)]%
        {wei2020label}
\bibfield{author}{\bibinfo{person}{Jun Wei}, \bibinfo{person}{Shuhui Wang},
  \bibinfo{person}{Zhe Wu}, \bibinfo{person}{Chi Su}, \bibinfo{person}{Qingming
  Huang}, {and} \bibinfo{person}{Qi Tian}.} \bibinfo{year}{2020}\natexlab{}.
\newblock \showarticletitle{Label decoupling framework for salient object
  detection}. In \bibinfo{booktitle}{\emph{Proceedings of the IEEE/CVF
  conference on computer vision and pattern recognition}}.
  \bibinfo{pages}{13025--13034}.
\newblock


\bibitem[Weyand et~al\mbox{.}(2020)]%
        {weyand2020GLDv2}
\bibfield{author}{\bibinfo{person}{T. Weyand}, \bibinfo{person}{A. Araujo},
  \bibinfo{person}{B. Cao}, {and} \bibinfo{person}{J. Sim}.}
  \bibinfo{year}{2020}\natexlab{}.
\newblock \showarticletitle{{Google Landmarks Dataset v2 - A Large-Scale
  Benchmark for Instance-Level Recognition and Retrieval}}. In
  \bibinfo{booktitle}{\emph{Proc. CVPR}}.
\newblock


\bibitem[Wu et~al\mbox{.}(2013)]%
        {wu2013rendering}
\bibfield{author}{\bibinfo{person}{Jiaze Wu}, \bibinfo{person}{Changwen Zheng},
  \bibinfo{person}{Xiaohui Hu}, {and} \bibinfo{person}{Fanjiang Xu}.}
  \bibinfo{year}{2013}\natexlab{}.
\newblock \showarticletitle{Rendering realistic spectral bokeh due to lens
  stops and aberrations}.
\newblock \bibinfo{journal}{\emph{The Visual Computer}}  \bibinfo{volume}{29}
  (\bibinfo{year}{2013}), \bibinfo{pages}{41--52}.
\newblock


\bibitem[Xian et~al\mbox{.}(2020)]%
        {Xian_2020_CVPR}
\bibfield{author}{\bibinfo{person}{Ke Xian}, \bibinfo{person}{Jianming Zhang},
  \bibinfo{person}{Oliver Wang}, \bibinfo{person}{Long Mai},
  \bibinfo{person}{Zhe Lin}, {and} \bibinfo{person}{Zhiguo Cao}.}
  \bibinfo{year}{2020}\natexlab{}.
\newblock \showarticletitle{Structure-Guided Ranking Loss for Single Image
  Depth Prediction}. In \bibinfo{booktitle}{\emph{The IEEE/CVF Conference on
  Computer Vision and Pattern Recognition (CVPR)}}.
\newblock


\bibitem[Xiao et~al\mbox{.}(2018)]%
        {xiaoDeepFocusLearnedImage2018f}
\bibfield{author}{\bibinfo{person}{Lei Xiao}, \bibinfo{person}{Anton
  Kaplanyan}, \bibinfo{person}{Alexander Fix}, \bibinfo{person}{Matthew
  Chapman}, {and} \bibinfo{person}{Douglas Lanman}.}
  \bibinfo{year}{2018}\natexlab{}.
\newblock \showarticletitle{{DeepFocus}: learned image synthesis for
  computational displays}.
\newblock \bibinfo{journal}{\emph{ACM Trans. Graph.}} \bibinfo{volume}{37},
  \bibinfo{number}{6} (\bibinfo{date}{Dec.} \bibinfo{year}{2018}),
  \bibinfo{pages}{1--13}.
\newblock
\showISSN{0730-0301, 1557-7368}
\urldef\tempurl%
\url{https://doi.org/10.1145/3272127.3275032}
\showDOI{\tempurl}


\bibitem[Xu et~al\mbox{.}(2017)]%
        {xu2017deep}
\bibfield{author}{\bibinfo{person}{Ning Xu}, \bibinfo{person}{Brian Price},
  \bibinfo{person}{Scott Cohen}, {and} \bibinfo{person}{Thomas Huang}.}
  \bibinfo{year}{2017}\natexlab{}.
\newblock \showarticletitle{Deep image matting}. In
  \bibinfo{booktitle}{\emph{Proceedings of the IEEE conference on computer
  vision and pattern recognition}}. \bibinfo{pages}{2970--2979}.
\newblock


\bibitem[Xu et~al\mbox{.}(2014)]%
        {xu2014depth}
\bibfield{author}{\bibinfo{person}{Shibiao Xu}, \bibinfo{person}{Xing Mei},
  \bibinfo{person}{Weiming Dong}, \bibinfo{person}{Xun Sun},
  \bibinfo{person}{Xukun Shen}, {and} \bibinfo{person}{Xiaopeng Zhang}.}
  \bibinfo{year}{2014}\natexlab{}.
\newblock \showarticletitle{Depth of field rendering via adaptive recursive
  filtering}.
\newblock In \bibinfo{booktitle}{\emph{SIGGRAPH Asia 2014 Technical Briefs}}.
  \bibinfo{pages}{1--4}.
\newblock


\bibitem[Yan et~al\mbox{.}(2015)]%
        {yan2015fast}
\bibfield{author}{\bibinfo{person}{Ling-Qi Yan}, \bibinfo{person}{Soham~Uday
  Mehta}, \bibinfo{person}{Ravi Ramamoorthi}, {and} \bibinfo{person}{Fredo
  Durand}.} \bibinfo{year}{2015}\natexlab{}.
\newblock \showarticletitle{Fast 4D sheared filtering for interactive rendering
  of distribution effects}.
\newblock \bibinfo{journal}{\emph{ACM Transactions on Graphics (TOG)}}
  \bibinfo{volume}{35}, \bibinfo{number}{1} (\bibinfo{year}{2015}),
  \bibinfo{pages}{1--13}.
\newblock


\bibitem[Yang et~al\mbox{.}(2017)]%
        {yang2017high}
\bibfield{author}{\bibinfo{person}{Chao Yang}, \bibinfo{person}{Xin Lu},
  \bibinfo{person}{Zhe Lin}, \bibinfo{person}{Eli Shechtman},
  \bibinfo{person}{Oliver Wang}, {and} \bibinfo{person}{Hao Li}.}
  \bibinfo{year}{2017}\natexlab{}.
\newblock \showarticletitle{High-resolution image inpainting using multi-scale
  neural patch synthesis}. In \bibinfo{booktitle}{\emph{Proceedings of the IEEE
  conference on computer vision and pattern recognition}}.
  \bibinfo{pages}{6721--6729}.
\newblock


\bibitem[Yang et~al\mbox{.}(2016)]%
        {yangVirtualDSLRHigh2016}
\bibfield{author}{\bibinfo{person}{Yang Yang}, \bibinfo{person}{Haiting Lin},
  \bibinfo{person}{Zhan Yu}, \bibinfo{person}{Sylvain Paris}, {and}
  \bibinfo{person}{Jingyi Yu}.} \bibinfo{year}{2016}\natexlab{}.
\newblock \showarticletitle{Virtual {DSLR}: {High} {Quality} {Dynamic}
  {Depth}-of-{Field} {Synthesis} on {Mobile} {Platforms}}.
\newblock \bibinfo{journal}{\emph{ei}} \bibinfo{volume}{28},
  \bibinfo{number}{18} (\bibinfo{date}{Feb.} \bibinfo{year}{2016}),
  \bibinfo{pages}{1--9}.
\newblock
\showISSN{2470-1173}
\urldef\tempurl%
\url{https://doi.org/10.2352/ISSN.2470-1173.2016.18.DPMI-031}
\showDOI{\tempurl}


\bibitem[Yu et~al\mbox{.}(2019)]%
        {yu2019free}
\bibfield{author}{\bibinfo{person}{Jiahui Yu}, \bibinfo{person}{Zhe Lin},
  \bibinfo{person}{Jimei Yang}, \bibinfo{person}{Xiaohui Shen},
  \bibinfo{person}{Xin Lu}, {and} \bibinfo{person}{Thomas~S Huang}.}
  \bibinfo{year}{2019}\natexlab{}.
\newblock \showarticletitle{Free-form image inpainting with gated convolution}.
  In \bibinfo{booktitle}{\emph{Proceedings of the IEEE/CVF international
  conference on computer vision}}. \bibinfo{pages}{4471--4480}.
\newblock


\bibitem[Zhang et~al\mbox{.}(2019b)]%
        {zhang2019depth}
\bibfield{author}{\bibinfo{person}{Benxuan Zhang}, \bibinfo{person}{Bin Sheng},
  \bibinfo{person}{Ping Li}, {and} \bibinfo{person}{Tong-Yee Lee}.}
  \bibinfo{year}{2019}\natexlab{b}.
\newblock \showarticletitle{Depth of field rendering using
  multilayer-neighborhood optimization}.
\newblock \bibinfo{journal}{\emph{IEEE Transactions on Visualization and
  Computer Graphics}} \bibinfo{volume}{26}, \bibinfo{number}{8}
  (\bibinfo{year}{2019}), \bibinfo{pages}{2546--2559}.
\newblock


\bibitem[Zhang et~al\mbox{.}(2020)]%
        {zhang2020path}
\bibfield{author}{\bibinfo{person}{Cheng Zhang}, \bibinfo{person}{Bailey
  Miller}, \bibinfo{person}{Kan Yan}, \bibinfo{person}{Ioannis Gkioulekas},
  {and} \bibinfo{person}{Shuang Zhao}.} \bibinfo{year}{2020}\natexlab{}.
\newblock \showarticletitle{Path-space differentiable rendering}.
\newblock \bibinfo{journal}{\emph{ACM transactions on graphics}}
  \bibinfo{volume}{39}, \bibinfo{number}{4} (\bibinfo{year}{2020}).
\newblock


\bibitem[Zhang et~al\mbox{.}(2019c)]%
        {zhang2019differential}
\bibfield{author}{\bibinfo{person}{Cheng Zhang}, \bibinfo{person}{Lifan Wu},
  \bibinfo{person}{Changxi Zheng}, \bibinfo{person}{Ioannis Gkioulekas},
  \bibinfo{person}{Ravi Ramamoorthi}, {and} \bibinfo{person}{Shuang Zhao}.}
  \bibinfo{year}{2019}\natexlab{c}.
\newblock \showarticletitle{A differential theory of radiative transfer}.
\newblock \bibinfo{journal}{\emph{ACM Transactions on Graphics (TOG)}}
  \bibinfo{volume}{38}, \bibinfo{number}{6} (\bibinfo{year}{2019}),
  \bibinfo{pages}{1--16}.
\newblock


\bibitem[Zhang et~al\mbox{.}(2019a)]%
        {zhangSyntheticDefocusLookahead2019}
\bibfield{author}{\bibinfo{person}{Xuaner Zhang}, \bibinfo{person}{Kevin
  Matzen}, \bibinfo{person}{Vivien Nguyen}, \bibinfo{person}{Dillon Yao},
  \bibinfo{person}{You Zhang}, {and} \bibinfo{person}{Ren Ng}.}
  \bibinfo{year}{2019}\natexlab{a}.
\newblock \showarticletitle{Synthetic defocus and look-ahead autofocus for
  casual videography}.
\newblock \bibinfo{journal}{\emph{ACM Trans. Graph.}} \bibinfo{volume}{38},
  \bibinfo{number}{4} (\bibinfo{date}{Aug.} \bibinfo{year}{2019}),
  \bibinfo{pages}{1--16}.
\newblock
\showISSN{0730-0301, 1557-7368}
\urldef\tempurl%
\url{https://doi.org/10.1145/3306346.3323015}
\showDOI{\tempurl}


\bibitem[Zhao et~al\mbox{.}(2020)]%
        {zhao2020physics}
\bibfield{author}{\bibinfo{person}{Shuang Zhao}, \bibinfo{person}{Wenzel
  Jakob}, {and} \bibinfo{person}{Tzu-Mao Li}.} \bibinfo{year}{2020}\natexlab{}.
\newblock \showarticletitle{Physics-based differentiable rendering: from theory
  to implementation}.
\newblock In \bibinfo{booktitle}{\emph{ACM siggraph 2020 courses}}.
  \bibinfo{pages}{1--30}.
\newblock


\bibitem[Zhu et~al\mbox{.}(2021)]%
        {zhu2021image}
\bibfield{author}{\bibinfo{person}{Manyu Zhu}, \bibinfo{person}{Dongliang He},
  \bibinfo{person}{Xin Li}, \bibinfo{person}{Chao Li}, \bibinfo{person}{Fu Li},
  \bibinfo{person}{Xiao Liu}, \bibinfo{person}{Errui Ding}, {and}
  \bibinfo{person}{Zhaoxiang Zhang}.} \bibinfo{year}{2021}\natexlab{}.
\newblock \showarticletitle{Image inpainting by end-to-end cascaded refinement
  with mask awareness}.
\newblock \bibinfo{journal}{\emph{IEEE Transactions on Image Processing}}
  \bibinfo{volume}{30} (\bibinfo{year}{2021}), \bibinfo{pages}{4855--4866}.
\newblock


\end{thebibliography}

\section{Appendix}
\subsection{Differentiabiliy}
Here are all the terms related to the derivatives in the Eqns.~\ref{Eq. diff_I}, ~\ref{Eq. diff_d} and ~\ref{Eq. diff_a}.

\begin{multline}
    \frac{\partial w(x, y)}{\partial d(x)} = \frac{A(x, y) K(x) a(x) \frac{\partial S(x, y)}{\partial d(x)} - S(x, y) K(x) a(x) \frac{\partial A(x)}{\partial d}}{A(x,y)^2} 
\end{multline}

where $\frac{\partial S(x, y)}{\partial d(x)}$ and $\frac{\partial S(x, y)}{\partial d(x)}$ are the following in practice: 
\begin{equation}
    \frac{\partial S(x, y)}{\partial d(x)} = \frac{0.3 e^{3(-(d(x)-d(y)))}}{(e^{3(-(d(x)-d(y)))} + 0.1)^2}
\end{equation}

\begin{multline}
    \frac{\partial O(x, y)}{\partial d(x)} = e^{-3 d(x)^2} \frac{-20}{(10(d(y)-d(x)- 0.1))^2} + \\ (- 0.5 - \tanh(10(d(y)-d(x)-0.1)))(-e^{-3d(x)^2}(-6 |d(x)| \text{sign}(y))) 
\end{multline}

\begin{equation}
    \frac{\partial w(x, y)}{\partial a(x)} = \frac{S(y, x) K(y)}{A(y)}  
\end{equation}

As mentioned in the paper, directly deriving and implementing the backward computation of \Name{} is complicated so we will release our code for reproducibility.

\subsection{Ray Tracing for Synthetic Benchmark}
Different rays hitting different points and collecting different colors is the main mechanism creating blurriness. The thin lens approximation equation ($1/f = 1/D_B + 1/D_I$) gives the recipe for computing the distance of an image $D_B$ to be in focus when the image plane $D_I$ is fixed.  Furthermore, suppose an image was placed at a different distance $D_B'$ (requiring a different image plane $D_I'$ to focus). In that case, the following equation gives the radius of circle-of-confusion ($r_c$) on the image plane: 
$$
r_c = R_a \frac{|D_I - D_I'|}{D_I'}
$$
This equation can also compute the distance $D_B'$ required for a blur of radius $r_c$. When the image billboards $I_1, I_2, \dots I_{n}$ are added to the scene to render, each image~$I_i$ is associated with a distance $D_i$ to the lens according to the amount of desired blurriness. Each image is then texture mapped to a rectangle that covers exactly the camera field of view when placed at distance $D_i$ (see Fig~\ref{fig:ray-lens-refraction}).  The vertices of the image billboards are therefore aligned on the sides of a pyramid in the scene.  If the aperture is small enough, the rendered image would be sharp as the camera is close to a pinhole camera.  When the lens has some size, the circle of confusion enlarges, which renders out-of-focus image billboards blurry.

\end{document}